\documentclass[12pt,a4paper]{article}
\usepackage[utf8]{inputenc}
\usepackage{authblk}
\usepackage[top=25truemm,bottom=25truemm,left=25truemm,right=25truemm,footskip=10truemm]{geometry}
\usepackage{booktabs}
\usepackage{amsmath}
\usepackage{amsthm}
\usepackage{float}
\usepackage{amssymb}
\usepackage{amsfonts}
\usepackage{scalefnt}
\usepackage{here}
\usepackage{indentfirst}
\usepackage{algorithmic}
\usepackage{algorithm}
\usepackage{setspace}
\usepackage{color}
\usepackage{enumitem}
\usepackage{graphicx}
\usepackage{subcaption}
\usepackage{lscape}
\usepackage{natbib}
\usepackage{hyperref}
\usepackage{cleveref}  
\usepackage{placeins}
\usepackage{tikz}
\usepackage{titlesec}
\usepackage{multirow}
\usepackage[normalem]{ulem}

\newtheorem{theorem}{Theorem}[section]
\newtheorem{proposition}[theorem]{Proposition}
\newtheorem{corollary}[theorem]{Corollary}
\newtheorem{lemma}[theorem]{Lemma}

\newtheorem{remark}{Remark}[section]

\newcommand{\PG}{\operatorname{PG}}
\newcommand{\expit}{\operatorname{expit}}
\newcommand{\dd}{\mathop{}\!\mathrm{d}}

\title{Efficient Gibbs Sampling in Cox Regression Models Using Composite Partial Likelihood and P\'olya-Gamma Augmentation}
\author[1]{Shu Tamano}
\author[1,*]{Yui Tomo}
\affil[1]{Department of Epidemiology, National Institute of Infectious Diseases, Japan Institute for Health Security, 1-23-1 Toyama, Shinjuku-Ku, Tokyo 162-0052, Japan}
\affil[*]{E-mail: tomoy@niid.go.jp}
\date{}

\begin{document}
\maketitle

\begin{abstract}
\noindent
The Cox regression models and their Bayesian extensions are widely used for time-to-event analysis. 
However, standard Bayesian approaches typically require baseline hazard modeling, and their full conditional distributions lack closed-form expressions, resulting in computational inefficiency and increased vulnerability to bias from baseline hazard misspecification.
To address these issues, we propose GS4Cox, a fully Gibbs sampler for Bayesian Cox regression models with four elements: (i) generalized Bayesian framework for avoiding baseline hazard specification, (ii) composite partial likelihood and (iii) P\'olya-Gamma augmentation for closed-form expressions of full conditional distributions, and (iv) affine posterior calibration via the open-faced sandwich adjustment for location and scale adjustment of the posterior distribution.
We prove asymptotic unbiasedness of the generalized Bayes estimator under composite partial likelihood and propose an affine posterior transformation that yields higher-order asymptotic agreement with the maximum partial likelihood estimator, while the posterior covariance matches the asymptotic target covariance.
We demonstrated that GS4Cox consistently outperformed existing sampling methods through numerical and real-data experiments.
\end{abstract}


\section{Introduction}
\label{sec:intro}

In time-to-event data analysis, the Cox regression model or the Cox proportional hazards model \citep{cox1972regression} is commonly used in various fields, such as medical research and quality assurance.
Parameter estimation is typically performed based on the partial likelihood function, which allows estimation without specifying the baseline hazard \citep{cox1975partial}. 
While the partial likelihood approach provides a useful framework for statistical inference on the covariate effects, it has limitations when applied to complex data structures such as hierarchical dependencies.
Moreover, it relies on point estimates and asymptotic distributions, which may be inadequate in the presence of high model uncertainty, such as in high-dimensional settings.
Bayesian extensions can provide a more flexible modeling framework for addressing hierarchical and other complex data structures by incorporating prior knowledge or assumptions on the parameters as priors \citep{ibrahim2001bayesian, chen2006posterior}.
These Bayesian Cox models have been studied under a variety of priors \citep{ibrahim1998prior, wu2018assessing, muse2022flexible}, and applied to large-scale datasets \citep{jung2023bayesian}. 
Furthermore, a systematic review has been conducted on their use in clinical trials \citep{brard2016bayesian}.

\citet{bissiri2016general} proposed a framework of general Bayesian inference, which allows using general loss functions instead of the full likelihood functions in Bayesian updating. 
This framework enables Bayesian approaches to clustering and robust estimation \citep{hashimoto2020robust,wade2023bayesian,rigon2023generalized,yonekura2023adaptation,chakraborty2024gibbs}.
The general Bayesian update also justifies the posterior distribution in the Cox regression model using the partial likelihood function instead of the full likelihood function \citep{mcgree2023general}. 

In the estimation process of Bayesian models, sampling algorithms such as Markov chain Monte Carlo (MCMC) methods are typically employed to draw samples from posterior distributions \citep{robert1999monte}.
The Gibbs sampling method is one of the simplest algorithms, in which each parameter is sampled conditionally on the others, allowing for efficient sampling when full conditional distributions are available in closed form \citep{geman1984stochastic,gelfand1990sampling}. 
The Metropolis-Hastings (MH) algorithm is a more general approach that constructs a Markov chain using a proposal distribution and an acceptance criterion, enabling sampling from posterior distributions even when full conditional distributions cannot be analytically obtained \citep{metropolis1953equation,hastings1970monte}. 
Furthermore, gradient-informed samplers have recently become standard in MCMC.
The Metropolis-adjusted Langevin algorithm (MALA) constructs proposals by a Langevin diffusion process using the gradient of the log-posterior and then applies a Metropolis-Hastings correction \citep{roberts1996exponential}.
Hamiltonian Monte Carlo (HMC) employs Hamiltonian dynamics to generate distant proposals that are accepted with high probability, and the No-U-Turn sampler (NUTS) adaptively selects HMC trajectory lengths \citep{duane1987hybrid, hoffman2014no}.
\cite{polson2013bayesian} proposed a P\'olya-Gamma augmentation scheme for logistic and negative binomial regression models.
In this scheme, when the priors are normal, the complete conditionals of the parameters can also be obtained as normal distributions.
Owing to this scheme, an efficient Gibbs sampling algorithm for logistic and negative binomial regression models can be developed.
Moreover, \cite{linderman2015dependent} showed that the P\'olya-Gamma augmentation scheme can be applied to the multinomial logit models with a stick-breaking expression of the likelihood.
\cite{neelon2019bayesian} proposed an efficient sampling algorithm for the zero-inflated negative binomial regression models based on the P\'olya-Gamma augmentation.
Furthermore, 
\citet{ren2025cox} proposed the Cox-P\'olya-Gamma (Cox-PG) algorithm, which uses the P\'olya-Gamma augmentation scheme in the sampling from the Cox regression model with the full likelihood function.
Their sampling algorithm uses a monotone spline approximation of the log-cumulative baseline hazard, P\'olya-Gamma augmentation with a negative-binomial approximation to the underlying Poisson process, and Metropolis-Hastings corrections to control the resulting bias.
The Cox-PG algorithm necessarily trades off sampling efficiency against approximation error.

In this study, we develop GS4Cox, an efficient fully Gibbs sampling algorithm for Cox regression built on four components:
(i) general Bayesian framework, (ii) composite partial likelihood, (iii) P\'olya-Gamma augmentation scheme, and (iv) affine posterior calibration.
First, we define the composite partial likelihood and show that the estimating function defined as the first derivative of the log-composite partial likelihood has expectation zero under Cox regression models, and that the estimator derived from this estimating function has weak consistency.
The general Bayesian framework using the composite partial likelihood as the loss function and the P\'olya-Gamma augmentation yields Gaussian full conditionals for the regression coefficients, enabling closed-form Gibbs updates.
To maintain comparability with the conventional partial-likelihood estimator, we then establish its asymptotic relationship with our composite partial likelihood estimator and derive an affine posterior calibration based on the open-faced sandwich adjustment \citep{shaby2014open-faced}.
This procedure adjusts both the location and the scale of the posterior distribution so that the resulting Bayes estimator has higher-order asymptotic agreement with the maximum partial likelihood estimator and the posterior covariance matches the sandwich limit.
Our GS4Cox algorithm requires no explicit baseline hazard modeling and no delicate tuning of proposals or step sizes; fully closed-form Gibbs updates deliver high sampling efficiency, and the affine calibration removes tuning parameter sensitivity.

The rest of the paper is organized as follows. 
In Section~\ref{sec:related-work}, we provide a review of the Cox regression model, the general Bayesian framework, and the P\'olya-Gamma augmentation scheme as applied to Cox regression.
In Section~\ref{sec:proposed-method}, we propose the statistical properties of our composite partial likelihood and present details of the proposed Gibbs sampling algorithm.
In Section~\ref{sec:num-experiments}, we perform numerical experiments.
In Section~\ref{sec:actual-experiment}, we apply the proposed method to real-world data.
Finally, in Section~\ref{sec:discussion}, we conclude the paper with a discussion.

\section{Related Work}
\label{sec:related-work}

In this section, we briefly review the Cox regression model and its general Bayesian formulation.
We then present the P\'olya-Gamma augmentation scheme for the logistic regression model.
Finally, we introduce a recent study by \cite{ren2025cox}, which applied the P\'olya-Gamma augmentation to construct an MCMC algorithm for Cox regression models.

\subsection{Cox Regression Models}
\label{subsec:cox-regression}

Assume that, for each subject $i\in \{1, \dots, n\}$, we observe the dataset $\mathcal{D}=\{\mathcal{D}_{i}\}_{i=1}^n=\{(T_i, \delta_i, \boldsymbol{X}_i)\}_{i=1}^n$, where $T_i\in \mathbb{R}$ denotes the observed time, defined as $T_i = \min(T_i^{\ast}, C_i^{\ast})$.
Here, $T_i^{\ast} \in \mathbb{R}$ represents the event time, and $C_i^{\ast} \in \mathbb{R}$ represents the censoring time.
The event indicator $\delta_i \in \{0, 1\}$ is defined as $\delta_i = \boldsymbol{1}\{T_i^{\ast}\leq C_i^{\ast}\}$, where $\boldsymbol{1}\{\cdot\}$ denotes the indicator function, taking the value $1$ if the condition inside the braces holds, and $0$ otherwise.
Additionally, $\boldsymbol{X}_i \in \mathbb{R}^p$ is the $p$-dimensional covariate vector.
We assume non-informative censoring, meaning that $T_i^{\ast}$ is conditionally independent of $C_i^{\ast}$ conditioned on $\boldsymbol{X}_i$ (i.e., $T_i^{\ast} \perp\!\!\!\perp C_i^{\ast}\mid \boldsymbol{X}_i$).

In the Cox regression model \citep{cox1972regression}, the hazard function for subject $i$ at time $t$ is given by $h(t\mid \boldsymbol{X}_i)=h_0(t)\exp(\boldsymbol{X}_i^\top\boldsymbol{\beta})$ where $\boldsymbol{\beta}\in\mathbb{R}^p$ is a $p$-dimensional vector of regression parameters, and $h_0(t)$ is a baseline hazard function.
For the dataset $\mathcal{D}$, the full likelihood is given as follows:
\begin{equation}
    L\left(\boldsymbol{\beta} \mid \mathcal{D}\right) = \prod_{i=1}^n \bigl\{\exp(\boldsymbol{X}_i^\top\boldsymbol{\beta})h_0(T_i)\bigr\}^{\delta_i}\exp\left\{-\exp(\boldsymbol{X}_i^\top\boldsymbol{\beta})\int_0^{T_i} h_0(\tau)\dd\tau\right\}
    .
    \label{eq:cox-likelihood}
\end{equation}
However, direct estimation of $\boldsymbol{\beta}$ is challenging due to the infinite dimensionality of $h_0(t)$.
To overcome this, \cite{cox1975partial} introduced the partial likelihood, which focuses on the relative hazard contributions of individuals at observed event times.
Define the risk set as the set-valued mapping $R: [0, \max_{1\le i \le n} T_{i}] \to 2^{\{1,\dots, n\}}$ and $R(t) := \{j\in \{1,\dots, n\}: T_{j} \ge t\}$.
Then, the conditional probability that individual $i$ fails at time $t$ is given by:
\begin{equation*}
    \Pr(i\text{ fails at }t\bigm|\text{a failure occurs at }t) = \frac{h_0(t)\exp(\boldsymbol{X}_i^\top\boldsymbol{\beta})}{\sum_{j\in R(t)}h_0(t)\exp(\boldsymbol{X}_j^\top\boldsymbol{\beta})}
    .
\end{equation*}
The baseline hazard $h_0(t)$ cancels out in this conditional probability.
Thus, the partial likelihood $L_{\mathrm{PL}}$ is formed by taking the product of these probabilities over all observed events, yielding
\begin{equation*}
    L_{\mathrm{PL}}(\boldsymbol{\beta}\mid \mathcal{D})
    := \prod_{i=1}^n \left\{\frac{\exp(\boldsymbol{X}_i^\top \boldsymbol{\beta})} {\sum_{j\in R(T_i)} \exp(\boldsymbol{X}_j^\top \boldsymbol{\beta})}\right\}^{\delta_i}
    .
\end{equation*}

\subsection{General Bayesian Framework in Cox Regression Models}
\label{subsec:general-bayes-cox}

In the standard Bayesian update, the posterior distribution for the parameter $\boldsymbol{\beta}$ is given by $\pi(\boldsymbol{\beta} \mid \mathcal{D}) \propto \pi(\boldsymbol{\beta}) \times \prod_{i=1}^n f(\mathcal{D}_i\mid \boldsymbol{\beta})$, where $\pi(\boldsymbol{\beta})$ denotes the prior distribution and $f(\mathcal{D}_i\mid \boldsymbol{\beta})$ denotes the likelihood contribution of the $i$th observation.
\cite{bissiri2016general} proposed a framework of general Bayesian inference, which allows using general loss functions instead of the likelihood functions.
Namely, the generalized Bayes posterior distribution is given by $\pi^{\ast}(\boldsymbol{\beta} \mid \mathcal{D}) \propto \pi(\boldsymbol{\beta})\times \exp\{-\eta \mathcal{L}(\mathcal{D} \mid \boldsymbol{\beta})\}$, where $\eta\in \mathbb{R}_{>0}$ is a learning rate parameter and $\mathcal{L}\left(\mathcal{D} \mid \boldsymbol{\beta}\right)$ denotes some loss function.
When $\eta = 1$ and the loss function $\mathcal{L}\left(\mathcal{D} \mid \boldsymbol{\beta}\right)$ is taken as the negative log-likelihood function, $\pi^{\ast}$ corresponds to the standard posterior distribution.
As $\eta$ decreases, the influence of the prior becomes stronger \citep{grunwald2017inconsistency}.
Thus, the learning rate adjusts the relative influence of the prior and the data on the posterior update.
A number of data-driven methods have been proposed for tuning the learning rate, such as calibrating credible intervals from the generalized posterior distribution to achieve nominal frequentist coverage probability \citep{syring2019calibrating, tanaka2024generalized}, employing sequential updating schemes to avoid hyper-compression (i.e., under model misspecification in a non-convex family, the Bayesian predictive mixture can lie outside the model and achieve a strictly smaller Kullback-Leibler divergence to the true data-generating distribution than any single model component) \citep{grunwald2017inconsistency}, using information-matching approach \citep{holmes2017assigning}, and applying asymptotic distribution-matching approach \citep{lyddon2019general}.
\cite{wu2023comparison} provides a comprehensive comparison of these representative methods.

By adopting the negative log-partial likelihood as the loss function, we can avoid explicit modeling of the baseline hazard function in the Cox regression model.
However, the full conditional distribution is generally not available in closed form.
Consequently, sampling from the posterior distribution is typically performed using the Metropolis-Hastings algorithm or Metropolis-within-Gibbs algorithm \citep{jalaluddin2000algorithm,kosorok2004robust,lambert2005bayesian,schuemie2022combining,ren2025cox}.

\subsection{P\'olya-Gamma Augmentation for the Logistic Model}
\label{subsec:pg-aug-logit}

\citet{polson2013bayesian} established the following integral identity:
\begin{equation*}
    \frac{\exp(\zeta)^a}{\bigl\{1+\exp(\zeta)\bigr\}^b} = 2^{-b}\exp(\kappa \zeta)\int_0^\infty \exp\left(-\frac{\omega \zeta^2}{2}\right)P(\omega \mid b, 0) \dd\omega,
\end{equation*}
where $\kappa=a-b/2$, $a\in \mathbb{R}$, $b\in \mathbb{R}_{>0}$, and $P(\omega\mid b, 0)$ denotes the density of a P\'olya-Gamma distribution $\PG(b,0)$.
The general $\PG(b,c)$ density is obtained by an exponential tilting of $\PG(b,0)$ density, and is given by $P(\omega\mid b, c) = \{\exp(-\omega c^2/2) / \mathbb{E}_{P(\omega \mid b,0)}[\exp(-\omega c^2/2)]\}P(\omega \mid b, 0)$.
Then, the density of $\omega \mid b, \zeta$ takes the form of $P(\omega\mid b, \zeta) = \{\exp(-\omega\zeta^2 / 2)P(\omega \mid b, 0)\} / \{\int \exp(-\omega\zeta^2/2)P(\omega\mid b, 0)\dd\omega\}$, and this shows $\omega \mid b, \zeta \sim \PG(b, \zeta)$. 

For a binary outcome $Y_i\in \{0, 1\}$, consider the logistic regression model $\Pr(Y_{i} = 1) = \exp(\boldsymbol{X}_i^\top\boldsymbol{\beta}) / \{1+\exp(\boldsymbol{X}_i^\top\boldsymbol{\beta})\}$ with a Gaussian prior $\boldsymbol{\beta} \sim \mathcal{N}_{p}(\boldsymbol{\mu}_0, \boldsymbol{\Sigma}_0)$, where $\Pr(\cdot)$ denotes probability, $\mathcal{N}_{p}(\boldsymbol{\mu}, \boldsymbol{\Sigma})$ denotes the $p$-variate normal distribution with mean $\boldsymbol{\mu} \in \mathbb{R}^{p}$, and covariance matrix $\boldsymbol{\Sigma} \in \mathbb{R}^{p\times p}$. 
By introducing P\'olya-Gamma auxiliary variables $\boldsymbol{\omega} = \{\omega_1, \dots, \omega_n\}$ and setting $\kappa_i = Y_i-1/2$, the complete conditional distribution of $\boldsymbol{\beta}$ can be expressed as $\pi(\boldsymbol{\beta}\mid \boldsymbol{\omega}, \{Y_i, \boldsymbol{X}_i\}_{i=1}^n)\propto \exp\{-2^{-1}(\boldsymbol{\beta}-\boldsymbol{\mu}^{\text{logit}})^\top(\boldsymbol{\Sigma}^{\text{logit}})^{-1}(\boldsymbol{\beta}-\boldsymbol{\mu}^{\text{logit}})\}$,
where $\boldsymbol{\Omega} = \operatorname{diag}(\omega_1,\dots, \omega_n)$, $\boldsymbol{\Sigma}^{\text{logit}}=(\boldsymbol{\Sigma}_0^{-1}+\boldsymbol{X}^\top\boldsymbol{\Omega} \boldsymbol{X})^{-1}$, $\boldsymbol{\mu}^{\text{logit}}=\boldsymbol{\Sigma}^{\text{logit}}[\boldsymbol{\Sigma}_0^{-1}\boldsymbol{\mu}_0 + \boldsymbol{X}^\top\boldsymbol{\kappa}]$, and $\boldsymbol{\kappa}=(\kappa_1, \dots, \kappa_n)^\top$.
Thus, conditioned on variables from $\PG(1, \boldsymbol{X}_i^\top\boldsymbol{\beta})$, the complete conditional distribution of $\boldsymbol{\beta}$ is a Gaussian distribution.
Moreover, \citet{polson2013bayesian} and \citet{linderman2015dependent} showed that this augmentation scheme can be extended to the multinomial and negative-binomial distributions.

\subsection{Cox-PG Algorithm}
\label{subsec:coxpg}

\citet{ren2025cox} extended P\'olya-Gamma augmentation to the Cox regression model and obtained a Gaussian-based Gibbs kernel equipped with a Metropolis-Hastings correction that targets the Cox proportional hazards likelihood in \eqref{eq:cox-likelihood}. 

Let $a(t) := \log \int_0^t h_0(\tau)\dd\tau$ and represent $a(t)$ by a monotone spline $a(t) = \boldsymbol{\xi}(t)^{\top}\boldsymbol{s}$ with positive slopes $\boldsymbol{s} \in \mathbb{R}_{>0}^{D}$.
With $\boldsymbol{\theta}=(\boldsymbol{s}^\top,\boldsymbol{\beta}^\top)^\top$ and $\boldsymbol{\gamma}_i= (\boldsymbol{\xi}(T_i)^\top,\boldsymbol{X}_i^\top)^\top$, the Cox proportional hazards likelihood contains the derivative factor $\{(\dd/\dd t)(\boldsymbol{\xi}(T_i)^{\top}\boldsymbol{s})\}^{\delta_i}$ and the term $\exp\{\delta_{i}\boldsymbol{\gamma}_{i}^{\top}\boldsymbol{\theta} - \exp(\boldsymbol{\gamma}_{i}^{\top}\boldsymbol{\theta})\}$, subject to the positivity constraint on $\boldsymbol{s}$.
To remove the derivative factor for events, introduce slice variables $\nu_i \mid\boldsymbol{\theta}\sim \operatorname{Unif}(0,(\dd/\dd t)(\boldsymbol{\xi}(T_i)^{\top}\boldsymbol{s}))$ for $\delta_{i} = 1$, where $\operatorname{Unif}(a,b)$ denotes the uniform distribution on the open interval $(a, b)$.
Under a histogram partition of time with knot sequence $0 = \tau_{0} < \tau_{1} < \cdots < \tau_{D}$, the derivative reduces to indicator bases, and the slice variables collapse to last order statistics $v_{j} = \max\{\nu_{i}: \delta_{i} = 1, T_{i}\in [\tau_{j-1}, \tau_{j}]\}$.
This yields the half-space $\mathcal{C}_{v} = \{\boldsymbol{s}: v_{j}\le s_{j}, \;j \in \{1,\dots, D\}\}$, which are enforced in truncated-Gaussian updates.

To obtain a conditionally Gaussian representation of the remaining term, place independent gamma frailties $z_{i} \sim\mathrm{Ga}(\varepsilon, \varepsilon)$ on the Poisson kernel.
Here $\mathrm{Ga}(a,b)$ is the gamma distribution with shape $a > 0$ and rate $b > 0$, density $f(z; a,b) = \{b^{a}/\Gamma(a)\}z^{a-1}\exp(-bz)$ for $z > 0$, and $\Gamma(a) = \int_{0}^{\infty} t^{a-1}\exp(-t) \dd t$ is Euler's gamma function.
A Mellin transform yields a negative-binomial kernel with $\psi_{i} = \boldsymbol{\gamma}_{i}^{\top}\boldsymbol{\theta} - \log \varepsilon$.
Applying the P\'olya-Gamma identity then introduces $\omega_{i}\sim \PG(\delta_{i}+\varepsilon, \psi_{i})$ and makes the complete-data log posterior quadratic in $\boldsymbol{\theta}$.
Conditional on $(\boldsymbol{\omega}, \boldsymbol{\nu})$ (or equivalently on the sufficient statistics $\boldsymbol{v}$), the complete conditional of $\boldsymbol{\theta}$ is multivariate normal truncated to $\{\boldsymbol{s} > 0\}\cap \mathcal{C}_{\boldsymbol{v}}$.

A full sweep of the sampler updates $\boldsymbol{\omega}$, the slice variables or their last-order sufficient statistics $\boldsymbol{v}$, and then $\boldsymbol{\theta}$.
Because finite $\varepsilon$ induces a negative-binomial approximation to the Cox proportional hazards likelihood, the algorithm appends an outer Metropolis-Hastings step that compares the Cox proportional hazards likelihood at the proposed and current states, thereby producing a Metropolis-within-Gibbs scheme that targets the Cox model while retaining Gaussian computation in the inner kernel.
In practice, approximation accuracy and mixing depend on the number of spline segments $D$ and the frailty hyperparameter $\varepsilon$: very fine  partitions or extremely small $\varepsilon$ can depress acceptance and slow exploration; explicit baseline representation enlarges the augmented state even when interest is primarily in $\boldsymbol{\beta}$; and aggressive knot placement may ill-condition the design.
Careful partitioning together with moderate $\varepsilon$ balances accuracy and efficiency.
In summary, \citet{ren2025cox} combine slice variables for the derivative term, gamma mixing to obtain a negative-binomial kernel, and P\'olya-Gamma tilting to recover Gaussian conditions, with an outer Metropolis-Hastings correction ensuring exact inference under the Cox proportional hazards likelihood.

\section{Proposed Method}
\label{sec:proposed-method}

In this section, we introduce a new Gibbs sampling algorithm for the Cox regression model, called GS4Cox, composed of four key elements: (i) general Bayesian framework, (ii) composite partial likelihood, (iii) P\'olya-Gamma augmentation scheme, and (iv) affine posterior calibration.

\subsection{General Bayesian Inference}
\label{subsec:general-bayes-for-cpl}

In accordance with the general Bayesian framework \citep{bissiri2016general}, and following the approach described in Section~\ref{subsec:general-bayes-cox}, we employ the negative log-partial likelihood as the loss function, thereby avoiding explicit specification of the baseline hazard function.
Similarly, our algorithm adopts the negative log-composite partial likelihood introduced in Section~\ref{subsec:cpl} as its loss function, thus obviating the need to specify the baseline hazard.

\subsection{Composite Partial Likelihood for the Cox Regression Models}
\label{subsec:cpl}

For each pair $(i,j)$ with $j\in R(T_i)\setminus \{i\}$, we define $p_{ij}(\boldsymbol{\beta})$ as follows:
\begin{equation*}
    p_{ij}(\boldsymbol{\beta}) := \expit\left\{\left(\boldsymbol{X}_i-\boldsymbol{X}_j\right)^\top \boldsymbol{\beta}\right\} = \frac{\exp\left\{\left(\boldsymbol{X}_i-\boldsymbol{X}_j\right)^\top \boldsymbol{\beta}\right\}}{1 + \exp\left\{\left(\boldsymbol{X}_i-\boldsymbol{X}_j\right)^\top \boldsymbol{\beta}\right\}},
\end{equation*}
which can be interpreted as the probability that subject $i$ experiences the event before subject $j$.
Using these probabilities, we define the composite partial likelihood as
\begin{equation}
    L_{\mathrm{CPL}}(\boldsymbol{\beta}\mid \mathcal{D}) := \prod_{i=1}^n \prod_{j\in R(T_i)\setminus \{i\}} p_{ij}(\boldsymbol{\beta})^{\delta_i},
    \label{eq:composite-partial-likelihood}
\end{equation}
which aligns with the definition of a composite likelihood as presented in \citet{lindsay1988composite} and \citet{varin2011overview}.
Correspondingly, we define the composite partial score function as
\begin{equation}
    S_{\mathrm{CPL}}(\boldsymbol{\beta}):= \frac{\partial}{\partial \boldsymbol{\beta}}\log L_{\mathrm{CPL}}(\boldsymbol{\beta}\mid \mathcal{D}) = \sum_{i=1}^n \sum_{j\in R(T_i)\setminus \{i\}} \delta_i \left\{1-p_{ij}(\boldsymbol{\beta})\right\}\bigl(\boldsymbol{X}_i - \boldsymbol{X}_j\bigr)
    .
    \label{eq:cpl-score-function}
\end{equation}

Let $\hat{\boldsymbol{\beta}}_{\mathrm{CPL}}$ denote the maximum composite partial likelihood estimator, and $\boldsymbol{\beta}_0$ denote the true value.
Let $P_{\boldsymbol{\beta}_0}$ denote the true data-generating distribution under the Cox proportional hazards model with true parameter $\boldsymbol{\beta}_0$, baseline hazard, covariate distribution, and noninformative censoring distribution.
See Section~\ref{app-sec:dgp} for more details of the data-generating process.
Expectations under $P_{\boldsymbol{\beta}_0}$ are written $\mathbb{E}_{P_{\boldsymbol{\beta}_0}}[\cdot]$.
The following propositions hold for the composite partial score function and the maximum composite partial likelihood estimator. 

\begin{proposition}\textbf{\textup{(Zero expectation of the composite partial score)}}
    Suppose a data-generating process under the Cox proportional hazards model with $\boldsymbol{\beta}_{0}$.
    Then, we have
    \begin{equation*}
        \mathbb{E}_{P_{\boldsymbol{\beta}_{0}}}[S_{\mathrm{CPL}}(\boldsymbol{\beta})] = 0
        .
    \end{equation*}
    \label{prop:expectation-zero}
\end{proposition}
\begin{proposition}\textbf{\textup{(Weak consistency of the maximum composite partial likelihood estimator)}}
    Under the regularity conditions \ref{item:reg-1}--\ref{item:reg-3} (provided in Section~\ref{app-subsec:regularity-conditions}), we have
    \begin{equation*}
        \hat{\boldsymbol{\beta}}_{\mathrm{CPL}} \overset{P_{\boldsymbol{\beta}_0}}{\longrightarrow} \boldsymbol{\beta}_0
        .
    \end{equation*}
    \label{prop:cpl-weak-consistency}
\end{proposition}
\noindent
See Sections~\ref{app-subsec:proof-of-expectation-zero} and \ref{app-subsec:proof-of-consistency} for the proofs of Proposition~\ref{prop:expectation-zero} and Proposition~\ref{prop:cpl-weak-consistency}, respectively.
Proposition~\ref{prop:expectation-zero} shows that the solution $\hat{\boldsymbol{\beta}}_{\mathrm{CPL}}$ to the estimating equation $S_{\mathrm{CPL}}(\hat{\boldsymbol{\beta}}_{\mathrm{CPL}})=0$ is asymptotically unbiased.
Proposition~\ref{prop:cpl-weak-consistency} presents that the $\boldsymbol{\beta}$ maximizing the composite partial likelihood converges in probability to the true value $\boldsymbol{\beta}_0$, just as the $\boldsymbol{\beta}$ maximizing the partial likelihood does.

Next, let
$\pi^{\ast}(\boldsymbol{\beta}\mid \mathcal{D}) \propto \exp\{\eta\,\ell_{n,\mathrm{CPL}}(\boldsymbol{\beta})\}\pi(\boldsymbol{\beta})$ 
denote the generalized Bayes posterior for the composite partial likelihood, where $\ell_{n, \mathrm{CPL}}$ is the empirical log-composite partial likelihood, and let
$\hat{\boldsymbol{\beta}}_{\mathrm{CPL,GB}} := \int \boldsymbol{\beta}\pi^{\ast}(\boldsymbol{\beta}\mid \mathcal{D})\dd\boldsymbol{\beta}$
its posterior mean.
Write $\bar{\ell}_{n,\mathrm{CPL}}(\boldsymbol{\beta}) := \binom{n}{2}^{-1}\ell_{n, \mathrm{CPL}}(\boldsymbol{\beta})$,
$\boldsymbol{J}_{0,\mathrm{CPL}} := -\mathbb{E}_{P_{\boldsymbol{\beta}_{0}}}[\nabla^2_{\boldsymbol{\beta}}\bar{\ell}_{n,\mathrm{CPL}}(\boldsymbol{\beta}_{0})] \succ 0$, and set
$\boldsymbol{H}_{0,\mathrm{CPL}} := \eta \boldsymbol{J}_{0,\mathrm{CPL}}$.
Then we have the following theorem.

\begin{theorem}\textbf{\textup{(Asymptotic bias rate of the generalized Bayes estimator under composite partial likelihood)}}
    Under the regularity conditions \ref{item:reg-1}--\ref{item:reg-8} (provided in Section~\ref{app-subsec:regularity-conditions}), we have
    \begin{equation*}
        \mathbb{E}_{P_{\boldsymbol{\beta}_{0}}}\left[\hat{\boldsymbol{\beta}}_{\mathrm{CPL,GB}}\right] - \boldsymbol{\beta}_{0} = O(n^{-1/2})
        .
    \end{equation*}
\label{thm:bayes-asym-unbiased}
\end{theorem}
\noindent
Theorem~\ref{thm:bayes-asym-unbiased} shows that the generalized Bayes estimator under the composite partial likelihood loss is asymptotically unbiased.

In addition, let $r_n$ denote the density of $\sqrt{n}(\boldsymbol{\beta}-\hat{\boldsymbol{\beta}}_{\mathrm{CPL,GB}})$ under $\pi^{\ast}(\cdot \mid \mathcal{D})$.
Then, the following corollary holds.

\begin{corollary}\textbf{\textup{(Bernstein--von Mises about the generalized Bayes posterior mean)}}
    Under the regularity conditions \ref{item:reg-1}--\ref{item:reg-7} (provided in Section~\ref{app-subsec:regularity-conditions}), we have
    \begin{equation*}
        \int_{\mathbb{R}^p}\left|r_n(x)-\mathcal N\!\left(x\mid 0, \boldsymbol{H}_{0, \mathrm{CPL}}^{-1}\right)\right|\dd x \longrightarrow 0
        .
    \end{equation*}
\label{corollary:bvm-for-cpl}
\end{corollary}
\noindent
Corollary~\ref{corollary:bvm-for-cpl} shows that a Bernstein--von Mises limit holds around the generalized Bayes posterior mean for the composite partial likelihood with covariance $\boldsymbol{H}_{0,\mathrm{CPL}}^{-1}$.

See Section~\ref{app-subsec:proof-of-bayes-asym-unbiased} for the proof of Theorem~\ref{thm:bayes-asym-unbiased}.
Propositions~\ref{prop:expectation-zero}, \ref{prop:cpl-weak-consistency}, and Theorem~\ref{thm:bayes-asym-unbiased} together justify using $\hat{\boldsymbol{\beta}}_{\mathrm{CPL,GB}}$ as a consistent and asymptotically unbiased estimator of $\boldsymbol{\beta}_0$.
We then develop our Gibbs sampling algorithm based on the composite partial likelihood function~\eqref{eq:composite-partial-likelihood}.

\begin{remark}\textbf{\textup{(Ties and censoring under the composite partial likelihood)}}
    The composite partial likelihood \eqref{eq:composite-partial-likelihood} is defined based on the probability that an event occurs for an individual before any other subject in the risk set at that time.
    Therefore, it can be directly applied even when multiple events occur at the same time.
    Similar to the partial likelihood, censoring is handled indirectly through the use of indicator functions, allowing it to be appropriately accounted for.
\end{remark}

\subsection{Posterior Representation of the Cox Regression Model via P\'olya-Gamma Augmentation}
\label{subsec:pg-for-cpl}

Applying the P\'olya-Gamma augmentation scheme to \eqref{eq:composite-partial-likelihood} yields the following conditional composite partial likelihood $L_{\mathrm{CPL}}(\boldsymbol{\beta}, \boldsymbol{\omega} \mid \mathcal{D}) \propto \prod_{i=1}^n \prod_{j\in R(T_i)\setminus \{i\}}\exp\{\kappa_{ij}\zeta_{ij}- (\omega_{ij}\zeta_{ij}^2)/2\}^{\delta_i}$, where $\zeta_{ij} = (\boldsymbol{X}_i-\boldsymbol{X}_j)^\top\boldsymbol{\beta}$, $\kappa_{ij}=1/2\;(\text{since } a=b=1)$, $\omega_{ij}\sim \PG(1, \zeta_{ij})$.
Defining the corresponding loss function as
\begin{equation*}
    \mathcal{L}_{\mathrm{CPL}}(\mathcal{D}; \boldsymbol{\beta}, \boldsymbol{\omega})
    :=-\log \left[\prod_{i=1}^n \prod_{j\in R(T_i)\setminus \{i\}}
    \exp\left(\kappa_{ij}\zeta_{ij}- \frac{\omega_{ij}\zeta_{ij}^2}{2}\right)^{\delta_i}\right]
    ,
\end{equation*}
and then we obtain the complete conditional for $\boldsymbol{\beta}$ as follows:
\begin{equation}
    \begin{split}
        \pi^{\ast}(\boldsymbol{\beta}\mid \boldsymbol{\omega}, \mathcal{D})
        &\propto \pi(\boldsymbol{\beta})\exp\bigl(-\eta \mathcal{L}_{\mathrm{CPL}}(\mathcal{D};\boldsymbol{\beta}, \boldsymbol{\omega})\bigr),\\
        &= \pi(\boldsymbol{\beta})\exp\left\{\eta \sum_{i=1}^n \sum_{j\in R(T_i)\setminus \{i\}}\delta_i\left(\kappa_{ij}\zeta_{ij}- \frac{\omega_{ij}\zeta_{ij}^2}{2}\right)\right\}
        .
    \end{split}
\label{eq:complete-conditional}
\end{equation}
Assuming a Gaussian prior $\pi(\boldsymbol{\beta}) = \mathcal{N}(\boldsymbol{\mu}_0,\boldsymbol{\Sigma}_0)$,
this conditional distribution takes the Gaussian form
\begin{equation*}
    \boldsymbol{\beta}\mid \boldsymbol{\omega}, \mathcal{D} \sim 
    \mathcal{N}\left(
    \boldsymbol{\Sigma}^{\ast}\left[\boldsymbol{\Sigma}_0^{-1}\boldsymbol{\mu}_0
    + \eta\sum_{i=1}^n\sum_{j\in R(T_i)\setminus \{i\}}\delta_i\kappa_{ij}\bigl(\boldsymbol{X}_i-\boldsymbol{X}_j\bigr)\right]
    ,\boldsymbol{\Sigma}^{\ast}
    \right)
    ,
\end{equation*}
where $\boldsymbol{\Sigma}^{\ast} = [\boldsymbol{\Sigma}_0^{-1} + \eta \sum_{i=1}^n\sum_{j\in R(T_i)\setminus \{i\}} \delta_i \omega_{ij}(\boldsymbol{X}_i-\boldsymbol{X}_j)(\boldsymbol{X}_i-\boldsymbol{X}_j)^\top]^{-1}$.
Thus, the complete conditional distributions for $\boldsymbol{\beta}$ can be expressed in a Gaussian closed form.

\subsection{Affine Posterior Calibration}
\label{subsec:apc-cpl}
The asymptotic covariance of a generalized Bayes posterior using a quasi-likelihood loss function (e.g., a negative log-composite partial likelihood) generally differs from the frequentist asymptotic sandwich covariance (see Propositions~\ref{app-prop:bvm-for-bissiri} and \ref{app-prop:z-asym}).
\citet{shaby2014open-faced} highlighted this gap and proposed the open-faced sandwich adjustment, a linear post-processing that aligns posterior draws to a chosen target covariance.
Accordingly, we derive an analogous affine posterior calibration, explicitly including a location adjustment, that re-centers the generalized Bayes posterior mean and maps its covariance to the target.

Let $\boldsymbol{V}_{\mathrm{target}} \succ 0$ be any fixed symmetric positive definite matrix, and let $\boldsymbol{H}_{0} \succ 0$ denote the population curvature of the generalized loss used in the update (See Section~\ref{app-subsec:proof-of-bayes-asym-unbiased} for formal definition of $\boldsymbol{H}_{0}$).
For any symmetric positive definite matrix $\boldsymbol{A}$, we write $\boldsymbol{A}^{1/2}$ for its unique symmetric positive definite square root, that is, the symmetric matrix satisfying $\boldsymbol{A}^{1/2}\boldsymbol{A}^{1/2} = \boldsymbol{A}$.
We also write $\boldsymbol{A}^{-1/2} = (\boldsymbol{A}^{1/2})^{-1}$.
We then define
$\boldsymbol{\Omega} := \boldsymbol{V}_{\mathrm{target}}^{1/2}\boldsymbol{H}_{0}^{1/2}$.
For draws $\{\boldsymbol{\beta}^{(m)}\}$ i.i.d.\ from $\pi^{\ast}(\cdot\mid \mathcal{D})$ and any data-measurable center $\boldsymbol{\beta}^{\ddagger} \in \mathbb{R}^p$, define
$\boldsymbol{\beta}_{\mathrm{APC}}^{(m)} := \boldsymbol{\beta}^{\ddagger} + \boldsymbol{\Omega}(\boldsymbol{\beta}^{(m)} - \hat{\boldsymbol{\beta}}_{\mathrm{GB}}),\;(m=0,\dots, M)$.
Under these definitions, the following proposition holds.

\begin{proposition}\textbf{\textup{(Affine posterior calibration to a target covariance in generalized Bayes posterior)}}
    Suppose the regularity conditions \ref{item:reg-1}--\ref{item:reg-8} (provided in Section~\ref{app-subsec:regularity-conditions}).
    Then, conditionally on the data, we have
    \begin{equation*}
        \sqrt{n}\left(\boldsymbol{\beta}_{\mathrm{APC}}^{(m)} - \boldsymbol{\beta}^{\ddagger}\right)
        \overset{d}{\longrightarrow} \mathcal{N}_{p}(\boldsymbol{0}, \boldsymbol{V}_{\mathrm{target}})
        .
    \end{equation*}
\label{prop:general-bayes-apc}
\end{proposition}
\noindent
Proposition~\ref{prop:general-bayes-apc} presents that the affine posterior calibration yields posterior draws centered at $\boldsymbol{\beta}^{\ddagger}$ with first-order asymptotic covariance equal to the chosen target $\boldsymbol{V}_{\mathrm{target}}$.

Let
$\boldsymbol{J}_{0,\mathrm{PL}} := - \mathbb{E}_{P_{\boldsymbol{\beta}_{0}}}[\nabla^2_{\boldsymbol{\beta}}\bar{\ell}_{n,\mathrm{PL}}(\boldsymbol{\beta}_{0})]$, 
$\boldsymbol{K}_{0,\mathrm{PL}} := \operatorname{Var}_{P_{\boldsymbol{\beta}_{0}}}[S_{\mathrm{PL}}(\boldsymbol{\beta_{0}})]$, 
$\boldsymbol{V}_{0,\mathrm{PL}} := \boldsymbol{J}_{0,\mathrm{PL}}^{-1}\boldsymbol{K}_{0,\mathrm{PL}}\boldsymbol{J}_{0,\mathrm{PL}}^{-1}$,
and $\boldsymbol{\Omega}_{0,\mathrm{PL}} := \boldsymbol{V}_{0,\mathrm{PL}}^{1/2}(\eta \boldsymbol{J}_{0,\mathrm{CPL}})^{1/2}$
.
Define
$\boldsymbol{\beta}^{\ddagger}_{\mathrm{PL}} := \hat{\boldsymbol{\beta}}_{\mathrm{GB}} - \boldsymbol{H}_{\mathrm{PL}}(\hat{\boldsymbol{\beta}}_{\mathrm{GB}})^{-1}S_{\mathrm{PL}}(\hat{\boldsymbol{\beta}}_{\mathrm{GB}})$
where $S_{\mathrm{PL}}(\boldsymbol{\beta})$ denotes the partial likelihood score and $\boldsymbol{H}_{\mathrm{PL}}(\boldsymbol{\beta})$ denotes the Hessian of the Cox log-partial likelihood at $\boldsymbol{\beta}$.
Then set
$\boldsymbol{\beta}^{(m)}_{\mathrm{APC}} := \boldsymbol{\beta}^{\ddagger}_{\mathrm{PL}} + \boldsymbol{\Omega}_{0,\mathrm{PL}}(\boldsymbol{\beta}^{(m)} - \hat{\boldsymbol{\beta}}_{\mathrm{GB}})$
.
Specializing Proposition~\ref{prop:general-bayes-apc} to the partial likelihood target yields the following.

\begin{corollary}\textbf{\textup{(Affine posterior calibration to the Cox partial likelihood target)}}
    Suppose the same assumptions as Proposition~\ref{prop:general-bayes-apc}.
    Then, conditionally on the data, we have
    \begin{equation*}
        \sqrt{n}\left(\boldsymbol{\beta}_{\mathrm{APC}}^{(m)} - \boldsymbol{\beta}^{\ddagger}_{\mathrm{PL}}\right) \overset{d}{\longrightarrow} \mathcal{N}_{p}(\boldsymbol{0}, \boldsymbol{V}_{0,\mathrm{PL}})
        ,
        \quad
        \boldsymbol{\beta}^{(m)}_{\mathrm{APC}} := \boldsymbol{\beta}^{\ddagger}_{\mathrm{PL}} + \boldsymbol{\Omega}_{0,\mathrm{PL}}(\boldsymbol{\beta}^{(m)} - \hat{\boldsymbol{\beta}}_{\mathrm{GB}})
        .
    \end{equation*}
\label{corollary:apc-for-gs4cox}
\end{corollary}
\noindent
Corollary~\ref{corollary:apc-for-gs4cox} shows that, conditional on the data, the calibrated draws share the same first-order asymptotic distribution as the Cox partial likelihood estimator.
See Section~\ref{app-subsec:proof-of-apc} for the proofs of Proposition~\ref{prop:general-bayes-apc} and Corollary~\ref{corollary:apc-for-gs4cox}.

\begin{remark}\textbf{\textup{(Learning rate cancellation in generalized Bayesian framework)}}
    Since $\boldsymbol{H}_{0, \mathrm{CPL}} = \eta \boldsymbol{J}_{0, \mathrm{CPL}}$,
    \begin{equation*}
        \boldsymbol{\Omega} \boldsymbol{H}_{0,\mathrm{CPL}}^{-1}\boldsymbol{\Omega}^\top = \boldsymbol{V}_{\mathrm{target}}^{1/2}(\eta \boldsymbol{J}_{0,\mathrm{CPL}})^{1/2}(\eta \boldsymbol{J}_{0,\mathrm{CPL}})^{-1}(\eta \boldsymbol{J}_{0,\mathrm{CPL}})^{1/2}\boldsymbol{V}_{\mathrm{target}}^{1/2} = \boldsymbol{V}_{\mathrm{target}}
        ,
    \end{equation*}
    so the effect of learning rate $\eta$ cancels exactly.
    Thus the affine posterior calibrated asymptotics do not depend on the learning rate.
\end{remark}

Define the partial likelihood sandwich covariance estimator
$\boldsymbol{V}_{\mathrm{PL}} = \hat{\boldsymbol{J}}_{\mathrm{PL}}^{-1}\hat{\boldsymbol{K}}_{\mathrm{PL}}\hat{\boldsymbol{J}}_{\mathrm{PL}}^{-1}$
with
$\hat{\boldsymbol{J}}_{\mathrm{PL}}:=-n^{-1}\boldsymbol{H}_{\mathrm{PL}}(\hat{\boldsymbol{\beta}}_{\mathrm{GB}})$
and $\hat{\boldsymbol{K}}_{\mathrm{PL}}:=n^{-1}\sum_{i=1}^n \delta_i\{\boldsymbol{X}_i-\bar{\boldsymbol{X}}(T_i;\hat{\boldsymbol{\beta}}_{\mathrm{GB}})\}\{\boldsymbol{X}_i-\bar{\boldsymbol{X}}(T_i;\hat{\boldsymbol{\beta}}_{\mathrm{GB}})\}^\top$,
where 
\begin{equation*}
    \bar{\boldsymbol{X}}\left(t;\hat{\boldsymbol{\beta}}_{\mathrm{GB}}\right)
    :=
    \frac{\sum_{j\in R(t)} \boldsymbol{X}_j \exp\left(\boldsymbol{X}_j^\top\hat{\boldsymbol{\beta}}_{\mathrm{GB}}\right)}{\sum_{j\in R(t)} \exp\left(\boldsymbol{X}_j^\top\hat{\boldsymbol{\beta}}_{\mathrm{GB}}\right)}
    .
\end{equation*}
In our implementation we set $\boldsymbol{V}_{\mathrm{target}} = \boldsymbol{V}_{\mathrm{PL}}$.

\subsection{Whole Algorithm}
\label{subsec:whole-algo-gs4cox}
Based on the discussion above, a Gibbs sampling algorithm for Cox regression models can be constructed using the Algorithm~\ref{alg:gs4cox}, named GS4Cox.

\begin{algorithm}[htbp]
   \caption{GS4Cox}
   \label{alg:gs4cox}
\begin{algorithmic}[1]
    \STATE {\bfseries Input:} Data $\mathcal{D}$; initial $\boldsymbol{\beta}^{(0)}$; prior hyperparameters $\boldsymbol{\mu}_0, \boldsymbol{\Sigma}_0$; learning rate $\eta$.
    \FOR{$m = 0,1,\dots, M-1$}
        \FOR{each subject $i \in \{1,\dots, n\}$}
            \FOR{each subject in risk set at time $T_i$ excluding $i$ (i.e., $j\in R(T_i)\setminus \{i\}$)}
                \STATE Compute $\zeta_{i,j}^{(m)} = (\boldsymbol{X}_{i} - \boldsymbol{X}_j)^\top \boldsymbol{\beta}^{(m)}$.
                \STATE Sample $\omega_{i,j}^{(m)} \sim \PG(1,\zeta_{i,j}^{(m)})$.
            \ENDFOR
        \ENDFOR
        \STATE Here $\kappa_{ij} = 1/2$. Update parameter $\boldsymbol{\beta}$
        \begin{equation*}
            \boldsymbol{\beta}^{(m +1)} \sim \mathcal{N}\Bigl(\boldsymbol{\mu}^{(m)},\boldsymbol{\Sigma}^{(m)}\Bigr)
        \end{equation*}
        with
        \begin{equation*}
            \boldsymbol{\Sigma}^{(m)} = \left[\boldsymbol{\Sigma}_0^{-1}+\eta\sum_{i=1}^n\sum_{j\in R(T_i)\setminus \{i\}}\delta_i\omega_{i,j}^{(m)}\bigl(\boldsymbol{X}_i-\boldsymbol{X}_j\bigr)\bigl(\boldsymbol{X}_i-\boldsymbol{X}_j\bigr)^\top\right]^{-1},
        \end{equation*}
        \begin{equation*}
            \boldsymbol{\mu}^{(m)} = \boldsymbol{\Sigma}^{(m)}\left[\boldsymbol{\Sigma}_0^{-1}\boldsymbol{\mu}_0
            +\eta\sum_{i=1}^n\sum_{j\in R(T_i)\setminus \{i\}}\delta_i\kappa_{ij}\bigl(\boldsymbol{X}_{i}-\boldsymbol{X}_j\bigr)\right],
        \end{equation*}
    \ENDFOR
    \STATE Compute posterior mean $\hat{\boldsymbol{\beta}}_{\mathrm{GB}} = (\sum_{k=1}^K \boldsymbol{\beta}^{(m^{\ast} + k)})/K$, where $m^{\ast}$ is the burn-in period.
    \STATE Compute empirical covariance $\hat{\boldsymbol{\Sigma}} = \sum_{k=1}^{K}(\boldsymbol{\beta}^{(m^{\ast} + k)} - \hat{\boldsymbol{\beta}}_{\mathrm{GB}})(\boldsymbol{\beta}^{(m^{\ast} + k)} - \hat{\boldsymbol{\beta}}_{\mathrm{GB}})^\top/(K-1)$.
    \STATE Compute the partial likelihood sandwich target $\boldsymbol{V}_{\mathrm{PL}}
    =\hat{\boldsymbol{J}}_{\mathrm{PL}}^{-1}\hat{\boldsymbol{K}}_{\mathrm{PL}}\hat{\boldsymbol{J}}_{\mathrm{PL}}^{-1}$.
    \STATE Set $\boldsymbol{V}_{\mathrm{target}}\leftarrow \boldsymbol{V}_{\mathrm{PL}}$.
    \STATE Compute the mapping $\boldsymbol{\Omega} = \boldsymbol{V}_{\mathrm{target}}^{1/2}\hat{\boldsymbol{\Sigma}}^{-1/2}$.
    \STATE Set $\boldsymbol{\beta}^{\ddagger} \leftarrow \hat{\boldsymbol{\beta}}_{\mathrm{GB}} - \boldsymbol{H}_{\mathrm{PL}}(\hat{\boldsymbol{\beta}}_{\mathrm{GB}})^{-1}S_{\mathrm{PL}}(\hat{\boldsymbol{\beta}}_{\mathrm{GB}})$
    \FOR{$m = 0,1,\dots, M$}
        \STATE Compute the $\boldsymbol{\beta}_{\mathrm{APC}}^{(m)}\gets \boldsymbol{\beta}^{\ddagger}+\boldsymbol\Omega\left(\boldsymbol\beta^{(m)}-\hat{\boldsymbol{\beta}}_{\mathrm{GB}}\right)$.
    \ENDFOR
\end{algorithmic}
\end{algorithm}
\noindent

Our GS4Cox algorithm leverages a composite partial likelihood within the general Bayesian framework, combined with P\'olya-Gamma augmentation to express the full conditional distribution of $\boldsymbol{\beta}$ in closed Gaussian form updates.
Important features of our algorithm are that we do not have to specify the baseline hazard or design Metropolis-Hastings proposal distributions.
Moreover, our algorithm does not depend on the selection of the learning rate.

\section{Numerical Experiments}
\label{sec:num-experiments}
We evaluated our GS4Cox on synthetic data against several alternative methods.
Our primary comparison metrics were the effective sample size (ESS), the effective sampling rate (ESR), defined as the effective sample size per second of runtime \citep{kass1998markov,polson2013bayesian}, and the Monte Carlo standard error (MCSE) \citep{geyer2011introduction,carpenter2017stan}.
The ESS is defined as $\operatorname{ESS} := (1/p)\sum_{q=1}^{p} \operatorname{ESS}_{q}$, $\operatorname{ESS}_{q} := (M - m^{\ast})/(1+2\sum_{l=1}^\infty \rho_q(l))$, where $M$ is the number of samples, $m^{\ast}$ is the burn-in period, and $\rho_q(l)$ is the $l$th autocorrelation of $\beta_q$.
The ESS quantifies the number of independent samples that an autocorrelated Markov chain is equivalent to, thereby reflecting the true amount of independent information available for inference.
In contrast, the ESR quantifies how quickly the MCMC can produce independent draws from the posterior distribution.
Higher values of ESS and ESR indicate that the obtained samples are nearly independent, leading to greater accuracy in statistical inference and more efficient use of computational resources.
Furthermore, the MCSE for each parameter is defined as $\operatorname{MCSE}_{q} := \operatorname{sd}_{q}/\sqrt{\operatorname{ESS}}_{q}$, where $\operatorname{sd}_{q}$ denotes the posterior standard deviation estimate for parameter $q$.
In this study, we report a single MCSE by averaging the parameter-wise MCSE (i.e., $\operatorname{MCSE} := (1/p)\sum_{q=1}^{p}\operatorname{MCSE}_{q}$).
The MCSE captures the Monte Carlo uncertainty of the posterior mean estimate.
Lower values of MCSE indicate more precise estimates, while higher values signal greater simulation error and less reliable posterior means.
We then computed the autocovariance function using the Fast Fourier Transform, and obtained the autocorrelation coefficient by normalizing by lag $0$ autocovariance for estimating $\operatorname{ESS}$.
Furthermore, we compared posterior means and credible intervals (CrI) with maximum partial likelihood estimates (MPLE) and their confidence intervals (CI).

Synthetic datasets were generated as follows.
First, $n$ independent covariate vectors were drawn from a standard normal distribution $\boldsymbol{X}_i \sim \mathcal{N}_p(\boldsymbol{0}, \boldsymbol{I}_{p})$, where $\boldsymbol{I}_{p}$ denotes the $p \times p$ identity matrix.
Using these covariates, the linear predictor was computed as $\boldsymbol{X}_i^\top \boldsymbol{\beta}_{0}$.
Survival times were drawn from an exponential distribution with a rate parameter that depends on the linear predictor.
Specifically, $T_{i}^{\ast} \sim \operatorname{Exp}(\exp(\boldsymbol{X}_i^\top\boldsymbol{\beta}_{0}))$, where $\operatorname{Exp}(\lambda)$ denotes the exponential distribution with rate parameter $\lambda$.
Censoring times were generated independently from another exponential distribution (i.e., $C_{i}^{\ast}\sim \operatorname{Exp}(1)$).
The observed time for each subject $i$ was defined as $T_i = \min(T_i^{\ast}, C_i^{\ast})$, and the event indicator $\delta_i$ was set as $\delta_{i} = \boldsymbol{1}\left\{T_{i}^{\ast} \le C_{i}^{\ast}\right\}$.

To simulate scenarios with tied events, we modified the above procedure by rounding.
After determining the continuous time as $\tilde{T}_i=\min(T_i^{\ast}, C^{\ast}_{i})$, we mapped it to the nearest point on the grid $\{0, r, 2r, \dots\}$ with a pre-specified rounding parameter $r > 0$ via $T_{i} = \lfloor \tilde{T}_{i}/r + 1/2\rfloor \times r$.
The event indicators were computed in the same manner as before.

\subsection{Baselines}
\label{subsec:num-baselines}
We compared our GS4Cox with four samplers implemented within the general Bayesian framework using the negative log-partial likelihood function as the loss function (methods (a) to (d)) and with one previous method (e): (a) the MH random-walk sampler with multivariate normal proposals whose covariance is based on the Hessian; (b) the HMC sampler based on leapfrog integration; (c) the NUTS, which adapts the trajectory length automatically; (d) the MALA, a gradient-informed Metropolis-Hastings method; (e) the Cox-PG algorithm proposed by \citet{ren2025cox}.

\subsection{Experimental Setup and Method Configuration}
\label{subsec:num-setup}
Unless stated otherwise, we used $n=300$, $\boldsymbol{\beta}_{0}=(\beta_1, \beta_2, \beta_3, \beta_4, \beta_5, \beta_6, \beta_7, \beta_8)^\top=(1.0, -1.0, 0.5, -0.5, 0.3, -0.3, 0.1, -0.1)^\top$, $1,000$ iterations with $500$ burn-in iterations, and without tied events.
Additionally, unless noted otherwise, we used a Gaussian prior $\boldsymbol{\beta} \sim \mathcal{N}_{p}(\boldsymbol{0}, 100\boldsymbol{I}_{p})$, and set the learning rate to $\eta = 1$.
To induce tied events when desired, we applied a rounding parameter of $r=0.001$.
In the presence of ties, we employed the Efron approximation \citep{efron1977efficiency}.

For MH, we used a multivariate normal random-walk proposal with covariance $-H_{\mathrm{PL}}(\boldsymbol{\beta})^{-1}$ evaluated at the current state.
For HMC, we used an identity mass matrix, standard-normal momenta, step size $0.01$, and $10$ leapfrog steps per iteration.
For NUTS, we set the initial step size to $0.01$ and the maximum tree depth to $10$.
For MALA, we used identity preconditioning and a Langevin step size of $0.01$.
For Cox-PG, we modeled the baseline log-cumulative hazard using an additive monotone spline with $D=5$ segments; set the frailty hyperparameter to $\varepsilon = 1$; enabled the Metropolis-Hastings calibration step; and used a slice sampler burn-in of $100$ for the spline block.
Our implementation closely followed the supplementary code provided by \citet{ren2025cox}.

Furthermore, the initial value for the Cox-PG algorithm was set using the maximum likelihood estimates, as described in \citet{ren2025cox}, while for other MCMC algorithms, including GS4Cox, the initial values were all set to $0$.

We reported the computational cost of our implementation in Section~\ref{app-subsec:computing-complexity}.

\subsection{Results}
\label{subsec:num-results}
Table~\ref{tab:numerical-data-efficiency} summarizes the evaluation metrics under two scenarios: one without tied events and the other with tied events introduced by rounding event times.
The results indicate that GS4Cox outperformed all alternative methods across all metrics in both settings.
In the no tied events setting, the ESS values for GS4Cox, HMC, and MALA were similar, but GS4Cox significantly outperformed the others in ESR.
This demonstrates that GS4Cox achieved a substantial improvement in execution time.

\begin{table}[htbp]
\caption{Evaluation metrics from synthetic data experiments under two scenarios: (i) without tied events; and (ii) tied events induced by rounding event times with rounding parameter $0.001$.
The compared methods are GS4Cox, MH, HMC, NUTS, MALA, and Cox-PG.}
\label{tab:numerical-data-efficiency}
\vskip 0.15in
\begin{center}
\begin{small}
\begin{sc}
\begin{tabular}{lrrrrrr}
\toprule
 &\multicolumn{3}{c}{without tied events}&\multicolumn{3}{c}{with tied events}\\
\cmidrule(lr){2-4}\cmidrule(lr){5-7}
 & ESS & ESR & MCSE & ESS & ESR & MCSE\\
\midrule
GS4Cox
& $263.21$ & $41.96$ & $0.0066$
& $311.19$ & $43.96$ & $0.0060$\\
MH     
& $25.10$  & $1.04$  & $0.0188$
& $30.63$  & $1.49$  & $0.0163$\\
HMC    
& $221.56$ & $3.47$  & $0.0073$
& $222.03$ & $4.01$  & $0.0073$\\
NUTS   
& $27.04$  & $0.18$  & $0.0165$ 
& $27.06$  & $0.20$  & $0.0164$\\
MALA   
& $201.16$ & $4.43$  & $0.0081$
& $214.75$ & $5.58$  & $0.0079$\\
Cox-PG
& $13.56$  & $1.56$  & $0.0208$
& $12.59$  & $1.44$  & $0.0074$\\
\bottomrule
\end{tabular}
\end{sc}
\end{small}
\end{center}
\vskip -0.1in
\end{table}

Table~\ref{tab:numerical-data-estimates-notie} compares the true parameter values with the MPLE and the posterior means obtained from the MCMC algorithms in the scenario without tied events.
For GS4Cox, as well as for MH, HMC, NUTS, and MALA, the posterior means were very close to the MPLE, whereas the posterior means obtained by Cox-PG were slightly farther from the MPLE.

\begin{table}[htbp]
\caption{
Parameter estimates from synthetic data experiments without tied events. 
The compared methods are MPLE, GS4Cox, MH, HMC, NUTS, MALA, and Cox-PG.
}
\label{tab:numerical-data-estimates-notie}
\vskip 0.15in
\begin{center}
\begin{small}
\begin{sc}
\begin{tabular}{lrrrrrrrr}
\toprule
& $\beta_{1}$ & $\beta_{2}$ & $\beta_{3}$ & $\beta_{4}$
& $\beta_{5}$ & $\beta_{6}$ & $\beta_{7}$ & $\beta_{8}$\\
\midrule
True Parameter
& $1.00$ & $-1.00$ & $0.50$ & $-0.50$
& $0.30$ & $-0.30$ & $0.10$ & $-0.10$
\\
MPLE
& $1.02$ & $-1.05$ & $0.51$ & $-0.56$
& $0.12$ & $-0.36$ & $0.08$ & $-0.10$
\\
GS4Cox
& $1.01$ & $-1.05$ & $0.51$ & $-0.56$
& $0.12$ & $-0.36$ & $0.08$ & $-0.10$
\\
MH
& $1.05$ & $-1.06$ & $0.51$ & $-0.56$
& $0.11$ & $-0.35$ & $0.07$ & $-0.09$
\\
HMC
& $1.01$ & $-1.04$ & $0.52$ & $-0.56$
& $0.13$ & $-0.35$ & $0.08$ & $-0.11$
\\
NUTS
& $1.00$ & $-1.05$ & $0.54$ & $-0.58$
& $0.08$ & $-0.41$ & $0.07$ & $-0.10$
\\
MALA
& $1.00$ & $-1.04$ & $0.52$ & $-0.54$
& $0.14$ & $-0.36$ & $0.08$ & $-0.10$
\\
Cox-PG
& $1.04$ & $-1.16$ & $0.55$ & $-0.60$
& $0.13$ & $-0.28$ & $-0.01$ & $-0.08$
\\
\bottomrule
\end{tabular}
\end{sc}
\end{small}
\end{center}
\vskip -0.1in
\end{table}

\begin{figure}[htb]
\vskip 0.2in
\begin{center}
\centerline{\includegraphics[width=0.65\textwidth]{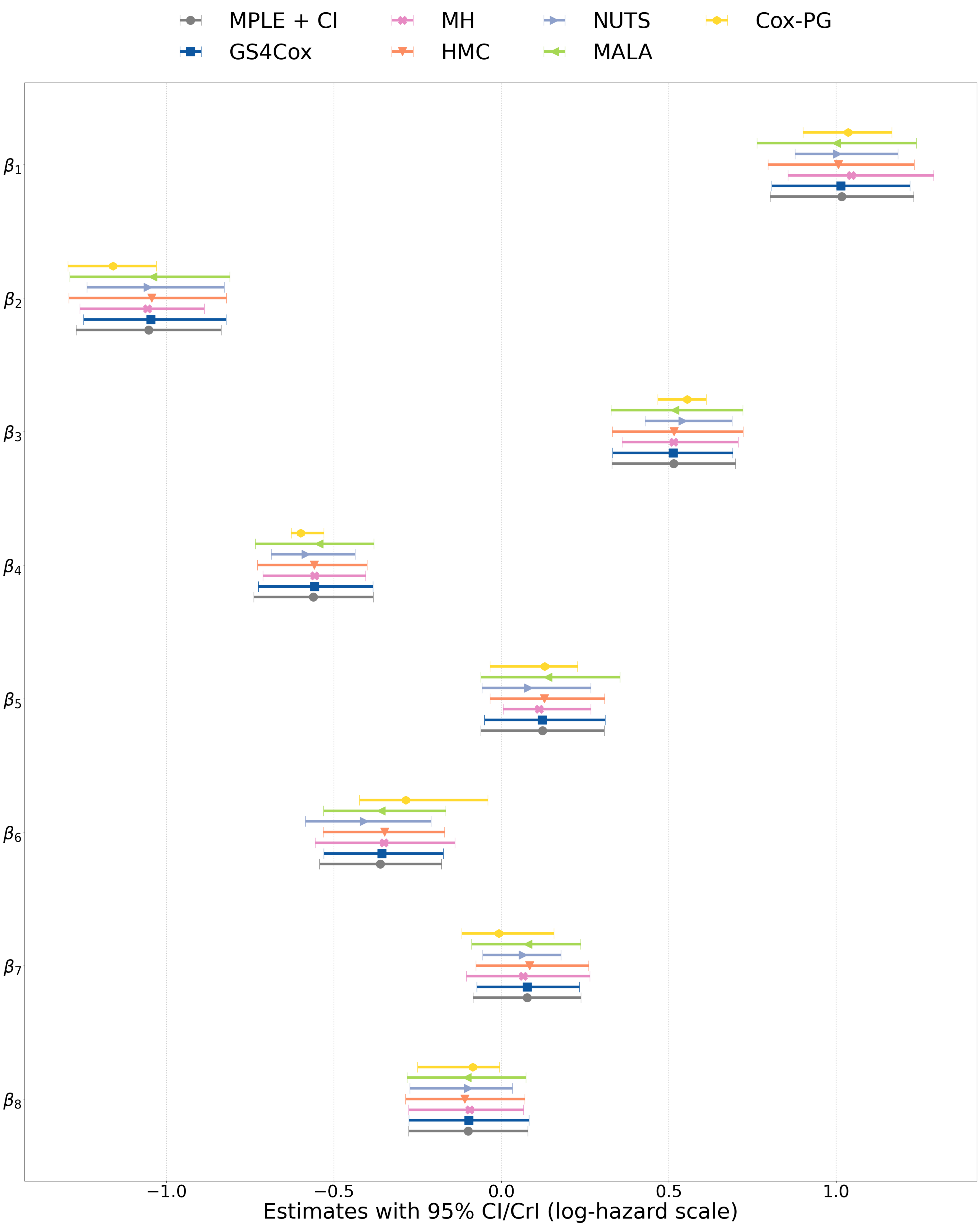}}
\caption{
Forest plot showing the regression coefficients and their corresponding 95\% CI and CrI from synthetic data experiments without tied events.
}
\label{fig:numerical-data-forest-plot}
\end{center}
\vskip -0.2in
\end{figure}

Figure~\ref{fig:numerical-data-forest-plot} shows the forest plot of the regression coefficients under the scenario without tied events.
The result indicates that the posterior means and CrIs obtained from GS4Cox were very close to the MPLE and its corresponding CIs.
HMC and MALA also show stable results, whereas MH, NUTS, and Cox-PG exhibit some variability in the interval widths compared with the CIs corresponding to the MPLE.
The actual values of the CIs and CrIs are provided in Table~\ref{app-tab:numerical-intervals-notie}.

\begin{figure}[t]
\vskip 0.2in
\begin{center}
\centerline{\includegraphics[width=\textwidth]{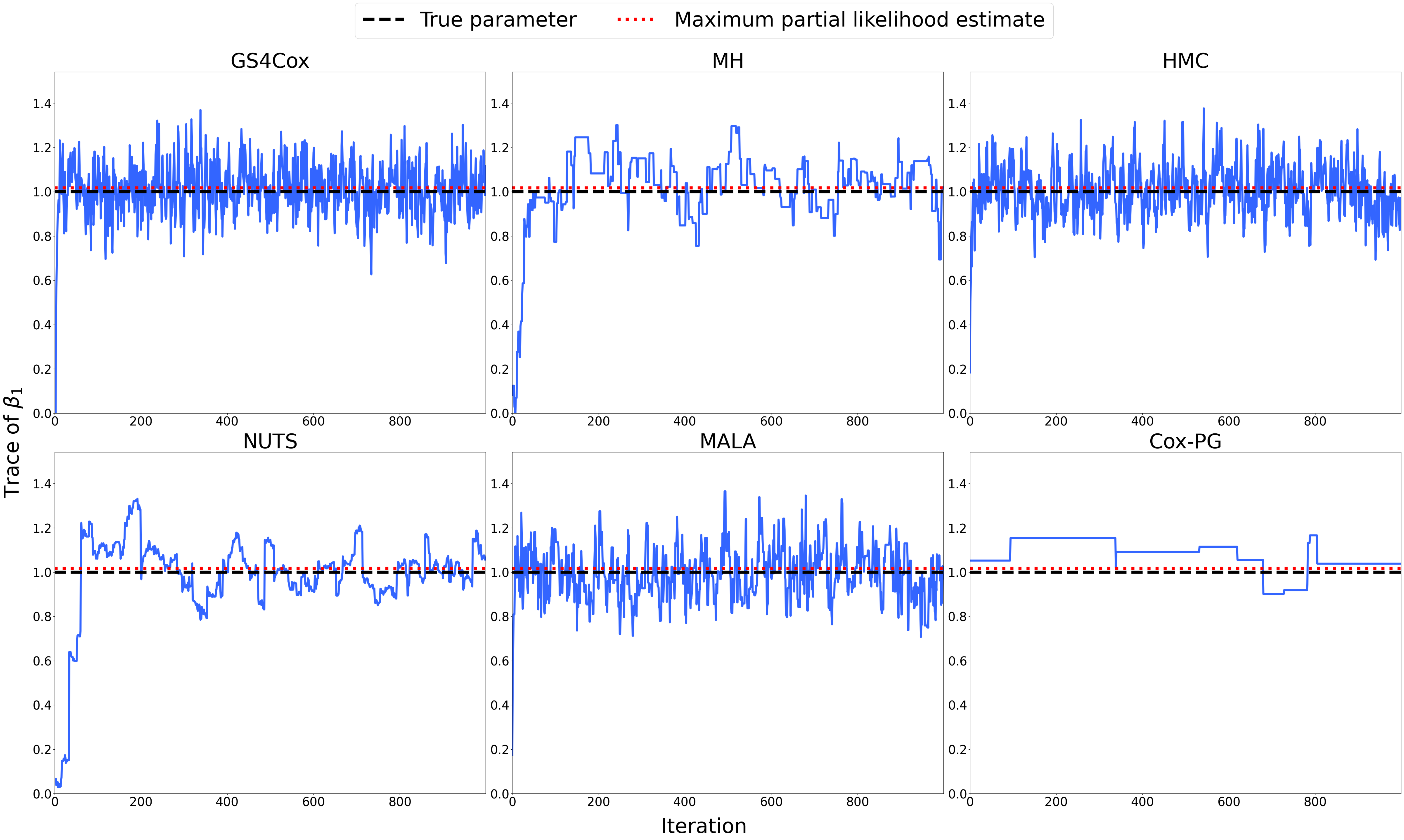}}
\caption{
Trace plots for log-hazard coefficient for $\beta_{1}$ from the numerical experiments without tied events.
The dashed lines indicate the true parameter values, and the dotted lines indicate the maximum partial likelihood estimates.
}
\label{supp-fig:numerical-trace-plot}
\end{center}
\vskip -0.2in
\end{figure}

Additionally, Figure~\ref{supp-fig:numerical-trace-plot} presents the trace plot for the regression parameter $\beta_{1}$.
This result indicates that GS4Cox, HMC, and MALA reached stationarity early in the sampling process.
In contrast, MH and NUTS did not reach stationarity within the common number of iterations ($1,000$) in this experimental setup.
Cox-PG, while producing values close to the MPLE at the initial step, also failed to reach stationarity within the same number of iterations.

Section~\ref{app-subsec:additional-numerical-results} reports the detailed numerical results for the scenarios discussed above and includes additional experiments conducted under different learning rate and sample size settings.

\section{Application to Actual Data}
\label{sec:actual-experiment}
We applied our GS4Cox to the lung dataset from the survival R package \citep{loprinzi1994prospective}.
This dataset describes the natural history of $228$ sequential patients with advanced lung cancer from the North Central Cancer Treatment Group, including the following $10$ variables:
institution code (inst), survival time in days (time), censoring status (status), age at study entry (age), patient's gender (sex), performance status (ph.ecog), Karnofsky performance status assessed by the physician (ph.karno) and the patient (pat.karno), daily caloric intake at study entry (meal.cal), weight loss over the past six months (wt.loss).
After excluding observations with missing values, $167$ of the original $228$ observations remained available for analysis, with $120$ events (deaths).
The Cox regression model was used to evaluate the hazard ratio of death, with adjustment for age, sex, ph.ecog, ph.karno, pat.karno, meal.cal and wt.loss.

As in Section~\ref{sec:num-experiments}, we used ESS, ESR, and MCSE as performance metrics for comparing the MCMC algorithms.
Additionally, we compared the posterior means and CrI with the MPLE and its CI.
Furthermore, as in Section~\ref{subsec:num-baselines}, we compared GS4Cox with MH, HMC, NUTS, MALA, and Cox-PG \citep{ren2025cox}.

Additionally, as in Section~\ref{subsec:num-setup}, each chain was run for $1,000$ iterations with a $500$-iteration burn-in.
We used a Gaussian prior $\boldsymbol{\beta}\sim \mathcal{N}_{p}(\boldsymbol{0}, 100\boldsymbol{I}_{p})$ and set the learning rate to $\eta = 1$.
When ties were present, the MPLE was computed using the Efron approximation.

As a preprocessing step, we standardized covariates for HMC, NUTS, MALA, and Cox-PG to ensure stable performance and harmonize tuning.
In contrast, to assess robustness to scaling, we additionally ran GS4Cox and Metropolis-Hastings on the unstandardized covariates.

\subsection{Results}
\label{subsec:actual-results}
\begin{figure}[t]
\vskip 0.2in
\begin{center}
\centerline{\includegraphics[width=\textwidth]{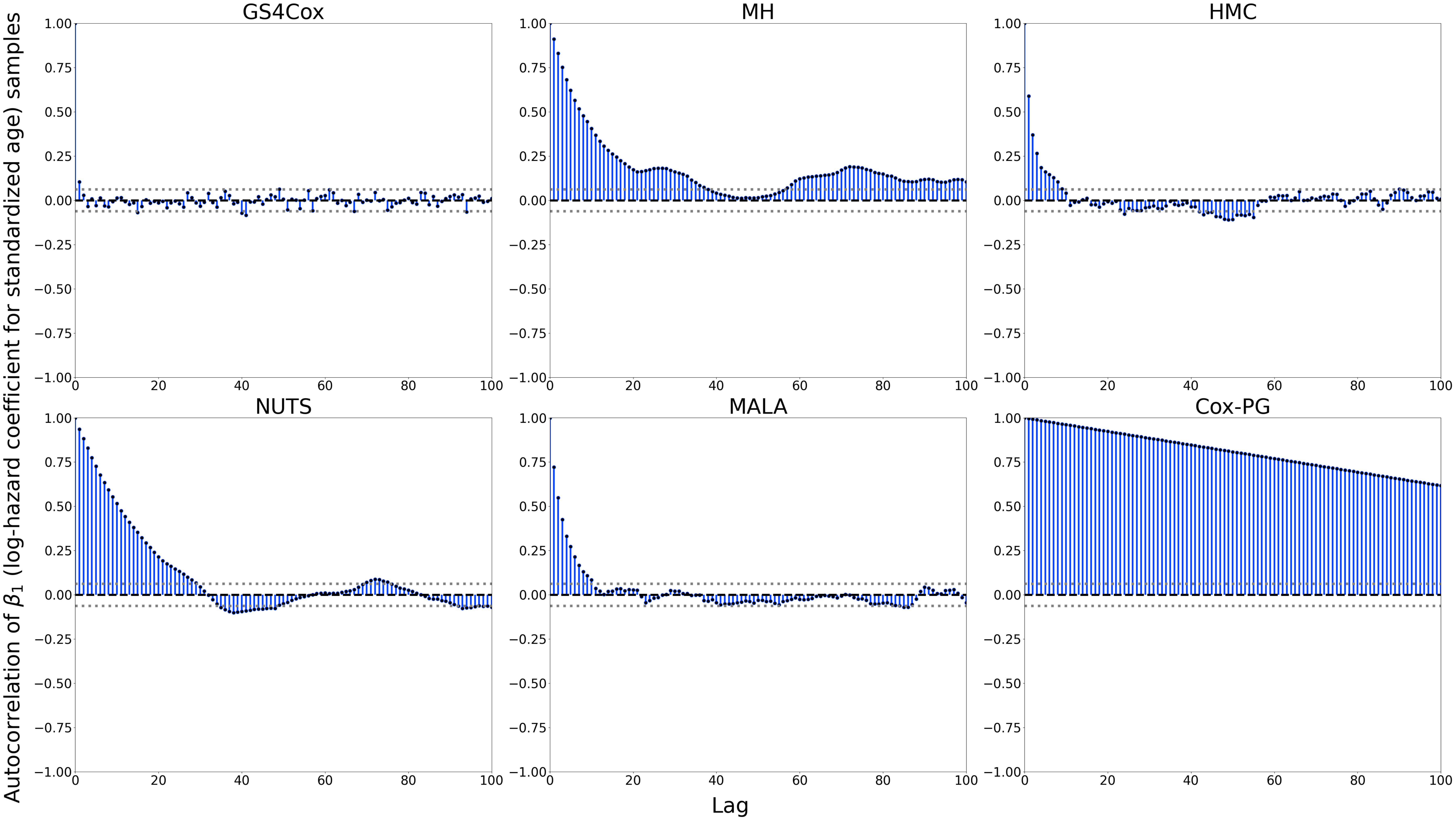}}
\caption{
The correlogram of the sample chains for the standardized age coefficient from the lung dataset experiments.
The dotted lines denote the $\pm 1.96/\sqrt{N}$ reference bounds under the white-noise null hypothesis (zero autocorrelation), where $N$ is the number of iterations.
}
\label{fig:real-correlogram}
\end{center}
\vskip -0.2in
\end{figure}

\begin{figure}[htb]
\vskip 0.2in
\begin{center}
\centerline{\includegraphics[width=0.65\textwidth]{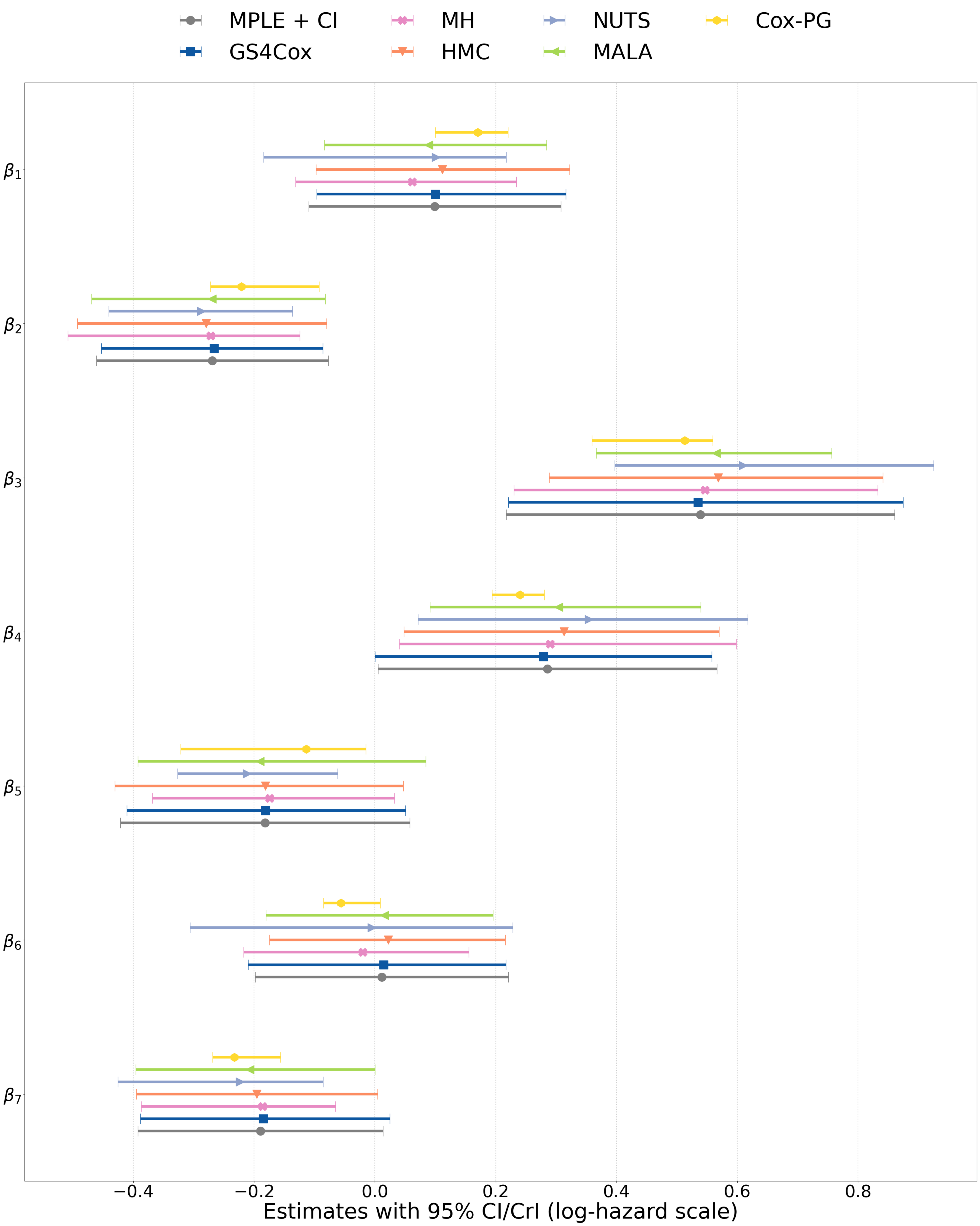}}
\caption{
Forest plot showing the regression coefficients and their corresponding 95\% CI and CrI from the lung dataset.
}
\label{fig:real-data-forest-plot}
\end{center}
\vskip -0.2in
\end{figure}

Table~\ref{tab:actual-data-efficiency} reports the evaluation metrics (ESS, ESR, and MCSE) in the real data analysis.
This shows that our method substantially outperformed the alternatives across all evaluation metrics.
Table~\ref{tab:actual-data-estimates} compares MPLE for each coefficient with posterior means from the MCMC algorithms.
As in the numerical experiments, GS4Cox, as well as MH, HMC, NUTS, and MALA, produced posterior means that were very close to the MPLE.
Figure~\ref{fig:real-correlogram} shows the correlogram for the age coefficient in the lung dataset and exhibited very low autocorrelation, suggesting that our method generated nearly independent draws.
Figure~\ref{fig:real-data-forest-plot} presents the forest plot for the regression coefficients in the lung dataset.
Consistent with the trends observed on our numerical experiments (Section~\ref{subsec:num-results}), this result indicates that the posterior means and CrIs from GS4Cox were very close to the MPLE and its corresponding CIs.
HMC and MALA also show stable performance, whereas MH, NUTS, and Cox-PG exhibit some variability in interval widths relative to the MPLE-based CIs.
The numerical values of all CIs and CrIs are reported in Table~\ref{app-tab:actual-data-intervals}.

Tables~\ref{tab:actual-data-efficiency-unst} and \ref{tab:actual-data-estimates-unst} present, for the unstandardized covariates, a comparison of the evaluation metrics and a comparison of posterior means with the MPLE.
These results indicate that our method maintained high sampling efficiency and accurate estimation even without standardization.

See Section~\ref{app-subsec:additional-actual-results} for the additional results of experiments on the lung data.

\begin{table}[htbp]
\caption{
Evaluation metrics from the lung data experiments.
The compared methods are GS4Cox, MH, HMC, NUTS, MALA, and Cox-PG.
}
\label{tab:actual-data-efficiency}
\vskip 0.15in
\begin{center}
\begin{small}
\begin{sc}
\begin{tabular}{lrrr}
\toprule
& ESS & ESR & MCSE \\
\midrule
GS4Cox & $672.78$ & $272.96$ & $0.0048$\\
MH     & $29.34$ & $1.69$ & $0.0209$\\
HMC    & $104.16$ & $2.33$ & $0.0125$\\
NUTS   & $16.38$ & $0.14$ & $0.0291$\\
MALA   & $84.58$ & $2.62$ & $0.0129$\\
Cox-PG & $11.15$ & $1.83$ & $0.0193$\\
\bottomrule
\end{tabular}
\end{sc}
\end{small}
\end{center}
\vskip -0.1in
\end{table}

\begin{table}[htbp]
\caption{
Coefficient estimates and posterior means (Bayesian methods) for the lung dataset.
The compared methods are MPLE, GS4Cox, MH, HMC, NUTS, MALA, and Cox-PG.
Coefficients: $\beta_{1}=$ age, $\beta_{2}=$ sex, $\beta_{3}=$ ph.ecog, $\beta_{4}=$ ph.karno, $\beta_{5}=$ pat.karno, $\beta_{6}=$ meal.cal, $\beta_{7}=$ wt.loss.
}
\label{tab:actual-data-estimates}
\vskip 0.15in
\begin{center}
\begin{small}
\begin{sc}
\begin{tabular}{lrrrrrrr}
\toprule
& $\beta_{1}$ & $\beta_{2}$ & $\beta_{3}$ & $\beta_{4}$
& $\beta_{5}$ & $\beta_{6}$ & $\beta_{7}$\\
\midrule
MPLE
& $0.10$ & $-0.27$ & $0.54$ & $0.29$
& $-0.18$ & $0.01$ & $-0.19$
\\
GS4Cox
& $0.10$ & $-0.27$ & $0.54$ & $0.28$
& $-0.18$ & $0.01$ & $-0.18$
\\
MH
& $0.06$ & $-0.27$ & $0.55$ & $0.29$
& $-0.17$ & $-0.02$ & $-0.19$
\\
HMC
& $0.11$ & $-0.28$ & $0.57$ & $0.31$
& $-0.18$ & $0.02$ & $-0.20$
\\
NUTS
& $0.10$ & $-0.29$ & $0.61$ & $0.36$
& $-0.21$ & $-0.00$ & $-0.22$
\\
MALA
& $0.09$ & $-0.27$ & $0.56$ & $0.30$
& $-0.19$ & $0.02$ & $-0.21$
\\
Cox-PG
& $0.17$ & $-0.22$ & $0.51$ & $0.24$
& $-0.11$ & $-0.06$ & $-0.23$
\\
\bottomrule
\end{tabular}
\end{sc}
\end{small}
\end{center}
\vskip -0.1in
\end{table}

\begin{table}[htb]
\caption{
Evaluation metrics from the lung data experiments using unstandardized covariates.
The compared methods are GS4Cox and MH.
}
\label{tab:actual-data-efficiency-unst}
\vskip 0.15in
\begin{center}
\begin{small}
\begin{sc}
\begin{tabular}{lrrr}
\toprule
& ESS & ESR & MCSE \\
\midrule
GS4Cox & $852.09$ & $350.92$ & $0.0031$\\
MH     & $34.05$  & $1.99$   & $0.0090$\\
\bottomrule
\end{tabular}
\end{sc}
\end{small}
\end{center}
\vskip -0.1in
\end{table}

\begin{table}[t]
\caption{
Coefficient estimates and posterior means (Bayesian methods) for the lung dataset using unstandardized covariates.
The compared methods are MPLE, GS4Cox, and MH.
Coefficients: $\beta_{1}=$ age, $\beta_{2}=$ sex, $\beta_{3}=$ ph.ecog, $\beta_{4}=$ ph.karno, $\beta_{5}=$ pat.karno, $\beta_{6}=$ meal.cal, $\beta_{7}=$ wt.loss.
}
\label{tab:actual-data-estimates-unst}
\vskip 0.15in
\begin{center}
\begin{small}
\begin{sc}
\begin{tabular}{lrrrrrrr}
\toprule
& $\beta_{1}$ & $\beta_{2}$ & $\beta_{3}$ & $\beta_{4}$
& $\beta_{5}$ & $\beta_{6}$ & $\beta_{7}$\\
\midrule
MPLE
& $0.01$ & $-0.55$ & $0.74$ & $0.02$
& $-0.01$ & $0.00$ & $-0.01$
\\
GS4Cox
& $0.01$ & $-0.55$ & $0.73$ & $0.02$
& $-0.01$ & $0.00$ & $-0.01$
\\
MH
& $0.01$ & $-0.48$ & $0.75$ & $0.02$
& $-0.01$ & $0.00$ & $-0.01$
\\
\bottomrule
\end{tabular}
\end{sc}
\end{small}
\end{center}
\vskip -0.1in
\end{table}

\section{Discussion}
\label{sec:discussion}
We developed a new Gibbs sampling algorithm for the Cox regression models, namely GS4Cox.
To construct the algorithm, we first define a composite partial likelihood and establish that it delivers a valid estimating equation for the Cox regression models.
Exploiting the fact that this composite partial likelihood can be written in a form analogous to a logistic regression likelihood, we embed it within the P\'olya-Gamma augmentation scheme and the general Bayesian framework to obtain closed-form Gaussian complete conditional distributions for the regression parameters.
We demonstrated that GS4Cox achieved substantially higher sampling efficiency and estimation accuracy than conventional MCMC methods (e.g., MH, HMC, NUTS, MALA) and the Cox-PG algorithm \citep{ren2025cox}.
Moreover, GS4Cox remains effective without requiring covariate standardization, which is especially valuable when interpretability of raw coefficients is important.
For example, in medicine and epidemiology, raw regression coefficients are interpreted as log-hazard ratios for a one-unit increase in the corresponding covariate.

\citet{ren2025cox} approximate the log-cumulative baseline hazard with splines and introduce Gamma frailties, producing a negative-binomial surrogate that requires Metropolis-Hastings corrections.
In contrast, we form a generalized posterior from a composite partial likelihood and apply an affine posterior calibration that re-centers the posterior on the partial-likelihood target and maps its covariance to the sandwich limit.
This yields a fully Gibbs procedure with no rejection steps, first-order location correction, and calibrated uncertainty.

The proposed composite partial likelihood is the one-vs-each lower bound to the partial likelihood in the sense discussed by \citet{titsias2016one-vs-each}.
Building on Propositions~\ref{prop:expectation-zero}, \ref{prop:cpl-weak-consistency}, it induces a mean-zero estimating equation and a consistent estimator;
Theorem~\ref{thm:bayes-asym-unbiased} shows the associated generalized posterior mean is asymptotically unbiased.
A known trade-off is that composite constructions discard some dependence, inflating Godambe variance relative to the partial likelihood.
We address this using an open-faced sandwich-style affine posterior calibration \citet{shaby2014open-faced} that both re-centers the draws to the partial likelihood target and maps their covariance to the sandwich limit, thereby calibrating point estimates and uncertainty while retaining a fully Gibbs algorithm.

When using a general Bayesian update, the learning rate governs posterior scale and therefore coverage.
Its selection remains debated.
Comparison studies such as \citet{wu2023comparison} review competing strategies, but many require substantial computation or exhibit instability across scenarios.
For example, the generalized posterior calibration algorithm of \citet{syring2019calibrating} iteratively tunes the learning rate until bootstrap coverage matches the nominal level, and \citet{grunwald2017inconsistency} consider grid search over $(0,1]$ with repeated refits;
both can be computationally intensive and susceptible to local minima.
Information-matching and asymptotic approaches \citep{holmes2017assigning,lyddon2019general} may also perform unevenly in practice.
\citet{collins2025efficient} propose an ostensibly learning rate-free scheme that combines generalized posterior calibration algorithm with the open-faced sandwich adjustment, but their mapping admits a universal solution $\boldsymbol{\Omega}=\boldsymbol{0}$, raising identifiability concerns in general settings.
Our contribution provides a different solution to the calibration problem.
By applying the open-faced sandwich adjustment to the generalized posterior under the P\'olya-Gamma augmentation, we show that the learning rate-dependent posterior variance can be exactly cancelled by the affine open-faced sandwich adjustments.
The resulting Gibbs sampler is therefore learning-rate invariant for purposes of uncertainty quantification.

In summary, GS4Cox delivers
(i) asymptotically unbiased point estimation relative to the partial likelihood benchmark, 
(ii) principled, partial likelihood-aligned uncertainty via open-faced sandwich,
and (iii) superior practical efficiency without requiring covariate standardization.


\section*{Code availability}
The Python implementation of proposed method and simulation experiments in this study is available at \url{https://github.com/shutech2001/GS4Cox}.

\bibliographystyle{apalike}
\bibliography{bibliography}

\newpage
\appendix
\onecolumn
\setcounter{figure}{0}
\setcounter{table}{0}
\renewcommand{\thefigure}{\thesection.\arabic{figure}}
\renewcommand{\thetable}{\thesection.\arabic{table}}
\section{Data-Generating Process under the Cox Proportional Hazards Model}
\label{app-sec:dgp}

Let $\{(T_i,\delta_i,\boldsymbol{X}_i)\}_{i=1}^n$ be independent and identically distributed samples from the true data-generating distribution.
For each subject $i$, define the counting process
\begin{equation*}
    N_i(t) = 
    \begin{cases}
        0, & t< T_i,\\
        1, & \text{$t\geq T_i$ and $\delta_i=1$},\\
        0, & \text{$t\geq T_i$ and $\delta_i=0$},
    \end{cases}
\end{equation*}
and the corresponding risk process
\begin{equation*}
    Y_i(t)=\boldsymbol{1}\{T_i\geq t\}.
\end{equation*}
Let $\mathcal{F}_{t^-}$ denote the filtration up to just before time $t$.
The conditional intensity of $N_i(t)$ is defined as
\begin{equation*}
    \begin{split}
        \lambda_i(t) &:= \lim_{\Delta t \to 0} \frac{1}{\Delta t} \Pr(N_i(t+\Delta t) - N_i(t)=1\mid \mathcal{F}_{t^-}),
    \end{split}
\end{equation*}
and $\dd N_i(t)$ denotes its increment over an infinitesimal time interval $[t, t+\dd t)$.
In the Cox regression model, the intensity can be expressed as
\begin{equation*}
    \lambda_i(t) = Y_i(t)h_0(t)\exp(\boldsymbol{X}_i^\top\boldsymbol{\beta}_{0}).
\end{equation*}
Let $\mathbb{E}_{P_{\boldsymbol{\beta}_{0}}}[\cdot]$ denote expectation with respect to the defined data-generating distribution.

\section{Omitted Proofs}
\label{app-sec:omitted-proof}

\subsection{Regularity Conditions}
\label{app-subsec:regularity-conditions}

Throughout the proofs, we define the empirical log-partial likelihood as
\begin{equation*}
    \ell_{n,\mathrm{PL}}\left(\boldsymbol{\beta}\right)
    :=
    \sum_{i=1}^n \delta_i\left(\boldsymbol{X}_i^\top\boldsymbol{\beta}-\log\sum_{j\in R(T_i)}\exp\left(\boldsymbol{X}_j^\top\boldsymbol{\beta}\right)\right)
    ,
\end{equation*}
with its averaged version as $\bar\ell_{n,\mathrm{PL}}(\boldsymbol{\beta}):=n^{-1}\ell_{n,\mathrm{PL}}(\boldsymbol{\beta})$.
Similarly, we define the empirical log-composite partial likelihood by
\begin{equation*}
    \begin{split}
        \ell_{n,\mathrm{CPL}}(\boldsymbol{\beta}) &:= \sum_{i=1}^n \sum_{j\in R(T_i)\setminus\{i\}} \delta_i \log p_{ij}(\boldsymbol{\beta}),\\
        &= \sum_{i<j}\left[\delta_i\boldsymbol{1}\{j\in R(T_i)\}\log p_{ij}(\boldsymbol{\beta})+ \delta_j\boldsymbol{1}\{i\in R(T_j)\} \log p_{ji}(\boldsymbol{\beta})\right],\\
        &= \sum_{i<j}k(\mathcal{D}_i, \mathcal{D}_j; \boldsymbol{\beta}),
    \end{split}
\end{equation*}
where $k(\mathcal{D}_{i}, \mathcal{D}_{j}; \boldsymbol{\beta}) := \delta_i\boldsymbol{1}\{j\in R(T_i)\}\log p_{ij}(\boldsymbol{\beta})+ \delta_j\boldsymbol{1}\{i\in R(T_j)\} \log p_{ji}(\boldsymbol{\beta})$, and its averaged form 
\begin{equation*}
    \bar\ell_{n,\mathrm{CPL}}(\boldsymbol{\beta}) :=\binom{n}{2}^{-1}\ell_{n,\mathrm{CPL}}(\boldsymbol{\beta})
    .
\end{equation*}

We further define the expected pairwise log contribution
\begin{equation*}
    U(\boldsymbol{\beta}):=\mathbb{E}_{P_{\boldsymbol{\beta}_0}}\left[k(\mathcal{D}_1,\mathcal{D}_2;\boldsymbol{\beta})\right]
    ,
\end{equation*}
and set
\begin{equation*}
    \ell_{\mathrm{CPL}}(\boldsymbol{\beta}):=\binom{n}{2}U(\boldsymbol{\beta})
    , 
    \quad
    \bar\ell_{\mathrm{CPL}}(\boldsymbol{\beta}):= U(\boldsymbol{\beta})
    .
\end{equation*}
Throughout, let $\boldsymbol{\beta}_{0}\in \mathbb{R}^{p}$ denote the true parameter.
Let $\ell(\mathcal{D}_i \mid \boldsymbol{\beta})$ be a generic loss contribution used in the general Bayesian inference.
Define the empirical loss and its average by
\begin{equation*}
    \mathcal{L}_{n}(\mathcal{D}\mid \boldsymbol{\beta})
    :=
    \sum_{i=1}^{n} \ell(\mathcal{D}_{i}\mid \boldsymbol{\beta}),
    \quad
    \bar{\mathcal{L}}_{n}(\boldsymbol{\beta})
    :=
    \frac{1}{n}\mathcal{L}_{n}(\mathcal{D}\mid \boldsymbol{\beta}).
\end{equation*}
We write $\mathcal{L}_{\mathcal{D}}(\boldsymbol{\beta}) := \bar{\mathcal{L}}_{n}(\boldsymbol{\beta})$ for the empirical average loss, and denote its population counterpart by
\begin{equation*}
    \mathcal{L}(\boldsymbol{\beta})
    :=
    \mathbb{E}_{P_{\boldsymbol{\beta}_0}}\left[\mathcal{L}_{\mathcal{D}}(\boldsymbol{\beta})\right]
    = 
    \mathbb{E}_{P_{\boldsymbol{\beta}_0}}\left[\ell(\mathcal{D}_1\mid \boldsymbol{\beta})\right].
\end{equation*}

We introduce a single estimating-function kernel $\psi_{\boldsymbol{\beta}}: \mathcal{D}^{m} \to \mathbb{R}^p$, assumed measurable and, when $m=2$, symmetric in its arguments.
Define the sample estimating function and its population counterpart by
\begin{equation*}
  \Psi_n(\boldsymbol{\beta})
  := \binom{n}{m}^{-1} \sum_{1\le i_1<\cdots<i_m\le n} \psi_{\boldsymbol{\beta}}(\mathcal{D}_{i_1},\dots,\mathcal{D}_{i_m}), 
\end{equation*}
and
\begin{equation*}
  \Psi(\boldsymbol{\beta})
  := \mathbb{E}_{P_{\boldsymbol{\beta}_0}}\left[\psi_{\boldsymbol{\beta}}(\mathcal{D}_1,\dots,\mathcal{D}_m)\right],  
\end{equation*}
respectively.

In our Cox analysis we also use the partial-likelihood score
$S_{\mathrm{PL}}(\boldsymbol{\beta}) := \nabla_{\boldsymbol{\beta}}\ell_{n,\mathrm{PL}}(\boldsymbol{\beta})$,
and for the composite partial likelihood developments we take $m=2$ with
\begin{equation*}
    \psi_{\boldsymbol{\beta}}(\mathcal{D}_i,\mathcal{D}_j)
    := \nabla_{\boldsymbol{\beta}}\,k(\mathcal{D}_i,\mathcal{D}_j;\boldsymbol{\beta}),
\end{equation*}
so that $\Psi_n(\boldsymbol{\beta})$ becomes the averaged composite score and
$\Psi(\boldsymbol{\beta}) = \nabla_{\boldsymbol{\beta}} U(\boldsymbol{\beta})$.

We then impose the following regularity conditions:
\begin{enumerate}[label=(RC\arabic*)]
    \item\label{item:reg-1}
        The parameter space $\mathcal{B} \subset \mathbb{R}^{p}$ is a nonempty compact set with nonempty interior, and the true parameter value $\boldsymbol{\beta}_{0}$ belongs to the interior of $\mathcal{B}$.
        The parameter $\boldsymbol{\beta}$ is restricted to $\mathcal{B}$.
        The prior density $\pi$ is $C^{2}$ and strictly positive on a neighborhood of $\boldsymbol{\beta}_{0}$.
    \item\label{item:reg-2}
        The model is identified via expected score (estimating) functions:
        \begin{equation*}
            \mathbb{E}_{P_{\boldsymbol{\beta}_{0}}}[S_{\mathrm{PL}}(\boldsymbol{\beta}_{0})] = 0
            ,
            \quad
            \mathbb{E}_{P_{\boldsymbol{\beta}_{0}}}[S_{\mathrm{PL}}(\boldsymbol{\beta})] \ne 0
            \text{ for }\boldsymbol{\beta} \ne \boldsymbol{\beta}_0
            ,
        \end{equation*}
        and, likewise, $\Psi(\boldsymbol{\beta}_{0}) = 0$ with $\Psi(\boldsymbol{\beta}) \ne 0$ for $\boldsymbol{\beta} \ne \boldsymbol{\beta}_0$.
    \item\label{item:reg-3}
        There exists a neighborhood $\mathcal{B}_{\mathcal{N}} \subset \mathcal{B}$ of $\boldsymbol{\beta}_{0}$ such that
        \begin{equation*}
            \bar{\ell}_{n, \mathrm{PL}}, \, 
            \bar{\ell}_{n, \mathrm{CPL}}, \, 
            \mathcal{L}_{\mathcal{D}}
            \in C^{3}(\mathcal{B}_{\mathcal{N}})
            ,
            \quad
            \sup_{\boldsymbol{\beta}\in \mathcal{B}_{\mathcal{N}}}\left\|\nabla^3_{\boldsymbol{\beta}}\bar{\ell}_{n, \cdot}(\boldsymbol{\beta})\right\| \le C
            ,
            \quad
            \sup_{\boldsymbol{\beta}\in \mathcal{B}_{\mathcal{N}}}\left\|\nabla^3_{\boldsymbol{\beta}}\mathcal{L}_{\mathcal{D}}(\boldsymbol{\beta})\right\| \le C
            ,
        \end{equation*}
        for some constant $C$ not depending on $n$.
    \item\label{item:reg-4}
        For $k=1,2$,
        \begin{equation*}
            \sup_{\boldsymbol{\beta}\in \mathcal{B}_{\mathcal{N}}}\left\|\nabla^k_{\boldsymbol{\beta}}\bar{\ell}_{n,\cdot}(\boldsymbol{\beta})\right\| = O_p(1)
            ,
        \end{equation*}
        and the derivatives converge uniformly to their expectations:
        \begin{equation*}
            \sup_{\boldsymbol{\beta}\in \mathcal{B}_{\mathcal{N}}}\left\|\nabla^2_{\boldsymbol{\beta}}\bar{\ell}_{n,\mathrm{PL}}(\boldsymbol{\beta}) + \boldsymbol{J}_{0, \mathrm{PL}}\right\| \overset{p}{\longrightarrow} 0
            ,
            \quad
            \partial_{\boldsymbol{\beta}}\Psi_{n}(\boldsymbol{\beta}_0)\overset{p}{\longrightarrow} \partial_{\boldsymbol{\beta}}\Psi(\boldsymbol{\beta}_{0})
            ,
        \end{equation*}
        where $\boldsymbol{J}_{0, \mathrm{PL}} := -\mathbb{E}_{P_{\boldsymbol{\beta}_{0}}}[\nabla^2_{\boldsymbol{\beta}}\bar{\ell}_{n,\mathrm{PL}}(\boldsymbol{\beta}_{0})]$.
    \item\label{item:reg-5}
        $\boldsymbol{J}_{0, \mathrm{PL}} := -\mathbb{E}_{P_{\boldsymbol{\beta}_{0}}}[\nabla^2_\beta \bar\ell_{n,\mathrm{PL}}(\beta_0)] \succ 0$.
        Moreover, for $\mathcal{L}_{\mathcal{D}}$,
        \begin{equation*}
            \boldsymbol{H}_{n} := \nabla^2_{\boldsymbol{\beta}}\mathcal{L}_{\mathcal{D}}(\boldsymbol{\beta}_0) \to \boldsymbol{H}_{0} \succ 0
            .
        \end{equation*}
    \item\label{item:reg-6}
        $\mathbb{E}_{P_{\boldsymbol{\beta}_{0}}}[\|S_{\mathrm{PL}}(\boldsymbol{\beta}_0)\|^2] < \infty$ and $\mathbb{E}_{P_{\boldsymbol{\beta}_{0}}}[\|\varphi_1(\boldsymbol{\beta}_0)\|^2] < \infty$, where
        \begin{equation*}
            \varphi_i(\boldsymbol{\beta}) := \delta_i \sum_{j\in R(T_{i})\setminus \{i\}}\left\{1-p_{ij}(\boldsymbol{\beta})\right\}\left(\boldsymbol{X}_{i} - \boldsymbol{X}_{j}\right)
            .
        \end{equation*}
    \item\label{item:reg-7}
        There exists a stationary point $\hat{\boldsymbol{\beta}}_n$ of $\mathcal{L}_{\mathcal{D}}$ with $\hat{\boldsymbol{\beta}}_n \overset{p}{\to} \boldsymbol{\beta}_0$.
        Moreover, for every $\epsilon > 0$,
        \begin{equation*}
            \inf_{\boldsymbol{\beta}\in \mathcal{B}\setminus B_\epsilon(\boldsymbol{\beta}_0)}
            \left\{\mathcal{L}(\boldsymbol{\beta}) - \mathcal{L}(\boldsymbol{\beta}_0) \right\} > 0
            ,
            \quad
            \sup_{\boldsymbol{\beta}\in\mathcal{B}}
            \left|\mathcal{L}_{\mathcal{D}}(\boldsymbol{\beta})-\mathcal{L}(\boldsymbol{\beta})\right|
            \overset{p}{\longrightarrow} 0
            ,
        \end{equation*}
        where $B_\epsilon(\boldsymbol{\beta}) := \{\boldsymbol{b}\in\mathcal{B}:\|\boldsymbol{b}-\boldsymbol{\beta}\|<\epsilon\}$ denotes the open Euclidean ball.
    \item\label{item:reg-8}
        $\sqrt{n}\Psi_{n}(\boldsymbol{\beta}_{0}) \to \mathcal{N}(0, \boldsymbol{K}_{0})$ with $\boldsymbol{K}_{0} = \operatorname{Var}_{P_{\boldsymbol{\beta}_{0}}}[\psi_{\boldsymbol{\beta}_{0}}(D_1)]$, where $D_1 := (\mathcal{D}_1,\ldots,\mathcal{D}_m)$ denotes an $m$-tuple of i.i.d.\ draws from $P_{\boldsymbol{\beta}_0}$ (in particular, when $m=1$, $D_1=\mathcal{D}_1$), and $\operatorname{Var}_{P}[\cdot]$ denotes variance under the probability measure $P$.
\end{enumerate}

\subsection{Proof of Proposition~\ref{prop:expectation-zero}}
\label{app-subsec:proof-of-expectation-zero}

\begin{proof}[Proof of Proposition~\ref{prop:expectation-zero}]
    Under the data-generating process defined in Section~\ref{app-sec:dgp}, we can express \eqref{eq:cpl-score-function} as the Riemann-Stieltjes integral representation \citep{hildebrandt1938definitions} as follows:
    \begin{equation*}
        \begin{split}
            S_{\mathrm{CPL}}(\boldsymbol{\beta}) &= \sum_{i=1}^n \sum_{j\in R(T_i)\setminus \{i\}} \delta_i \left\{1-p_{ij}(\boldsymbol{\beta})\right\}\left(\boldsymbol{X}_i-\boldsymbol{X}_j\right)
            ,
            \\
            &= \sum_{i=1}^n \sum_{j\neq i} \int_{0}^\tau Y_j(t) \left\{1-p_{ij}(\boldsymbol{\beta})\right\}\left(\boldsymbol{X}_i-\boldsymbol{X}_j\right)\dd N_i(t),
        \end{split}
    \end{equation*}
    Using \citet[Lem.~7.2.V]{daley2003introduction}, the expectation can be expressed as
    \begin{equation*}
        \begin{split}
            \mathbb{E}_{P_{\boldsymbol{\beta}_{0}}}\left[S_{\mathrm{CPL}}(\boldsymbol{\beta})\right] 
            &= \sum_{i=1}^n\sum_{j\neq i}\mathbb{E}_{P_{\boldsymbol{\beta}_{0}}}\left[\int_{0}^\tau Y_j(t) \left\{1-p_{ij}(\boldsymbol{\beta})\right\}\left(\boldsymbol{X}_i-\boldsymbol{X}_j\right)\dd N_i(t)\right],\\
            &= \sum_{i=1}^n\sum_{j\neq i}\mathbb{E}_{P_{\boldsymbol{\beta}_{0}}}\left[\int_{0}^\tau Y_j(t) \left\{1-p_{ij}(\boldsymbol{\beta})\right\}\left(\boldsymbol{X}_i-\boldsymbol{X}_j\right)\lambda_i(t)\dd t\right],\\
            &= \sum_{i=1}^n\sum_{j\neq i}\mathbb{E}_{P_{\boldsymbol{\beta}_{0}}}\left[\int_{0}^\tau Y_j(t) \left\{1-p_{ij}(\boldsymbol{\beta})\right\}\left(\boldsymbol{X}_i-\boldsymbol{X}_j\right)Y_i(t)h_0(t)\exp(\boldsymbol{X}_i^\top\boldsymbol{\beta}_{0})\dd t\right].
        \end{split}
    \end{equation*}
    Define the integrand
    \begin{equation*}
        A_{ij}(t) := Y_j(t) \left\{1-p_{ij}(\boldsymbol{\beta})\right\}\left(\boldsymbol{X}_i-\boldsymbol{X}_j\right)Y_i(t)h_0(t)\exp(\boldsymbol{X}_i^\top\boldsymbol{\beta}_{0}).
    \end{equation*}
    Since $A_{ji}(t)=Y_j(t)Y_i(t)(\boldsymbol{X}_j-\boldsymbol{X}_i)\{1-p_{ji}(\boldsymbol{\beta})\}h_0(t)\exp(\boldsymbol{X}_j^\top\boldsymbol{\beta}_{0})=-A_{ij}(t)$, summation over all pairs $i\neq j$ collapses:
    \begin{equation*}
        \sum_{i\neq j} A_{ij}(t) = \sum_{\{i,j\}}\left\{A_{ij}(t)+A_{ji}(t)\right\} = 0.
    \end{equation*}
    Therefore, $\mathbb{E}_{P_{\boldsymbol{\beta}_{0}}}[S_{\mathrm{CPL}}(\boldsymbol{\beta})]=0$.
\end{proof}

\subsection{Proof of Proposition~\ref{prop:cpl-weak-consistency}}
\label{app-subsec:proof-of-consistency}

\begin{proof}[Proof of Proposition~\ref{prop:cpl-weak-consistency}]
    Under the regularity conditions~\ref{item:reg-1} and \ref{item:reg-3}, since for each $\boldsymbol{X}=\{\boldsymbol{X}_1,\dots, \boldsymbol{X}_n\}\in \mathcal{X}^n$ the function $\ell_{n,\mathrm{CPL}}(\boldsymbol{\beta})$ is continuous on $\mathcal{B}$, a maximum composite partial likelihood estimator $\hat{\boldsymbol{\beta}}_{\mathrm{CPL}}\in \mathcal{B}$ exists for every $\boldsymbol{X} \in \mathcal{X}^n$.
    
    Setting $\bar{\ell}_{n,\mathrm{CPL}}(\boldsymbol{\beta})=U_n^{(k)}(\boldsymbol{\beta})$, and noting that $k$ is manifestly a symmetric kernel, we can regard $U_n^{(k)}(\boldsymbol{\beta})$ as a $U$-statistic \citep{hoeffding1948class}.
    Consequently, \citet[Thm.~A in \S5.4]{serfling2009approximation} implies
    \begin{equation*}
        \sup_{\boldsymbol{\beta}\in \mathcal{B}}|U_n^{(k)}(\boldsymbol{\beta})-U(\boldsymbol{\beta})| \overset{P_{\boldsymbol{\beta}_0}}{\longrightarrow}0,
        \quad
        n\to\infty,
    \end{equation*}
    which can equivalently be written as
    \begin{equation*}
        \lim_{n\to\infty} P_{\boldsymbol{\beta}_0}\left(\sup_{\boldsymbol{\beta}\in \mathcal{B}}\left|U_n^{(k)}(\boldsymbol{\beta}) - U(\boldsymbol{\beta})\right| > \epsilon\right) = 0,
        \quad
        \forall \epsilon > 0.
    \end{equation*}
    By regularity condition~\ref{item:reg-2}, the population estimating function $\Psi(\boldsymbol{\beta}) = \nabla_{\boldsymbol{\beta}}U(\boldsymbol{\beta})$ vanished if and only if $\boldsymbol{\beta} = \boldsymbol{\beta}_{0}$.
    Moreover, each term $\log p_{ij}(\boldsymbol{\beta})$ is concave in $\boldsymbol{\beta}$, so $U$ is concave on $\mathcal{B}$ and $\boldsymbol{\beta}_{0}$ is the unique maximizer of $U$.
    Hence
    \begin{equation*}
        \ell_{\mathrm{CPL}}(\boldsymbol{\beta}_{0})
        =
        \binom{n}{2}U(\boldsymbol{\beta}_{0})
        >
        \binom{n}{2}U(\boldsymbol{\beta})
        =
        \ell_{\mathrm{CPL}}(\boldsymbol{\beta})
        ,
        \quad
        \text{for every $\boldsymbol{\beta}\neq \boldsymbol{\beta}_{0}$}
        .
    \end{equation*}
    In particular, for any $\epsilon > 0$,
    \begin{equation*}
        \ell_{\mathrm{CPL}}(\boldsymbol{\beta}_{0})
        >
        \sup_{\boldsymbol{\beta}\in \mathcal{B}: \|\boldsymbol{\beta} - \boldsymbol{\beta}_{0}\| > \epsilon} \ell_{\mathrm{CPL}}(\boldsymbol{\beta})
        ,
    \end{equation*}
    so there exists $\delta > 0$ such that
    \begin{equation*}
        \sup_{\boldsymbol{\beta}\in \mathcal{B}: \|\boldsymbol{\beta}-\boldsymbol{\beta}_0\|>\epsilon} \bigl[\ell_{\mathrm{CPL}}(\boldsymbol{\beta})-\ell_{\mathrm{CPL}}(\boldsymbol{\beta}_0)\bigr] < -\delta.
    \end{equation*}
    Next, we define the set
    \begin{equation*}
        \mathcal{C}_n(\delta) := \left\{\boldsymbol{X} \in \mathcal{X}^n: \sup_{\boldsymbol{\beta}\in\mathcal{B}}\left|\ell_{n, \mathrm{CPL}}(\boldsymbol{\beta})-\ell_{\mathrm{CPL}}(\boldsymbol{\beta})\right|\leq \frac{\delta}{2}\right\}.
    \end{equation*}
    On $\mathcal{C}_n(\delta)$, for any $\boldsymbol{\beta}\in \mathcal{B}$ with $\|\boldsymbol{\beta}-\boldsymbol{\beta}_0\| > \epsilon$, we have
    \begin{equation*}
        \ell_{n,\mathrm{CPL}}(\boldsymbol{\beta}) - \ell_{\mathrm{CPL}}(\boldsymbol{\beta})\leq \frac{\delta}{2}, \quad
        \ell_{\mathrm{CPL}}(\boldsymbol{\beta})-\ell_{\mathrm{CPL}}(\boldsymbol{\beta}_0) < -\delta.
    \end{equation*}
    Adding these yields
    \begin{equation*}
        \left\|\boldsymbol{\beta}-\boldsymbol{\beta}_0\right\|>\epsilon \Longrightarrow \ell_{n,\mathrm{CPL}}(\boldsymbol{\beta})-\ell_{\mathrm{CPL}}(\boldsymbol{\beta}_0) < -\frac{\delta}{2}.
    \end{equation*}
    Taking contrapositives gives
    \begin{equation*}
        \ell_{n, \mathrm{CPL}}(\boldsymbol{\beta})-\ell_{\mathrm{CPL}}(\boldsymbol{\beta}_0)\geq -\frac{\delta}{2}\Rightarrow \left\|\boldsymbol{\beta}-\boldsymbol{\beta}_0\right\|\leq \epsilon.
    \end{equation*}
    By definition, $\hat{\boldsymbol{\beta}}_{\mathrm{CPL}}=\arg\sup_{\boldsymbol{\beta}\in \mathcal{B}}\ell_{n,\mathrm{CPL}}(\boldsymbol{\beta})=\arg\sup_{\boldsymbol{\beta}\in\mathcal{B}}U_n^{(k)}(\boldsymbol{\beta})$, so equivalently,
    \begin{equation*}
        \hat{\boldsymbol{\beta}}_{\mathrm{CPL}} = \arg\sup_{\boldsymbol{\beta}\in \mathcal{B}}\bigl[U_n^{(k)}(\boldsymbol{\beta})-U(\boldsymbol{\beta})\bigr].
    \end{equation*}
    On $\mathcal{C}_n(\delta)$, it follows that
    \begin{equation*}
        \ell_{n,\mathrm{CPL}}\left(\hat{\boldsymbol{\beta}}_{\mathrm{CPL}}\right)-\ell_{\mathrm{CPL}}\left(\boldsymbol{\beta}_0\right) \geq -\frac{\delta}{2},
    \end{equation*}
    and hence $\|\hat{\boldsymbol{\beta}}_{\mathrm{CPL}}-\boldsymbol{\beta}_0\|\leq \epsilon$.
    Thus, for any sufficiently small $\epsilon>0$ there exists $\delta > 0$ such that $\boldsymbol{X} \in \mathcal{C}_n(\delta) \Rightarrow \|\hat{\boldsymbol{\beta}}_{\mathrm{CPL}}-\boldsymbol{\beta}_0\|\leq \epsilon$.
    Equivalently,
    \begin{equation*}
        \mathcal{C}_n(\delta) \subset \left\{\boldsymbol{X} \in \mathcal{X}^n : \left\|\hat{\boldsymbol{\beta}}_{\mathrm{CPL}}-\boldsymbol{\beta}_0\right\|\leq \epsilon\right\}\subset\mathcal{X}^n.
    \end{equation*}
    Taking $P_{\boldsymbol{\beta}_0}$-probabilities and using uniform convergence ($P_{\boldsymbol{\beta}_0}\bigl(\mathcal{C}_n(\delta)\bigr)\to 1$) yields
    \begin{equation*}
        P_{\boldsymbol{\beta}_0}\left(\|\hat{\boldsymbol{\beta}}_{\mathrm{CPL}}-\boldsymbol{\beta}_0\|\leq \epsilon\right)\to 1,
        \quad
        n\to \infty,
    \end{equation*}
    thereby establishing $\hat{\boldsymbol{\beta}}_{\mathrm{CPL}}\overset{P_{\boldsymbol{\beta}_0}}{\longrightarrow}\boldsymbol{\beta}_0$.
\end{proof}

\subsection{Proof of Theorem~\ref{thm:bayes-asym-unbiased}}
\label{app-subsec:proof-of-bayes-asym-unbiased}

\citet{miller2021asymptotic} establish the asymptotic behavior of generalized posteriors that are induced by a broad class of generalized likelihoods, which strictly includes the loss-based posteriors \citep{bissiri2016general}.

Let $q_{n}$ denote the density of
$\sqrt{n}(\boldsymbol{\beta} - \hat{\boldsymbol{\beta}}_{n})$
when
$\boldsymbol{\beta} \sim \pi^{\ast}$
, and
$\mathcal{N}(x\mid 0, \boldsymbol{H}_{0}^{-1})$
denote the density of
$\mathcal{N}(0, \boldsymbol{H}_{0}^{-1})$
.
Define
$\boldsymbol{J}_{0} := -\mathbb{E}_{P_{\boldsymbol{\beta}_{0}}}[\nabla_{\boldsymbol{\beta}}^2\mathcal{L}_{\mathcal{D}}(\boldsymbol{\beta}_{0})]$
and set $\boldsymbol{H}_{0} := \eta \boldsymbol{J}_{0}$
.
With this notation in place, specializing \citet[Thm.~4]{miller2021asymptotic} to the generalized Bayes posterior based on the loss $\mathcal{L}(\mathcal D\mid \boldsymbol{\beta})$ and prior $\pi(\boldsymbol{\beta})$ yields the following result.
\begin{proposition}\textbf{\textup{(Bernstein--von Mises Theorem for generalized Bayes posteriors)}}
    Assume the regularity conditions~\ref{item:reg-1}--\ref{item:reg-7}.
    Then,
    \begin{equation*}
        \int_{\mathbb{R}^{p}}\left|q_{n}(x) - \mathcal{N}(x\mid 0, \boldsymbol{H}_{0}^{-1})\right| \dd x \to 0
        ,
        \quad
        n \to \infty
        .
    \end{equation*}
    That is, $q_{n}$ converges to $\mathcal{N}(0, \boldsymbol{H}_{0}^{-1})$ in total variation.
\label{app-prop:bvm-for-bissiri}
\end{proposition}
\noindent
Proposition~\ref{app-prop:bvm-for-bissiri} shows that the generalized Bayes posterior satisfies a Bernstein--von Mises limit around $\hat{\boldsymbol{\beta}}_{n}$ with covariance $\boldsymbol{H}_{0}^{-1}$.

Next, let $\hat{\boldsymbol{\beta}}_{\mathrm{GB}} = \int \boldsymbol{\beta}\pi^{\ast}(\dd \boldsymbol{\beta})$
denote the generalized Bayes posterior mean and $r_n$ be the density of
$\sqrt{n}\,(\boldsymbol{\beta}-\hat{\boldsymbol{\beta}}_{\mathrm{GB}})$
when
$\boldsymbol{\beta}\sim\pi^{\ast}$
.
Then Proposition~\ref{app-prop:bvm-for-bissiri} leads to the following corollary.
\begin{corollary}\textbf{\textup{(Bernstein--von Mises centered at the generalized Bayes posterior mean)}}
    Under the regularity conditions of Proposition~\ref{app-prop:bvm-for-bissiri} and, in addition, the prior density $\pi$ is strictly positive and continuous in a neighborhood of $\boldsymbol{\beta}_0$.
    Then,
    \begin{equation*}
        \int_{\mathbb{R}^p} \left|r_{n}(x) - \mathcal{N}(x\mid 0, \boldsymbol{H}_{0}^{-1})\right| \dd x \to 0
        ,
        \quad
        n \to \infty
        .
    \end{equation*}
    That is, $r_{n}$ converges to $\mathcal{N}(0, \boldsymbol{H}_{0}^{-1})$ in total variation.
    \label{app-corollary:bvm-for-bissiri-post-mean}
\end{corollary}
\noindent
Corollary~\ref{app-corollary:bvm-for-bissiri-post-mean} establishes that centering at the generalized Bayes posterior mean yields the same Bernstein--von Mises limit with covariance $\boldsymbol{H}_{0}^{-1}$.

\begin{proof}[Proof of Corollary~\ref{app-corollary:bvm-for-bissiri-post-mean}]
    Let $\hat{\boldsymbol{\beta}}_{n}$ be the stationary point of $f_{n}(\boldsymbol{\beta}) = \eta \mathcal{L}(\mathcal{D}\mid \boldsymbol{\beta})$ from Proposition~\ref{app-prop:bvm-for-bissiri}, with $\hat{\boldsymbol{\beta}}_n\to\boldsymbol{\beta}_0$ and the separation property in regularity condition~\ref{item:reg-7}.
    Introduce the change of variables
    \begin{equation*}
        \boldsymbol{u} := \sqrt{n}\left(\boldsymbol{\beta} - \hat{\boldsymbol{\beta}}_{n}\right)
        ,
        \quad
        \boldsymbol{\beta} = \hat{\boldsymbol{\beta}}_{n} + \frac{\boldsymbol{u}}{\sqrt{n}}
        .
    \end{equation*}
    By a second-order Taylor expansion of $\mathcal{L}_{n}$ at $\hat{\boldsymbol{\beta}}_{n}$ and the uniform boundedness of third derivatives of $\bar{\mathcal{L}}_{n}$ on $\mathcal{B}_{\mathcal{N}}$ by \ref{item:reg-3}, for each fixed $M<\infty$ there exists a deterministic sequence $\varepsilon_{n}(M)\downarrow 0$ and a remainder $R_n(u)$ such that
    \begin{equation*}
        \mathcal{L}_{n}\left(\hat{\boldsymbol{\beta}}_n+\frac{\boldsymbol{u}}{\sqrt{n}}\right)
        = \mathcal{L}_{n}\left(\hat{\boldsymbol{\beta}}_{n}\right) + \frac{1}{2n}\boldsymbol{u}^{\top} \nabla^2 \bar{\mathcal{L}}_{n}\left(\hat{\boldsymbol{\beta}}_n\right)\boldsymbol{u} + R_n(\boldsymbol{u})
    \end{equation*}
    with
    \begin{equation*}
        \sup_{\|\boldsymbol{u}\|\le M} \left|R_{n}(\boldsymbol{u})\right|
        \le \varepsilon_{n}(M)\|\boldsymbol{u}\|^{2}
        ,
        \quad
        \text{i.e. }
        \sup_{\|\boldsymbol{u}\|\le M} \left|R_{n}(\boldsymbol{u})\right|
        = o(1)\|\boldsymbol{u}\|^{2}
        .
    \end{equation*}

    Next, expand $\log\pi$ at $\hat{\boldsymbol{\beta}}_{n}$ using $\pi \in C^2(\mathcal{B}_{\mathcal{N}})$:
    \begin{equation*}
        \log\pi\left(\hat{\boldsymbol{\beta}}_{n} + \frac{\boldsymbol{u}}{\sqrt{n}}\right)
        = \log\pi\left(\hat{\boldsymbol{\beta}}_{n}\right)
        + \frac{1}{\sqrt{n}}\left\{\nabla\log\pi(\hat{\boldsymbol{\beta}}_{n})\right\}^{\top}\boldsymbol{u}
        + O\left(\frac{\|\boldsymbol{u}\|^2}{n}\right)
        .
    \end{equation*}
    Since $\hat{\boldsymbol{\beta}}_{n}\to \boldsymbol{\beta}_{0}$ and $\nabla\log\pi$ is continuous, $\nabla\log\pi(\hat{\boldsymbol{\beta}}_{n}) = \nabla\log\pi(\boldsymbol{\beta}_{0}) + o_p(1)$.
    Writing
    \begin{equation*}
        \boldsymbol{a}
        :=
        \nabla\log\pi(\boldsymbol{\beta}_{0})
        ,
        \quad
        \boldsymbol{\delta}_{n}
        :=
        \nabla\log\pi(\hat{\boldsymbol{\beta}}_{n}) - \boldsymbol{a}
        =
        o_p(1)
        ,
    \end{equation*}
    we may rewrite the expansion as
    \begin{equation*}
        \log\pi\left(\hat{\boldsymbol{\beta}}_{n} + \frac{\boldsymbol{u}}{\sqrt{n}}\right)
        = \log\pi\left(\hat{\boldsymbol{\beta}}_{n}\right) + \frac{1}{\sqrt{n}}\boldsymbol{a}^{\top}\boldsymbol{u} + \rho_{n}(\boldsymbol{u})
        ,
    \end{equation*}
    where
    \begin{equation}
    \label{app-eq:rep-rho}
        \rho_{n}(\boldsymbol{u})
        =
        \frac{1}{\sqrt{n}}\boldsymbol{\delta}_{n}^{\top}\boldsymbol{u} + O\left(\frac{\|\boldsymbol{u}\|^{2}}{n}\right)
        .
    \end{equation}
    For each fixed $M<\infty$ we then have
    \begin{equation*}
        \sup_{\|\boldsymbol{u}\|\le M}\rho_{n}(\boldsymbol{u})
        \le
        \frac{M}{\sqrt{n}}\|\boldsymbol{\delta}_n\| + C\frac{M^{2}}{n}
        =
        O_p\left(\frac{1}{\sqrt{n}}\right)
        =
        o_p(1)
        ,
    \end{equation*}
    for some constant $C>0$ independent of $n$ and $\boldsymbol{u}$.

    Therefore, the generalized Bayes posterior in the $\boldsymbol{u}$-coordinates has density (up to normalization)
    \begin{equation*}
        \pi^{\ast}_{n}(\boldsymbol{u})
        \propto
        \exp\left\{-\frac{1}{2} \boldsymbol{u}^{\top} (\eta \boldsymbol{J}_{0})\boldsymbol{u} + \frac{1}{\sqrt{n}}\boldsymbol{a}^{\top}\boldsymbol{u} + \rho_{n}(\boldsymbol{u})\right\}
        .
    \end{equation*}
    Let $\boldsymbol{H}_0 := \eta \boldsymbol{J}_{0}$ and define the Gaussian reference law
    \begin{equation*}
        P_{n}^{0}:= \mathcal{N}(\boldsymbol{\mu}_{n}, \boldsymbol{H}_{0}^{-1})
        ,
        \quad
        \boldsymbol{\mu}_{n} := \boldsymbol{H}_{0}^{-1}\frac{\boldsymbol{a}}{\sqrt{n}}
        .
    \end{equation*}
    Then we can write $\pi_{n}^{\ast}(\dd \boldsymbol{u})\propto \exp\{\rho_{n}(\boldsymbol{u})\}P_{n}^{0}(\dd \boldsymbol{u})$.
    Since the quadratic term $\boldsymbol{u}^{\top} \boldsymbol{H}_{0} \boldsymbol{u}/2$ yields Gaussian tails under $P_{n}^{0}$ and $\sup_{\|\boldsymbol{u}\|\le M}|\rho_{n}(\boldsymbol{u})| = o_{p}(1)$ for each fixed $M$, a standard truncation argument implies
    \begin{equation*}
        \mathbb{E}_{P_{n}^{0}}\left[\exp\{\rho_{n}(\boldsymbol{u})\} - 1\right]
        =
        o_p(1)
        .
    \end{equation*}

    Moreover, using the expansion $\exp(\rho) - 1 = \rho + g(\rho)$ with $|g(\rho)|\le C\rho^{2}$ for $\rho$ in a neighborhood of $0$, we obtain
    \begin{equation*}
        \begin{split}
            \mathbb{E}_{P_{n}^0}\left[(\boldsymbol{u}-\boldsymbol{\mu}_{n})\{\exp\{\rho_{n}(\boldsymbol{u})\} - 1\}\right]
            &= \mathbb{E}_{P_{n}^0}\left[(\boldsymbol{u}-\boldsymbol{\mu}_{n})\rho_{n}(\boldsymbol{u})\right]
            +
            \mathbb{E}_{P_n^0}\left[(\boldsymbol{u}-\boldsymbol{\mu}_n)g(\rho_n(\boldsymbol{u}))\right]
            ,
            \\
            &= A_n + B_n
            .
        \end{split}
    \end{equation*}
    The remainder term satisfies
    \begin{equation*}
        |B_n|
        \le
        C\mathbb{E}_{P_{n}^{0}}\left[\|\boldsymbol{u} - \boldsymbol{\mu}_{n}\|\rho_{n}(\boldsymbol{u})^{2}\right]
        .
    \end{equation*}
    By \eqref{app-eq:rep-rho} with $\boldsymbol{\delta}_n = o_{p}(1)$ and the fact that $\boldsymbol{u}\sim \mathcal{N}(\boldsymbol{\mu}_{n}, \boldsymbol{H}_{0}^{-1})$ has uniformly bounded moments, it follows that $\mathbb{E}_{P_{n}^{0}}[\rho_n(\boldsymbol{u})^{2}] = o_p(n^{-1})$ and $\mathbb{E}_{P_{n}^{0}}[\|\boldsymbol{u} - \boldsymbol{\mu}_{n}\|\rho_n(\boldsymbol{u})^2] = O_p(n^{-1})$.
    Hence, by Cauchy-Schwarz,
    \begin{equation*}
        |A_{n}|
        \le
        \left\{\mathbb{E}_{P_{n}^{0}}\left[\|\boldsymbol{u} - \boldsymbol{\mu}_{n}\|^{2}\right]\right\}^{1/2}
        \left\{\mathbb{E}_{P_{n}^{0}}\left[\rho_{n}(\boldsymbol{u})^{2}\right]\right\}^{1/2}
        = o_p(n^{-1/2})
        .
    \end{equation*}
    because $\mathbb{E}_{P_{n}^{0}}\|\boldsymbol{u} - \boldsymbol{\mu}_n\|^{2} = \operatorname{tr}(\boldsymbol{H}_{0}^{-1}) < \infty$ where $\operatorname{tr}(\cdot)$ denotes the matrix trace.
    Together with the bound on $B_{n}$ we deduce
    \begin{equation*}
        \mathbb{E}_{P_{n}^{0}}\left[(\boldsymbol{u} - \boldsymbol{\mu}_{n})\{\exp\{\rho_{n}(\boldsymbol{u})\} - 1\}\right]
        =
        A_n + B_n
        =
        o_p(n^{-1/2})
        .
    \end{equation*}

    Consequently,
    \begin{equation*}
        \begin{split}
            \mathbb{E}_{\pi_n^\ast}\left[\boldsymbol{u}\right]
            &= \frac{\mathbb{E}_{P_n^0}\left[\boldsymbol{u}\exp\{\rho_n(\boldsymbol{u})\}\right]}{\mathbb{E}_{P_n^0}\left[\exp\{\rho_n(\boldsymbol{u})\}\right]}
            ,
            \\
            &= \boldsymbol{\mu}_n + \frac{\mathbb{E}_{P_n^0}\left[(\boldsymbol{u}-\boldsymbol{\mu}_n)\{\exp\{\rho_n(\boldsymbol{u})\}-1\}\right]}{1+\mathbb{E}_{P_n^0}\left[\exp\{\rho_n(\boldsymbol{u})\}-1\right]}
            ,
            \\
            &= \boldsymbol{\mu}_n + o_p\left(n^{-1/2}\right)
            .
        \end{split}
    \end{equation*}
    Returning to the original scale,
    \begin{equation}
        \begin{split}
            \hat{\boldsymbol{\beta}}_{\mathrm{GB}}
            &:= \int \boldsymbol{\beta}\pi_n^{\ast}(\dd \boldsymbol{\beta})
            ,\\
            &= \hat{\boldsymbol{\beta}}_n + \frac{1}{\sqrt{n}}\mathbb{E}_{\pi_n^\ast}[\boldsymbol{u}]
            ,
            \\
            &= \hat{\boldsymbol{\beta}}_n + \frac{1}{n}\boldsymbol{H}_{0}^{-1}\boldsymbol{a} + o_p\!\left(\frac{1}{n}\right)
            ,
            \\
            &= \hat{\boldsymbol{\beta}}_n + \frac{1}{n}\boldsymbol{H}_0^{-1}\nabla\log\pi(\boldsymbol{\beta}_0) + o_p\!\left(\frac{1}{n}\right)
            ,
        \end{split}
    \label{app-eq:general-bayes-post-mean-expansion}
    \end{equation}
    hence 
    \begin{equation}
        \left\|\hat{\boldsymbol{\beta}}_{\mathrm{GB}}-\hat{\boldsymbol{\beta}}_n\right\|=o_p(n^{-1/2})
        .
    \label{app-eq:norm-post-and-z-est}
    \end{equation}

    Now apply the triangle inequality for total variation.
    For probability measures $P$ and $Q$ with densities $p$ and $q$ with respect to a common dominating measure, we write the total variation distance as $\|P - Q\|_{\mathrm{TV}} := \|p - q\|_{L^{1}}/2$.
    Then,
    \begin{equation*}
        \begin{split}
            &\left\|\pi^{\ast}_{n}-\mathcal{N}\left(\hat{\boldsymbol{\beta}}_{\mathrm{GB}},(n\boldsymbol{H}_0)^{-1}\right)\right\|_{\mathrm{TV}}\\
            &\quad\le
            \left\|\pi_n^{\ast}-\mathcal{N}\!\left(\hat{\boldsymbol{\beta}}_n,(n\boldsymbol{H}_0)^{-1}\right)\right\|_{\mathrm{TV}}
            +\left\|\mathcal{N}\left(\hat{\boldsymbol{\beta}}_n,(n\boldsymbol{H}_0)^{-1}\right)-\mathcal{N}\left(\hat{\boldsymbol{\beta}}_{\mathrm{GB}},(n\boldsymbol{H}_0)^{-1}\right)\right\|_{\mathrm{TV}}
            .
        \end{split}
    \end{equation*}
    The first term tends to $0$ by Proposition~\ref{app-prop:bvm-for-bissiri}.
    For the second term, use Pinsker's inequality and the KL divergence for normal distributions with equal covariance:
    \begin{equation*}
        \left\|\mathcal{N}(\boldsymbol{\mu}_1,\boldsymbol{\Sigma}_n)-\mathcal{N}(\boldsymbol{\mu}_2,\boldsymbol{\Sigma}_n)\right\|_{\mathrm{TV}}
        \le
        \frac{1}{2}\sqrt{(\boldsymbol{\mu}_1-\boldsymbol{\mu}_2)^\top\boldsymbol{\Sigma}_n^{-1}(\boldsymbol{\mu}_1-\boldsymbol{\mu}_2)}
        ,\quad
        \boldsymbol{\Sigma}_n=(n\boldsymbol{H}_0)^{-1}
        ,
    \end{equation*}
    so the bound equals $2^{-1}\sqrt{n(\hat{\boldsymbol{\beta}}_n-\hat{\boldsymbol{\beta}}_{\mathrm{GB}})^\top H_0(\hat{\boldsymbol{\beta}}_n-\hat{\boldsymbol{\beta}}_{\mathrm{GB}})}=o_p(1)$ because $\hat{\boldsymbol{\beta}}_{\mathrm{GB}}-\hat{\boldsymbol{\beta}}_n=O_p(n^{-1})$.
    Therefore,
    \begin{equation*}
        \left\|\pi_n^{\ast}-\mathcal{N}\left(\hat{\boldsymbol{\beta}}_{\mathrm{GB}},(n\boldsymbol{H}_0)^{-1}\right)\right\|_{\mathrm{TV}} \longrightarrow 0
        .
    \end{equation*}

    Finally, because total variation is invariant under invertible affine transformations applied to both measures, by applying the map $T(\boldsymbol{\beta})=\sqrt{n}(\boldsymbol{\beta}-\hat{\boldsymbol{\beta}}_{\mathrm{GB}})$ to both distributions, the pushforward density $r_n$ of $\sqrt{n}(\boldsymbol{\beta}-\hat{\boldsymbol{\beta}}_{\mathrm{GB}})$ satisfies
    \begin{equation*}
        \int_{\mathbb{R}^p}\left|r_n(x)-\mathcal{N}(x\mid 0,\boldsymbol{H}_0^{-1})\right|\dd x = 2\left\|\pi_n^{\ast}-\mathcal{N}\left(\hat{\boldsymbol{\beta}}_{\mathrm{GB}},(n\boldsymbol{H}_0)^{-1}\right)\right\|_{\mathrm{TV}}
        \longrightarrow
        0
        ,
    \end{equation*}
    which completes the proof.
\end{proof}

\begin{proof}[Proof of Theorem~\ref{thm:bayes-asym-unbiased}]
    Under the assumptions of Corollary~\ref{app-corollary:bvm-for-bissiri-post-mean}, from \eqref{app-eq:general-bayes-post-mean-expansion} and \eqref{app-eq:norm-post-and-z-est}, it follows that
    \begin{equation*}
        \hat{\boldsymbol{\beta}}_{\mathrm{CPL, GB}} - \hat{\boldsymbol{\beta}}_{\mathrm{CPL},n} = \frac{1}{n}\boldsymbol{H}_{0,\mathrm{CPL}}^{-1}\nabla\log\pi(\boldsymbol{\beta}_{0}) + o_p\left(\frac{1}{n}\right)
        ,
        \quad
        \left\|\hat{\boldsymbol{\beta}}_{\mathrm{CPL, GB}} - \hat{\boldsymbol{\beta}}_{\mathrm{CPL},n}\right\| = o_p(n^{-1/2})
        ,
    \end{equation*}
    where $\hat{\boldsymbol{\beta}}_{\mathrm{CPL},n}$ is a stationary point of the empirical log-composite partial likelihood.

    Standard $Z$-estimation theory applied to the composite partial likelihood score implies
    \begin{equation*}
        \sqrt{n}\left(\hat{\boldsymbol{\beta}}_{n} - \boldsymbol{\beta}_{0}\right) \overset{d}{\longrightarrow} \mathcal{N}(\boldsymbol{0}, \boldsymbol{J}_{0,\mathrm{CPL}}^{-1}\boldsymbol{K}_{0,\mathrm{CPL}}\boldsymbol{J}_{0,\mathrm{CPL}}^{-1})
        ,
    \end{equation*}
    where $\overset{d}{\to}$ denotes convergence in distribution.
    Hence $\mathbb{E}_{P_{\boldsymbol{\beta}_{0}}}[\hat{\boldsymbol{\beta}}_{\mathrm{CPL},n} - \boldsymbol{\beta}_{0}] = O(n^{-1/2})$.
    Taking $P_{\boldsymbol{\beta}_0}$-expectations in the display above and combining the orders gives
    \begin{equation*}
        \mathbb{E}_{P_{\boldsymbol{\beta}_{0}}}\left[\hat{\boldsymbol{\beta}}_{\mathrm{CPL,GB}}\right] - \boldsymbol{\beta}_{0} = \mathbb{E}_{P_{\boldsymbol{\beta}_{0}}}\left[\hat{\boldsymbol{\beta}}_{\mathrm{CPL},n} - \boldsymbol{\beta}_{0}\right] + \frac{1}{n}\boldsymbol{H}_{0}^{-1}\nabla\log\pi(\boldsymbol{\beta}_{0}) + o\left(\frac{1}{n}\right)
        = O(n^{-1/2})
        .
    \end{equation*}
\end{proof}
 
\subsection{Proofs of Proposition~\ref{prop:general-bayes-apc} and Corollary~\ref{corollary:apc-for-gs4cox}}
\label{app-subsec:proof-of-apc}

Let $\hat{\boldsymbol{\beta}}_{n}$ be any (measurable) solution to the estimating equation $\Psi_{n}(\boldsymbol{\beta})=0$, and suppose $\hat{\boldsymbol{\beta}}_{n}\overset{p}{\longrightarrow}\boldsymbol{\beta}_{0}$.
Define
\begin{equation*}
    \boldsymbol{J}_{0} := -\nabla_{\boldsymbol{\beta}}\Psi(\boldsymbol{\beta}_{0})
    ,
    \quad
    \boldsymbol{K}_{0}:=\operatorname{Var}_{P_{\boldsymbol{\beta}_{0}}}\left[\psi_{\boldsymbol{\beta}_{0}}(D_{1})\right]
    ,
    \;
    \text{ and }
    \;
    \boldsymbol{G}_{0} := \boldsymbol{J}_{0}^{-1}\boldsymbol{K}_{0}\boldsymbol{J}_{0}^{-1}
    .
\end{equation*}

We recall the standard central limit theorem for $Z$-estimators.
We state it here in our notation for later use (see, e.g., \citet[Thm.~5.41]{vandervaart1998asymptotic}).
\begin{proposition}\textbf{\textup{(Asymptotic normality of $Z$-estimators)}}
\label{app-prop:z-asym}
    Suppose the regularity conditions~\ref{item:reg-2}, \ref{item:reg-4}, \ref{item:reg-5}, and \ref{item:reg-8}.
    Then
    \begin{equation*}
        \sqrt{n}\left(\hat{\boldsymbol{\beta}}_{n} - \boldsymbol{\beta}_{0}\right)
        \overset{d}{\longrightarrow}
        \mathcal{N}(\boldsymbol{0}, \boldsymbol{G}_{0})
        .
    \end{equation*}
\end{proposition}

\begin{lemma}\textbf{\textup{(Laplace equivalence of posterior mean and MAP)}}
    Under the usual local quadratic conditions for Laplace approximation, the posterior mean $\hat{\boldsymbol{\beta}}_{\mathrm{GB}} = \mathbb{E}_{\pi^{\ast}}[\boldsymbol{\beta}]$ and the maximum a posteriori (MAP) satisfy
    \begin{equation*}
        \hat{\boldsymbol{\beta}}_{\mathrm{GB}} - \hat{\boldsymbol{\beta}}_{\mathrm{MAP}} = O(n^{-1})
        ,
    \end{equation*}
    so they are $\sqrt{n}$-equivalent \citep{tierney1986accurate}.
\label{app-lemma:diff-post-mean-map}
\end{lemma}

We then establish the following lemma.

\begin{lemma}\textbf{\textup{(One-step Newton correction from composite partial likelihood to partial likelihood)}}
    Suppose the regularity conditions~\ref{item:reg-1}--\ref{item:reg-6} are satisfied.
    Then, we have 
    \begin{equation*}
        \hat{\boldsymbol{\beta}}_{\mathrm{PL}}=\hat{\boldsymbol{\beta}}_{\mathrm{CPL}}-H_{\mathrm{PL}}(\boldsymbol{\hat{\beta}}_{\mathrm{CPL}})^{-1}S_{\mathrm{PL}}(\hat{\boldsymbol{\beta}}_{\mathrm{CPL}})
        +O_{p}(n^{-1})
        .
    \end{equation*}
    \label{app-lemma:correction-term}
\end{lemma}

\begin{proof}[Proof of Lemma~\ref{app-lemma:correction-term}]
    By regularity condition~\ref{item:reg-1}, $\hat{\boldsymbol{\beta}}_{\mathrm{PL}}$ is an interior stationary point of the partial likelihood, hence $S_{\mathrm{PL}}(\hat{\boldsymbol{\beta}}_{\mathrm{PL}})=\boldsymbol{0}$.
    Moreover, by regularity conditions~\ref{item:reg-2}, \ref{item:reg-4}, \ref{item:reg-5}, and \ref{item:reg-6}, both $\hat{\boldsymbol{\beta}}_{\mathrm{PL}}$ and $\hat{\boldsymbol{\beta}}_{\mathrm{CPL}}$ are $\sqrt{n}$-consistent for $\boldsymbol{\beta}_0$ and lie in the neighborhood $\mathcal{B}_{\mathcal N}$ with probability tending to one;
    in particular, $\|\hat{\boldsymbol{\beta}}_{\mathrm{PL}}-\hat{\boldsymbol{\beta}}_{\mathrm{CPL}}\|=O_p(n^{-1/2})$.

    Because $S_{\mathrm{PL}}(\hat{\boldsymbol{\beta}}_{\mathrm{PL}})=\boldsymbol{0}$, perform a mean-value Taylor expansion of $S_{\mathrm{PL}}(\boldsymbol{\beta})$ around $\boldsymbol{\beta}=\hat{\boldsymbol{\beta}}_{\mathrm{CPL}}$.
    Let $\tilde{\boldsymbol{\beta}}$ lie on the line segment joining $\hat{\boldsymbol{\beta}}_{\mathrm{CPL}}$ and $\hat{\boldsymbol{\beta}}_{\mathrm{PL}}$.
    By regularity condition~\ref{item:reg-3}, the Taylor expansion with remainder is valid:
    \begin{equation}
      \boldsymbol{0} = S_{\mathrm{PL}}(\hat{\boldsymbol{\beta}}_{\mathrm{PL}})
        = S_{\mathrm{PL}}(\hat{\boldsymbol{\beta}}_{\mathrm{CPL}})
          + H_{\mathrm{PL}}(\hat{\boldsymbol{\beta}}_{\mathrm{CPL}})
            (\hat{\boldsymbol{\beta}}_{\mathrm{PL}}-\hat{\boldsymbol{\beta}}_{\mathrm{CPL}})
          + R_n,
      \label{app-eq:mean-value-expansion}
    \end{equation}
    where $H_{\mathrm{PL}}(\boldsymbol{\beta}):=\nabla_{\boldsymbol{\beta}} S_{\mathrm{PL}}(\boldsymbol{\beta}) = \nabla_{\boldsymbol{\beta}}^2 \ell_{n,\mathrm{PL}}(\boldsymbol{\beta})$ and
    \begin{equation*}
        R_n = \frac{1}{2}(\hat{\boldsymbol{\beta}}_{\mathrm{PL}}-\hat{\boldsymbol{\beta}}_{\mathrm{CPL}})^\top
        \left[\nabla^2 S_{\mathrm{PL}}(\tilde{\boldsymbol{\beta}})\right]
        (\hat{\boldsymbol{\beta}}_{\mathrm{PL}}-\hat{\boldsymbol{\beta}}_{\mathrm{CPL}}).
    \end{equation*}

    Note that $\nabla_{\boldsymbol{\beta}}^{2} S_{\mathrm{PL}}=\nabla_{\boldsymbol{\beta}}^{3}\ell_{n,\mathrm{PL}}$.
    Hence $n^{-1}\nabla_{\boldsymbol{\beta}}^{2} S_{\mathrm{PL}} = \nabla_{\boldsymbol{\beta}}^{3}\bar{\ell}_{n,\mathrm{PL}}$.
    By regularity condition~\ref{item:reg-3}, the third derivatives of the normalized log-likelihood are uniformly bounded on $\mathcal{B}_{\mathcal N}$, so
    \begin{equation*}
        \sup_{\boldsymbol{\beta}\in\mathcal{B}_{\mathcal N}}\left\|n^{-1}\nabla_{\boldsymbol{\beta}}^{2} S_{\mathrm{PL}}(\boldsymbol{\beta})\right\| = \sup_{\boldsymbol{\beta}\in\mathcal{B}_{\mathcal N}} \left\|\nabla_{\boldsymbol{\beta}}^{3}\bar{\ell}_{n,\mathrm{PL}}(\boldsymbol{\beta})\right\| = O(1)
        ,
    \end{equation*}
    which implies $\|\nabla_{\boldsymbol{\beta}}^{2} S_{\mathrm{PL}}(\tilde{\boldsymbol{\beta}})\|=O_p(n)$.
    Together with $\|\hat{\boldsymbol{\beta}}_{\mathrm{PL}}-\hat{\boldsymbol{\beta}}_{\mathrm{CPL}}\|=O_p(n^{-1/2})$, we obtain
    \begin{equation*}
        R_{n} = O_{p}(n^{-1/2})O_{p}(n)O_{p}(n^{-1/2}) = O_p(1)
        .
    \end{equation*}

    Rearranging \eqref{app-eq:mean-value-expansion} and multiplying by the inverse Hessian gives
    \begin{equation*}
        \hat{\boldsymbol{\beta}}_{\mathrm{PL}}-\hat{\boldsymbol{\beta}}_{\mathrm{CPL}}
        = -H_{\mathrm{PL}}(\hat{\boldsymbol{\beta}}_{\mathrm{CPL}})^{-1}\,S_{\mathrm{PL}}(\hat{\boldsymbol{\beta}}_{\mathrm{CPL}})
        -H_{\mathrm{PL}}(\hat{\boldsymbol{\beta}}_{\mathrm{CPL}})^{-1}R_n
        .
    \end{equation*}
    By regularity condition~\ref{item:reg-4} with $k=2$,
    \begin{equation*}
        \sup_{\boldsymbol{\beta}\in\mathcal{B}_{\mathcal N}}\left\|n^{-1} H_{\mathrm{PL}}(\boldsymbol{\beta})\right\|
        = \sup_{\boldsymbol{\beta}\in\mathcal{B}_{\mathcal N}}
        \left\|\nabla_{\boldsymbol{\beta}}^{2}\bar{\ell}_{n,\mathrm{PL}}(\boldsymbol{\beta})\right\|
        = O_p(1)
        ,
    \end{equation*}
    so $\|H_{\mathrm{PL}}(\tilde{\boldsymbol{\beta}})\|=O_p(n)$.
    Moreover, by regularity conditions~\ref{item:reg-4} and \ref{item:reg-5}, $-n^{-1}H_{\mathrm{PL}}(\boldsymbol{\beta}) \overset{p}{\to} J \succ 0$ uniformly on $\mathcal{B}_{\mathcal N}$, which implies the invertibility of $H_{\mathrm{PL}}(\hat{\boldsymbol{\beta}}_{\mathrm{CPL}})$ with $H^{-1}_{\mathrm{PL}}(\hat{\boldsymbol{\beta}}_{\mathrm{CPL}}) = O_p(n^{-1})$.
    Consequently,
    \begin{equation*}
        H_{\mathrm{PL}}(\hat{\boldsymbol{\beta}}_{\mathrm{CPL}})^{-1} R_n = O_p(n^{-1})\, O_p(1) = O_p(n^{-1})
        ,
    \end{equation*}
    and therefore
    \begin{equation*}
        \hat{\boldsymbol{\beta}}_{\mathrm{PL}}
        = \hat{\boldsymbol{\beta}}_{\mathrm{CPL}}
        - H_{\mathrm{PL}}(\hat{\boldsymbol{\beta}}_{\mathrm{CPL}})^{-1} S_{\mathrm{PL}}(\hat{\boldsymbol{\beta}}_{\mathrm{CPL}})
        + O_{p}(n^{-1})
        .
    \end{equation*}
\end{proof}

Using the above results, we prove Proposition~3 as follows.

\begin{proof}[Proof of Proposition~\ref{prop:general-bayes-apc}]
    Let $\Pi_n^{\ast}:=\pi^\ast(\cdot\mid\mathcal D)$ denote the generalized Bayes posterior and set
    $\boldsymbol{Z}_n:=\sqrt{n}(\boldsymbol{\beta}-\hat{\boldsymbol{\beta}}_{\mathrm{GB}})$ for
    $\boldsymbol{\beta}\sim\Pi_n^{\ast}$.
    Define $\Omega := \boldsymbol{V}_{\mathrm{target}}^{1/2}\boldsymbol{H}_{0}^{1/2}$.
    By Corollary~\ref{app-corollary:bvm-for-bissiri-post-mean}, the distribution $Q_n$ of $\boldsymbol{Z}_n$ under $\Pi_n^{\ast}$ satisfies
    \begin{equation*}
        \left\|Q_{n} - \Phi_{0}\right\|_{\mathrm{TV}} \to 0
        ,
    \end{equation*}
    where $\Phi_{0} = \mathcal{N}(\boldsymbol{0}, \boldsymbol{H}_{0}^{-1})$.
    
    Consider the linear projection $T(z):=\boldsymbol{\Omega}z$ and write $Q_n^{\boldsymbol{\Omega}}:=Q_n\circ T^{-1}$ and $\Phi_0^{\boldsymbol{\Omega}}:=\Phi_0\circ T^{-1}$.
    Total variation is nonexpansive under measurable mappings, hence
    \begin{equation*}
        \left\|Q_{n}^{\boldsymbol{\Omega}} - \Phi_{0}^{\boldsymbol{\Omega}}\right\|_{\mathrm{TV}}
        \le
        \left\|Q_{n} - \Phi_{0}\right\|_{\mathrm{TV}}
        \longrightarrow
        0
        .
    \end{equation*}
    Since $T(Z) \sim \mathcal{N}(\boldsymbol{0}, \boldsymbol{\Omega}\boldsymbol{H}_{0}^{-1}\boldsymbol{\Omega}^{\top})$ whenever $\boldsymbol{Z} \sim \Phi_0$, and
    \begin{equation*}
        \boldsymbol{\Omega}\boldsymbol{H}_{0}^{-1}\boldsymbol{\Omega}^{\top} = \boldsymbol{V}_{\mathrm{target}}^{1/2}\boldsymbol{H}_{0}^{1/2}\boldsymbol{H}_{0}^{-1}\boldsymbol{H}_{0}^{1/2}\boldsymbol{V}_{\mathrm{target}}^{1/2} = \boldsymbol{V}_{\mathrm{target}}
        ,
    \end{equation*}
    we have $\Phi_{0}^{\boldsymbol{\Omega}} = \mathcal{N}(\boldsymbol{0}, \boldsymbol{V}_{\mathrm{target}})$.
    
    By the definition of the affine posterior calibrated draw,
    \begin{equation*}
        \boldsymbol{\beta}_{\mathrm{APC}}^{(m)} := \boldsymbol{\beta}^{\ddagger} + \boldsymbol{\Omega}\left(\boldsymbol{\beta}^{(m)} - \hat{\boldsymbol{\beta}}_{\mathrm{GB}}\right)
    \end{equation*}
    with $\boldsymbol{\beta}^{(m)}\sim\Pi_{n}^{\ast}$, so
    \begin{equation*}
        \sqrt{n}\left(\boldsymbol{\beta}_{\mathrm{APC}}^{(m)} - \boldsymbol{\beta}^{\ddagger}\right)
        = \boldsymbol{\Omega}\sqrt{n}\left(\boldsymbol{\beta}^{(m)} - \hat{\boldsymbol{\beta}}_{\mathrm{GB}}\right)
        = T(\boldsymbol{Z}_{n})
        .
    \end{equation*}
    Therefore the posterior distribution of $\sqrt{n}(\boldsymbol{\beta}_{\mathrm{APC}}^{(m)}-\boldsymbol{\beta}^\ddagger)$ converges in total variation to $\mathcal N_p(\boldsymbol{0},\boldsymbol{V}_{\mathrm{target}})$.
    This completes the proof.
\end{proof}

\begin{proof}[Proof of Corollary~\ref{corollary:apc-for-gs4cox}]
    Apply Proposition~\ref{prop:general-bayes-apc} with $\boldsymbol{V}_{\mathrm{target}} = \boldsymbol{V}_{\mathrm{PL}}$ and $\boldsymbol{\Omega} = \boldsymbol{\Omega}_{\mathrm{PL}}$ to obtain the covariance calibration.
    For the centering, Lemmas~\ref{app-lemma:diff-post-mean-map} and \ref{app-lemma:correction-term}, we replace $\hat{\boldsymbol{\beta}}_{\mathrm{CPL}}$ with $\hat{\boldsymbol{\beta}}_{\mathrm{CPL,GB}}$ at $O_p(n^{-1})$, yielding $\boldsymbol{\beta}^{\ddagger} = \hat{\boldsymbol{\beta}}_{\mathrm{PL}} + O_p(n^{-1})$.
\end{proof}

\section{Detailed Experimental Settings and Results}
\label{app-sec:detail-experiments}

\subsection{Computational Complexity}
\label{app-subsec:computing-complexity}

Let $e(t)$ be the number of events at time $t$, $E$ be the number of unique events, and $Q=\sum_{t} R(t)e(t)$ be the number of all pairs, where $R(t)$ is the size of the risk set at time $t$.
In the pathological case, where every subject experiences an event at a distinct time, we have $E=n$ and $Q = n(n-1)/2 = O(n^{2})$.

\subsubsection{GS4Cox}
\label{app-subsubsec:computing-complexity-gs4cox}

Before sampling begins, we sort the event times and build each risk set via binary search in $O(n\log n)$.
We then enumerate all $Q$ subject pair contrasts once, and assemble the design matrix $\boldsymbol{D} = \{\boldsymbol{X}_i-\boldsymbol{X}_j\}_{(i,j)}\in \mathbb{R}^{Q\times p}$, at a cost of $O(Qp)$ in time and $O(Qp)$ in memory.
Within each Gibbs iteration, computing the linear predictor $\zeta_{ij} = (\boldsymbol{X}_i-\boldsymbol{X}_j)^\top\boldsymbol{\beta}$ for all pairs costs $O(Qp)$, and sampling the $Q$ P\'olya-Gamma variables $\omega_{ij}\sim \PG(1, \zeta_{ij})$ costs $O(Q)$.
Forming the weighted scatter matrix $\boldsymbol{D}^\top(\boldsymbol{D}\odot\boldsymbol{\omega})$ requires $O(Qp^2)$ operations, and the subsequent Cholesky factorization and Gaussian draw in $\mathbb{R}^{p}$ contribute $O(p^{3})$ flops.
When $p\ll Q$, the per-iteration complexity is dominated by $O(Qp^2)$, after a pre-computation of $\boldsymbol{D}$.

\subsubsection{Metropolis-Hastings (MH) Algorithm}
\label{app-subsubsec:computing-complexity-mh}

The MH routine uses an analytic score and Hessian of the Cox partial log likelihood to construct a multivariate normal (Laplace) proposal.
Computing the score and Hessian requires $O(n\log n + np^2)$, inverting the Hessian takes $O(p^3)$.
The Metropolis correction evaluates the log-partial likelihood at the current and proposed states;
each such evaluation costs $O(n\log n + np)$ using the same cumulative-sum machinery.
Thus one MH iteration costs $O(n\log n + np^{2} + p^{3})$, and, for $p\ll n$, the complexity is dominated by the $O(np^{2})$ term arising from the Hessian computation.

\subsubsection{Hamiltonian Monte Carlo (HMC)}
\label{app-subsubsec:computing-complexity-hmc}

Each leapfrog step requires one gradient of the (generalized) Cox log-partial posterior.
Using the standard cumulative-sum scheme over sorted event times, a single gradient evaluation costs $O(n\log n + np)$:
$O(n\log n)$ to obtain the time ordering and $O(np)$ for the reverse weighted cumulative moments $(S_0,S_1)$ and the event-block aggregates.
With $L$ leapfrog steps per iteration, the integrator performs $L+1$ gradients, so the Hamiltonian dynamics incur a cost of $O((L+1)(n\log n + np))$.
The Metropolis correction step requires two additional evaluations of the log-partial likelihood, adding $2O(n\log n + np)$.
Hence the per-iteration complexity is $O((L+3)(n\log n + np))$, which simplifies to $O(Lnp)$ when the initial sort is cached and $n\gg p$.

\subsubsection{No-U-Turn Sampler (NUTS)}
\label{app-subsubsec:computing-complexity-nuts}

NUTS adaptively builds a binary tree of leapfrog steps until the no-U-turn criterion is violated.
Let $U$ denote the random number of leapfrog steps in a given iteration.
As in HMC, each leapfrog step requires one gradient evaluation of cost $O(n\log n + np)$, and each base node introduces a log-density evaluation of the same order.
Consequently, a single NUTS iteration costs $O(U(n\log n + np))$.
With a maximum tree depth $D_{\max}$, we have the worst-case per-iteration cost is $O(1024(n\log n + np))$, while typical runs use substantially smaller $U$.

\subsubsection{Metropolis-Adjusted Langevin Algorithm (MALA)}
\label{app-subsubsec:computing-complexity-mala}

Each MALA update evaluates the gradient of the log-posterior at the current state to form a Langevin proposal, and again at the proposed state to compute the reverse proposal density.
The Metropolis ratio also requires two log-partial likelihood evaluations.
With the cumulative-sum scheme, one gradient is $O(n\log n + np)$ and one log-likelihood is $O(n\log n + np)$.
Thus a full MALA iteration uses two gradients and two log-likelihoods, for a total cost $O(4(n\log n + np))=O(n\log n + np)$, up to a constant.

\subsubsection{Cox-P\'olya-Gamma (Cox-PG) Algorithm}
\label{app-subsubsec:computing-complexity-coxpg}

Let $J$ denote the number of spline segments for the baseline log-cumulative hazard (we use $J=D=5$), and define $d = J + 1+ p + M_M$ as the total number of fixed and random effects in the augmented linear predictor (intercept $+p$ covariates $+J$ spline coefficients $+M_M$ random-effect coefficients, if present).
Preprocessing the monotone-spline design and indicators costs $O(nJ)$.

Within each MCMC iteration, sampling $n$ P\'olya-Gamma variables is $O(n)$, and forming the weighted normal equations $Q=M^\top(\Omega M)+A$ costs $O(nd^2)$.
A Cholesky factorization of $Q$ requires $O(d^3)$ operations.
Computing the posterior mean and, when needed, the full inverse via triangular solves costs $O(d^2)$ and $O(d^3)$, respectively.
The two conditional multivariate normal updates the truncated multivariate normal draw for the $J$-dimensional spline block together contribute $O(d^3)$ and $O(J^3)$.
The optional Metropolis-Hastings calibration step evaluates the objective twice at a cost of $O(nd)$.
Collecting term, the per-iteration complexity of the Cox-PG algorithm is $O(nd^2 + d^3)$, which is typically dominated by $O(n(p+J+M_M)^2)$ in regimes where $n\gg d$.
The memory usage is $O(nd)$ to store the design matrix $M$ and $O(d^2)$ for the Cholesky factors and related matrix quantities.

\subsection{Additional Numerical Experiment Results}
\label{app-subsec:additional-numerical-results}

First, Table~\ref{app-tab:numerical-intervals-notie} presents the 95\% confidence intervals and credible intervals corresponding to the setting without tied events, as presented in Figure~\ref{fig:numerical-data-forest-plot}.
These results quantitatively demonstrate that the credible intervals computed by our GS4Cox and HMC methods are very close to the confidence intervals corresponding to the maximum partial likelihood estimator (MPLE).

\begin{table}[htbp]
\caption{
95\% confidence intervals and credible intervals (Bayesian methods) from synthetic data experiments without tied events.
The compared methods are MPLE, GS4Cox, MH, HMC, NUTS, MALA, and Cox-PG.
}
\label{app-tab:numerical-intervals-notie}
\vskip 0.15in
\begin{center}
\begin{small}
\begin{sc}
\begin{tabular}{lrrrr}
\toprule
& $\beta_{1}$ & $\beta_{2}$ & $\beta_{3}$ & $\beta_{4}$
\\
\midrule
MPLE
& $[0.80, 1.23]$ & $[-1.27, -0.84]$
& $[0.33, 0.70]$ & $[-0.74, -0.38]$
\\
GS4Cox
& $[0.81, 1.22]$ & $[-1.25, -0.82]$
& $[0.33, 0.69]$ & $[-0.73, -0.33]$
\\
MH
& $[0.86, 1.29]$ & $[-1.26, -0.89]$
& $[0.36, 0.71]$ & $[-0.71, -0.41]$
\\
HMC
& $[0.80, 1.23]$ & $[-1.29, -0.82]$
& $[0.33, 0.72]$ & $[-0.73, -0.40]$
\\
NUTS
& $[0.88, 1.18]$ & $[-1.24, -0.83]$
& $[0.43, 0.69]$ & $[-0.69, -0.44]$
\\
MALA
& $[0.76, 1.24]$ & $[-1.29, -0.81]$
& $[0.33, 0.72]$ & $[-0.73, -0.38]$
\\
Cox-PG
& $[0.90, 1.17]$ & $[-1.29, -1.03]$
& $[0.47, 0.61]$ & $[-0.63, -0.53]$
\\
\hline
\hline
& $\beta_{5}$ & $\beta_{6}$ & $\beta_{7}$ & $\beta_{8}$
\\
\midrule
MPLE
& $[-0.06, 0.31]$ & $[-0.54, -0.18]$
& $[-0.08, 0.24]$ & $[-0.28, 0.08]$
\\
GS4Cox
& $[-0.05, 0.31]$ & $[-0.53, -0.17]$
& $[-0.07, 0.23]$ & $[-0.28, 0.08]$
\\
MH
& $[0.01, 0.27]$ & $[-0.56, -0.14]$
& $[-0.10, 0.26]$ & $[-0.28, 0.07]$
\\
HMC
& $[-0.03, 0.31]$ & $[-0.53, -0.17]$
& $[-0.08, 0.26]$ & $[-0.29, 0.07]$
\\
NUTS
& $[-0.06, 0.27]$ & $[-0.59, -0.21]$
& $[-0.06, 0.18]$ & $[-0.27, 0.03]$
\\
MALA
& $[-0.06, 0.35]$ & $[-0.53, -0.17]$
& $[-0.09, 0.24]$ & $[-0.28, 0.07]$
\\
Cox-PG
& $[-0.03, 0.23]$ & $[-0.42, -0.04]$
& $[-0.12, 0.16]$ & $[-0.25, -0.01]$
\\
\bottomrule
\end{tabular}
\end{sc}
\end{small}
\end{center}
\vskip -0.1in
\end{table}

\begin{figure}[htbp]
\vskip 0.2in
\begin{center}
\centerline{\includegraphics[width=\textwidth]{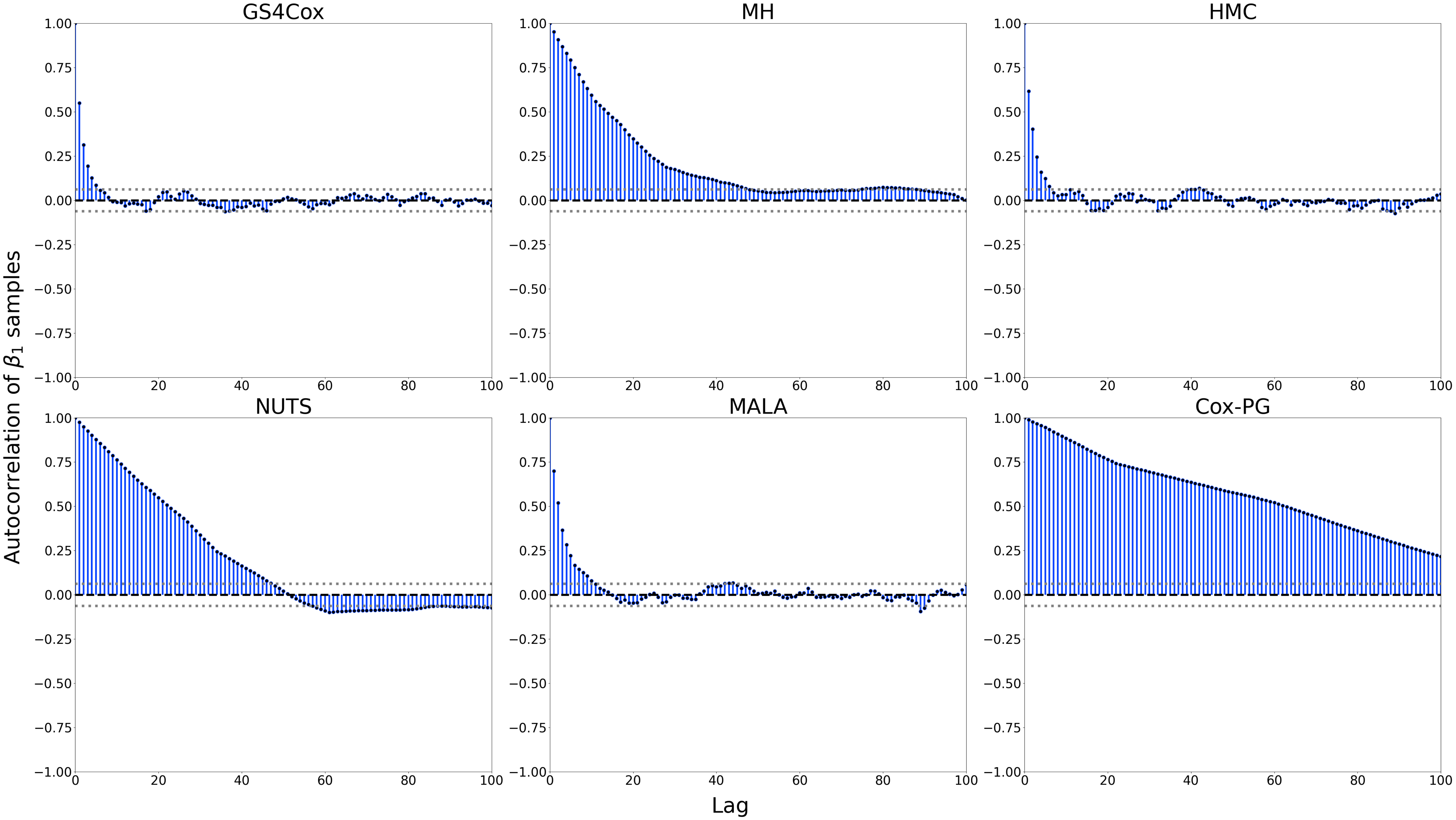}}
\caption{
The correlogram of sample chains for the log-hazard coefficient $\beta_{1}$ from the numerical experiments with no tied events.
The dotted lines denote the $\pm 1.96/\sqrt{N}$ reference bounds under the white-noise null hypothesis (zero autocorrelation), where $N$ is the number of iterations.
}
\label{app-fig:numerical-correlogram}
\end{center}
\vskip -0.2in
\end{figure}

Additionally, Figure~\ref{app-fig:numerical-correlogram} shows the correlogram for the regression parameter $\beta_{1}$ of the numerical experiments with no tied events.
This result indicates that GS4Cox, HMC, and MALA sampled with low autocorrelation, whereas MH, NUTS, and Cox-PG showed high autocorrelation and did not produce independent samples.

Tables~\ref{app-tab:numerical-data-estimates-tie} and \ref{app-tab:numerical-intervals-tie} display the estimates, posterior means, and the corresponding 95\% confidence intervals and credible intervals for the setting with existing tied events, another scenario to evaluate sampling efficiency, as described in Table~\ref{tab:numerical-data-efficiency}.
These results suggest that GS4Cox and HMC are able to calculate posterior means and credible intervals that closely match those of MPLE, even in the presence of tied events.

\begin{table}[htbp]
\caption{
Parameter estimates from synthetic data experiments with tied events. 
The compared methods are MPLE, GS4Cox, MH, HMC, NUTS, MALA, and Cox-PG.
}
\label{app-tab:numerical-data-estimates-tie}
\vskip 0.15in
\begin{center}
\begin{small}
\begin{sc}
\begin{tabular}{lrrrrrrrr}
\toprule
& $\beta_{1}$ & $\beta_{2}$ & $\beta_{3}$ & $\beta_{4}$
& $\beta_{5}$ & $\beta_{6}$ & $\beta_{7}$ & $\beta_{8}$\\
\midrule
True Parameter
& $1.00$ & $-1.00$ & $0.50$ & $-0.50$
& $0.30$ & $-0.30$ & $0.10$ & $-0.10$
\\
MPLE
& $1.02$ & $-1.06$ & $0.52$ & $-0.56$
& $0.12$ & $-0.36$ & $0.08$ & $-0.10$
\\
GS4Cox
& $1.01$ & $-1.05$ & $0.51$ & $-0.56$
& $0.12$ & $-0.35$ & $0.08$ & $-0.10$
\\
MH
& $1.00$ & $-1.09$ & $0.54$ & $-0.57$
& $0.07$ & $-0.38$ & $0.07$ & $-0.07$
\\
HMC
& $1.00$ & $-1.04$ & $0.51$ & $-0.56$
& $0.13$ & $-0.35$ & $0.08$ & $-0.11$
\\
NUTS
& $1.00$ & $-1.05$ & $0.54$ & $-0.58$
& $0.08$ & $-0.41$ & $0.06$ & $-0.10$
\\
MALA
& $1.00$ & $-1.04$ & $0.52$ & $-0.54$
& $0.14$ & $-0.36$ & $0.08$ & $-0.10$
\\
Cox-PG
& $1.05$ & $-1.12$ & $0.49$ & $-0.55$
& $0.08$ & $-0.33$ & $0.06$ & $-0.12$
\\
\bottomrule
\end{tabular}
\end{sc}
\end{small}
\end{center}
\vskip -0.1in
\end{table}

\begin{table}[htbp]
\caption{
95\% confidence intervals and credible intervals (Bayesian methods) from synthetic data experiments with tied events.
The compared methods are MPLE, GS4Cox, MH, HMC, NUTS, MALA, and Cox-PG.
}
\label{app-tab:numerical-intervals-tie}
\vskip 0.15in
\begin{center}
\begin{small}
\begin{sc}
\begin{tabular}{lrrrr}
\toprule
& $\beta_{1}$ & $\beta_{2}$ & $\beta_{3}$ & $\beta_{4}$
\\
\midrule
MPLE
& $[0.81, 1.23]$ & $[-1.27, -0.84]$
& $[0.33, 0.70]$ & $[-0.74, -0.38]$
\\
GS4Cox
& $[0.80, 1.22]$ & $[-1.27, -0.82]$
& $[0.32, 0.69]$ & $[-0.73, -0.39]$
\\
MH
& $[0.83, 1.19]$ & $[-1.35, -0.83]$
& $[0.39, 0.66]$ & $[-0.74, -0.41]$
\\
HMC
& $[0.79, 1.23]$ & $[-1.29, -0.82]$
& $[0.33, 0.72]$ & $[-0.73, -0.40]$
\\
NUTS
& $[0.87, 1.18]$ & $[-1.24, -0.83]$
& $[0.43, 0.69]$ & $[-0.69, -0.44]$
\\
MALA
& $[0.77, 1.24]$ & $[-1.29, -0.81]$
& $[0.33, 0.72]$ & $[-0.73, -0.38]$
\\
Cox-PG
& $[1.03, 1.11]$ & $[-1.17, -1.10]$
& $[0.49, 0.53]$ & $[-0.58, -0.49]$
\\
\hline
\hline
& $\beta_{5}$ & $\beta_{6}$ & $\beta_{7}$ & $\beta_{8}$
\\
\midrule
MPLE
& $[-0.06, 0.31]$ & $[-0.55, -0.18]$
& $[-0.08, 0.24]$ & $[-0.28, 0.08]$
\\
GS4Cox
& $[-0.06, 0.31]$ & $[-0.53, -0.17]$
& $[-0.07, 0.23]$ & $[-0.28, 0.08]$
\\
MH
& $[-0.09, 0.21]$ & $[-0.52, -0.22]$
& $[-0.02, 0.22]$ & $[-0.22, 0.08]$
\\
HMC
& $[-0.03, 0.31]$ & $[-0.53, -0.17]$
& $[-0.08, 0.26]$ & $[-0.29, 0.07]$
\\
NUTS
& $[-0.06, 0.27]$ & $[-0.58, -0.21]$
& $[-0.06, 0.18]$ & $[-0.27, 0.03]$
\\
MALA
& $[-0.06, 0.35]$ & $[-0.53, -0.16]$
& $[-0.09, 0.24]$ & $[-0.28, 0.08]$
\\
Cox-PG
& $[0.08, 0.10]$ & $[-0.41, -0.31]$
& $[-0.06, 0.08]$ & $[-0.13, -0.09]$
\\
\bottomrule
\end{tabular}
\end{sc}
\end{small}
\end{center}
\vskip -0.1in
\end{table}

Table~\ref{app-tab:numerical-efficiency-sample} summarizes the sampling efficiency metrics for sample sizes of $100$ and $1,000$, and Tables~\ref{app-tab:numerical-estimates-sample-100}--\ref{app-tab:numerical-intervals-sample-1000} report the posterior means and 95\% credible intervals for each scenario.
For the sample size of $1,000$, MALA and Cox-PG did not update their initial values, so their results are marked with "$-$".
The results in Table~\ref{app-tab:numerical-efficiency-sample} suggest that, for smaller sample sizes ($100$), our method significantly outperformed others in terms of sampling efficiency.
However, with larger sample sizes ($1,000$), HMC showed the highest sampling efficiency.
This can be attributed to the fact that GS4Cox requires computation of pairwise linear predictors and sampling of P\'olya-Gamma auxiliary variables, which introduces additional computational cost as the sample size grows.
However, as shown in Tables~\ref{app-tab:numerical-estimates-sample-100}--\ref{app-tab:numerical-intervals-sample-1000}, our method consistently produces posterior means and credible intervals that closely align with MPLE and its corresponding confidence intervals, regardless of sample size.
In particular, for small sample sizes, while the credible intervals from other alternative methods diverge from MPLE, GS4Cox still achieves high accuracy in its estimates.

\begin{table}[htbp]
\caption{Evaluation metrics from synthetic data experiments under two scenarios: (i) sample size is $100$; and (ii) sample size is $1,000$.
The compared methods are GS4Cox, MH, HMC, NUTS, MALA, and Cox-PG.
}
\label{app-tab:numerical-efficiency-sample}
\vskip 0.15in
\begin{center}
\begin{small}
\begin{sc}
\begin{tabular}{lrrrrrr}
\toprule
 &\multicolumn{3}{c}{$n=100$}&\multicolumn{3}{c}{$n=1,000$}\\
\cmidrule(lr){2-4}\cmidrule(lr){5-7}
 & ESS & ESR & MCSE & ESS & ESR & MCSE\\
\midrule
GS4Cox
& $257.07$ & $242.26$ & $0.0131$
& $233.75$ & $2.28$ & $0.0034$\\
MH     
& $28.91$ & $3.48$ & $0.0314$
& $22.96$ & $0.41$ & $0.0103$\\
HMC    
& $60.09$ & $2.86$ & $0.0276$
& $1857.25$ & $9.96$ & $0.0014$\\
NUTS
& $14.67$ & $0.26$ & $0.0559$
& $36.33$ & $0.10$ & $0.0083$\\
MALA   
& $50.87$ & $3.46$ & $0.0290$
& $-$ & $-$ & $-$\\
Cox-PG
& $26.60$ & $3.01$ & $0.0322$
& $-$ & $-$ & $-$\\
\bottomrule
\end{tabular}
\end{sc}
\end{small}
\end{center}
\vskip -0.1in
\end{table}

\begin{table}[htbp]
\caption{
Parameter estimates from synthetic data experiments with a sample size of $100$.
The compared methods are MPLE, GS4Cox, MH, HMC, NUTS, MALA, and Cox-PG.
}
\label{app-tab:numerical-estimates-sample-100}
\vskip 0.15in
\begin{center}
\begin{small}
\begin{sc}
\begin{tabular}{lrrrrrrrr}
\toprule
& $\beta_{1}$ & $\beta_{2}$ & $\beta_{3}$ & $\beta_{4}$
& $\beta_{5}$ & $\beta_{6}$ & $\beta_{7}$ & $\beta_{8}$\\
\midrule
True Parameter
& $1.00$ & $-1.00$ & $0.50$ & $-0.50$
& $0.30$ & $-0.30$ & $0.10$ & $-0.10$
\\
MPLE
& $1.24$ & $-1.00$ & $0.34$ & $-0.93$
& $0.31$ & $-0.65$ & $0.20$ & $0.31$
\\
GS4Cox
& $1.24$ & $-1.00$ & $0.34$ & $-0.90$
& $0.32$ & $-0.64$ & $0.19$ & $0.30$
\\
MH
& $1.31$ & $-1.03$ & $0.41$ & $-0.94$
& $0.34$ & $-0.63$ & $0.20$ & $0.27$
\\
HMC
& $1.20$ & $-0.94$ & $0.35$ & $-0.92$
& $0.32$ & $-0.61$ & $0.23$ & $0.27$
\\
NUTS
& $1.31$ & $-0.93$ & $0.36$ & $-0.92$
& $0.32$ & $-0.65$ & $0.26$ & $0.32$
\\
MALA
& $1.20$ & $-0.94$ & $0.36$ & $-0.91$
& $0.32$ & $-0.63$ & $0.21$ & $0.27$
\\
Cox-PG
& $0.98$ & $-0.92$ & $0.29$ & $-0.78$
& $0.21$ & $-0.55$ & $0.19$ & $0.31$
\\
\bottomrule
\end{tabular}
\end{sc}
\end{small}
\end{center}
\vskip -0.1in
\end{table}

\begin{table}[htbp]
\caption{
95\% confidence intervals and credible intervals (Bayesian methods) from synthetic data experiments with a sample size of $100$.
The compared methods are MPLE, GS4Cox, MH, HMC, NUTS, MALA, and Cox-PG.
}
\label{app-tab:numerical-intervals-sample-100}
\vskip 0.15in
\begin{center}
\begin{small}
\begin{sc}
\begin{tabular}{lrrrr}
\toprule
& $\beta_{1}$ & $\beta_{2}$ & $\beta_{3}$ & $\beta_{4}$
\\
\midrule
MPLE
& $[0.73, 1.76]$ & $[-1.46, -0.54]$
& $[0.07, 0.62]$ & $[-1.34, -0.52]$
\\
GS4Cox
& $[0.73, 1.77]$ & $[-1.45, -0.49]$
& $[0.08, 0.64]$ & $[-1.31, -0.50]$
\\
MH
& $[0.89, 1.65]$ & $[-1.52, -0.72]$
& $[0.15, 0.63]$ & $[-1.34, -0.52]$
\\
HMC
& $[0.63, 1.63]$ & $[-1.41, -0.57]$
& $[0.09, 0.66]$ & $[-1.25, -0.54]$
\\
NUTS
& $[0.94, 1.96]$ & $[-1.32, -0.52]$
& $[0.24, 0.53]$ & $[-1.38, -0.63]$
\\
MALA
& $[0.70, 1.60]$ & $[-1.38, -0.53]$
& $[0.10, 0.65]$ & $[-1.24, -0.56]$
\\
Cox-PG
& $[0.54, 1.34]$ & $[-1.21, -0.55]$
& $[0.06, 0.49]$ & $[-1.01, -0.50]$
\\
\hline
\hline
& $\beta_{5}$ & $\beta_{6}$ & $\beta_{7}$ & $\beta_{8}$
\\
\midrule
MPLE
& $[0.02, 0.61]$ & $[-0.99, -0.31]$
& $[-0.08, 0.49]$ & $[-0.01, 0.63]$
\\
GS4Cox
& $[0.05, 0.61]$ & $[-0.98, -0.30]$
& $[-0.10, 0.49]$ & $[-0.00, 0.60]$
\\
MH
& $[0.10, 0.60]$ & $[-0.92, -0.31]$
& $[-0.03, 0.46]$ & $[0.01, 0.52]$
\\
HMC
& $[0.01, 0.60]$ & $[-0.95, -0.26]$
& $[-0.01, 0.49]$ & $[-0.02, 0.57]$
\\
NUTS
& $[0.00, 0.53]$ & $[-1.03, -0.16]$
& $[0.05, 0.63]$ & $[-0.03, 0.62]$
\\
MALA
& $[0.03, 0.64]$ & $[-0.98, -0.29]$
& $[-0.03, 0.47]$ & $[-0.03, 0.56]$
\\
Cox-PG
& $[-0.15, 0.65]$ & $[-0.91, -0.26]$
& $[-0.12, 0.53]$ & $[0.08, 0.57]$
\\
\bottomrule
\end{tabular}
\end{sc}
\end{small}
\end{center}
\vskip -0.1in
\end{table}

\begin{table}[htbp]
\caption{
Parameter estimates from synthetic data experiments with a sample size of $1,000$.
The compared methods are MPLE, GS4Cox, MH, HMC, NUTS, MALA, and Cox-PG.
}
\label{app-tab:numerical-estimates-sample-1000}
\vskip 0.15in
\begin{center}
\begin{small}
\begin{sc}
\begin{tabular}{lrrrrrrrr}
\toprule
& $\beta_{1}$ & $\beta_{2}$ & $\beta_{3}$ & $\beta_{4}$
& $\beta_{5}$ & $\beta_{6}$ & $\beta_{7}$ & $\beta_{8}$\\
\midrule
True Parameter
& $1.00$ & $-1.00$ & $0.50$ & $-0.50$
& $0.30$ & $-0.30$ & $0.10$ & $-0.10$
\\
MPLE
& $1.06$ & $-0.99$ & $0.61$ & $-0.54$
& $0.38$ & $-0.27$ & $0.17$ & $-0.06$
\\
GS4Cox
& $1.06$ & $-0.99$ & $0.61$ & $-0.54$
& $0.38$ & $-0.26$ & $0.17$ & $-0.06$
\\
MH
& $1.07$ & $-0.98$ & $0.60$ & $-0.55$
& $0.37$ & $-0.28$ & $0.18$ & $-0.06$
\\
HMC
& $1.06$ & $-0.98$ & $0.61$ & $-0.54$
& $0.38$ & $-0.26$ & $0.17$ & $-0.06$
\\
NUTS
& $1.05$ & $-0.98$ & $0.62$ & $-0.53$
& $0.37$ & $-0.26$ & $0.16$ & $-0.06$
\\
MALA
& $-$ & $-$ & $-$ & $-$
& $-$ & $-$ & $-$ & $-$
\\
Cox-PG
& $-$ & $-$ & $-$ & $-$
& $-$ & $-$ & $-$ & $-$
\\
\bottomrule
\end{tabular}
\end{sc}
\end{small}
\end{center}
\vskip -0.1in
\end{table}

\begin{table}[htbp]
\caption{
95\% confidence intervals and credible intervals (Bayesian methods) from synthetic data experiments with a sample size of $1,000$.
The compared methods are MPLE, GS4Cox, MH, HMC, NUTS, MALA, and Cox-PG.
}
\label{app-tab:numerical-intervals-sample-1000}
\vskip 0.15in
\begin{center}
\begin{small}
\begin{sc}
\begin{tabular}{lrrrr}
\toprule
& $\beta_{1}$ & $\beta_{2}$ & $\beta_{3}$ & $\beta_{4}$
\\
\midrule
MPLE
& $[0.94, 1.18]$ & $[-1.09, -0.88]$
& $[0.51, 0.71]$ & $[-0.63, -0.44]$
\\
GS4Cox
& $[0.94, 1.17]$ & $[-1.10, -0.88]$
& $[0.51, 0.71]$ & $[-0.63, -0.44]$
\\
MH
& $[0.98, 1.16]$ & $[-1.11, -0.90]$
& $[0.52, 0.69]$ & $[-0.64, -0.43]$
\\
HMC
& $[0.94, 1.19]$ & $[-1.10, -0.87]$
& $[0.51, 0.71]$ & $[-0.64, -0.45]$
\\
NUTS
& $[0.95, 1.15]$ & $[-1.11, -0.88]$
& $[0.52, 0.71]$ & $[-0.60, -0.44]$
\\
MALA
& $-$ & $-$ & $-$ & $-$
\\
Cox-PG
& $-$ & $-$ & $-$ & $-$
\\
\hline
\hline
& $\beta_{5}$ & $\beta_{6}$ & $\beta_{7}$ & $\beta_{8}$
\\
\midrule
MPLE
& $[0.28, 0.47]$ & $[-0.35, -0.18]$
& $[0.08, 0.25]$ & $[-0.15, 0.03]$
\\
GS4Cox
& $[0.28, 0.46]$ & $[-0.34, -0.17]$
& $[0.08, 0.26]$ & $[-0.14, 0.04]$
\\
MH
& $[0.27, 0.47]$ & $[-0.36, -0.21]$
& $[0.12, 0.25]$ & $[-0.17, 0.04]$
\\
HMC
& $[0.29, 0.48]$ & $[-0.34, -0.18]$
& $[0.07, 0.25]$ & $[-0.16, 0.04]$
\\
NUTS
& $[0.27, 0.47]$ & $[-0.34, -0.17]$
& $[0.07, 0.25]$ & $[-0.15, 0.03]$
\\
MALA
& $-$ & $-$ & $-$ & $-$
\\
Cox-PG
& $-$ & $-$ & $-$ & $-$
\\
\bottomrule
\end{tabular}
\end{sc}
\end{small}
\end{center}
\vskip -0.1in
\end{table}

Furthermore, Table~\ref{app-tab:numerical-efficiency-lr} summarizes the sampling efficiency metrics for different learning rates in the general Bayesian inference framework, and Tables~\ref{app-tab:numerical-estimates-lr0.1}--\ref{app-tab:numerical-intervals-lr10} report the posterior means and 95\% credible intervals for each scenario.
For a learning rate of $10$, MALA showed no updates from the initial values, and thus its result is marked with a "$-$".
Table~\ref{app-tab:numerical-efficiency-lr} shows that, at small learning rates, our method attains substantially higher sampling efficiency.
In contrast, at larger learning rates, HMC performs best according to the sampling efficiency metrics.
Moreover, the sampling efficiency of our method is essentially insensitive to the choice of learning rate over the range examined.
The findings in Tables~\ref{app-tab:numerical-estimates-lr0.1}--\ref{app-tab:numerical-intervals-lr10} provide important insights into the role of learning rate in general Bayesian inference.
When the learning rate is small, all samplers except GS4Cox showed very wide credible intervals, while for larger learning rate, the credible intervals became very narrow.
This indicates a common issue with samplers based on general Bayesian inference, including HMC, where the sampling efficiency heavily depends on the learning rate.
On the other hand, our GS4Cox method remained almost unaffected by changes in learning rate and continues to produce posterior means and credible intervals that were very close to MPLE and its corresponding confidence intervals.

\begin{table}[htbp]
\caption{Evaluation metrics from synthetic data experiments under two scenarios: (i) learning rate is $0.1$; and (ii) learning rate is $10.0$.
The compared methods are GS4Cox, MH, HMC, NUTS, and MALA.
.}
\label{app-tab:numerical-efficiency-lr}
\vskip 0.15in
\begin{center}
\begin{small}
\begin{sc}
\begin{tabular}{lrrrrrr}
\toprule
 &\multicolumn{3}{c}{learning rate $\eta = 0.1$}&\multicolumn{3}{c}{learning rate $\eta = 10.0$}\\
\cmidrule(lr){2-4}\cmidrule(lr){5-7}
 & ESS & ESR & MCSE & ESS & ESR & MCSE\\
\midrule
GS4Cox
& $289.93$ & $46.13$ & $0.0060$
& $299.34$ & $47.51$ & $0.0066$\\
MH     
& $23.28$ & $0.96$ & $0.0628$
& $22.00$ & $0.91$ & $0.0073$\\
HMC    
& $23.37$ & $0.36$ & $0.0604$
& $2076.63$ & $32.46$ & $0.0007$\\
NUTS   
& $15.23$ & $0.08$ & $0.0576$
& $55.08$ & $0.46$ & $0.0044$\\
MALA   
& $22.92$ & $0.51$ & $0.0617$
& $-$ & $-$ & $-$\\
\bottomrule
\end{tabular}
\end{sc}
\end{small}
\end{center}
\vskip -0.1in
\end{table}

\begin{table}[htbp]
\caption{
Parameter estimates from synthetic data experiments where a learning rate is set to $0.1$.
The compared methods are MPLE, GS4Cox, MH, HMC, NUTS, and MALA
}
\label{app-tab:numerical-estimates-lr0.1}
\vskip 0.15in
\begin{center}
\begin{small}
\begin{sc}
\begin{tabular}{lrrrrrrrr}
\toprule
& $\beta_{1}$ & $\beta_{2}$ & $\beta_{3}$ & $\beta_{4}$
& $\beta_{5}$ & $\beta_{6}$ & $\beta_{7}$ & $\beta_{8}$\\
\midrule
True Parameter
& $1.00$ & $-1.00$ & $0.50$ & $-0.50$
& $0.30$ & $-0.30$ & $0.10$ & $-0.10$
\\
MPLE
& $1.02$ & $-1.05$ & $0.51$ & $-0.56$
& $0.12$ & $-0.36$ & $0.08$ & $-0.10$
\\
GS4Cox
& $1.01$ & $-1.05$ & $0.51$ & $-0.56$
& $0.12$ & $-0.36$ & $0.08$ & $-0.10$
\\
MH
& $1.14$ & $-1.08$ & $0.59$ & $-0.62$
& $0.06$ & $-0.37$ & $0.13$ & $-0.02$
\\
HMC
& $0.97$ & $-0.98$ & $0.56$ & $-0.58$
& $0.17$ & $-0.25$ & $0.16$ & $-0.22$
\\
NUTS
& $0.84$ & $-1.05$ & $0.42$ & $-0.55$
& $0.20$ & $-0.39$ & $-0.02$ & $-0.12$
\\
MALA
& $0.98$ & $-1.00$ & $0.56$ & $-0.57$
& $0.17$ & $-0.25$ & $0.16$ & $-0.22$
\\
\bottomrule
\end{tabular}
\end{sc}
\end{small}
\end{center}
\vskip -0.1in
\end{table}

\begin{table}[htbp]
\caption{
95\% confidence intervals and credible intervals (Bayesian methods) from synthetic data experiments where a learning rate is set to $0.1$.
The compared methods are MPLE, GS4Cox, MH, HMC, NUTS, and MALA.
}
\label{app-tab:numerical-intervals-lr0.1}
\vskip 0.15in
\begin{center}
\begin{small}
\begin{sc}
\begin{tabular}{lrrrr}
\toprule
& $\beta_{1}$ & $\beta_{2}$ & $\beta_{3}$ & $\beta_{4}$
\\
\midrule
MPLE
& $[0.80, 1.23]$ & $[-1.27, -0.84]$
& $[0.33, 0.70]$ & $[-0.74, -0.38]$
\\
GS4Cox
& $[0.81, 1.22]$ & $[-1.25, -0.82]$
& $[0.34, 0.70]$ & $[-0.74, -0.39]$
\\
MH
& $[0.50, 1.83]$ & $[-1.73, -0.51]$
& $[-0.09, 1.29]$ & $[-0.98, -0.06]$
\\
HMC
& $[0.19, 1.51]$ & $[-1.71, -0.55]$
& $[0.06, 1.08]$ & $[-0.99, -0.05]$
\\
NUTS
& $[0.39, 1.31]$ & $[-1.41, -0.67]$
& $[0.18, 0.68]$ & $[-1.02, 0.06]$
\\
MALA
& $[0.25, 1.56]$ & $[-1.72, -0.57]$
& $[0.05, 1.09]$ & $[-0.99, -0.03]$
\\
\hline
\hline
& $\beta_{5}$ & $\beta_{6}$ & $\beta_{7}$ & $\beta_{8}$
\\
\midrule
MPLE
& $[-0.06, 0.31]$ & $[-0.54, -0.18]$
& $[-0.08, 0.24]$ & $[-0.28, 0.08]$
\\
GS4Cox
& $[-0.06, 0.31]$ & $[-0.53, -0.18]$
& $[-0.09, 0.24]$ & $[-0.27, 0.08]$
\\
MH
& $[-0.55, 0.55]$ & $[-0.97, 0.14]$
& $[-0.38, 0.98]$ & $[-0.48, 0.55]$
\\
HMC
& $[-0.44, 0.72]$ & $[-0.78, 0.40]$
& $[-0.36, 0.56]$ & $[-0.76, 0.34]$
\\
NUTS
& $[-0.57, 0.72]$ & $[-0.78, -0.03]$
& $[-0.37, 0.27]$ & $[-0.39, 0.13]$
\\
MALA
& $[-0.47, 0.72]$ & $[-0.78, 0.36]$
& $[-0.37, 0.57]$ & $[-0.77, 0.31]$
\\
\bottomrule
\end{tabular}
\end{sc}
\end{small}
\end{center}
\vskip -0.1in
\end{table}

\begin{table}[htbp]
\caption{
Parameter estimates from synthetic data experiments where a learning rate is set to $10$.
The compared methods are MPLE, GS4Cox, MH, HMC, NUTS, and MALA.
}
\label{app-tab:numerical-estimates-lr10}
\vskip 0.15in
\begin{center}
\begin{small}
\begin{sc}
\begin{tabular}{lrrrrrrrr}
\toprule
& $\beta_{1}$ & $\beta_{2}$ & $\beta_{3}$ & $\beta_{4}$
& $\beta_{5}$ & $\beta_{6}$ & $\beta_{7}$ & $\beta_{8}$\\
\midrule
True Parameter
& $1.00$ & $-1.00$ & $0.50$ & $-0.50$
& $0.30$ & $-0.30$ & $0.10$ & $-0.10$
\\
MPLE
& $1.02$ & $-1.05$ & $0.51$ & $-0.56$
& $0.12$ & $-0.36$ & $0.08$ & $-0.10$
\\
GS4Cox
& $1.01$ & $-1.05$ & $0.51$ & $-0.56$
& $0.12$ & $-0.36$ & $0.08$ & $-0.10$
\\
MH
& $1.01$ & $-1.05$ & $0.52$ & $-0.57$
& $0.12$ & $-0.35$ & $0.07$ & $-0.10$
\\
HMC
& $1.02$ & $-1.05$ & $0.51$ & $-0.56$
& $0.12$ & $-0.36$ & $0.08$ & $-0.10$
\\
NUTS
& $1.01$ & $-1.05$ & $0.51$ & $-0.57$
& $0.13$ & $-0.37$ & $0.07$ & $-0.09$
\\
MALA
& $-$ & $-$ & $-$ & $-$
& $-$ & $-$ & $-$ & $-$
\\
\bottomrule
\end{tabular}
\end{sc}
\end{small}
\end{center}
\vskip -0.1in
\end{table}

\begin{table}[htbp]
\caption{
95\% confidence intervals and credible intervals (Bayesian methods) from synthetic data experiments where a learning rate is set to $10$.
The compared methods are MPLE, GS4Cox, MH, HMC, NUTS, and MALA.
}
\label{app-tab:numerical-intervals-lr10}
\vskip 0.15in
\begin{center}
\begin{small}
\begin{sc}
\begin{tabular}{lrrrr}
\toprule
& $\beta_{1}$ & $\beta_{2}$ & $\beta_{3}$ & $\beta_{4}$
\\
\midrule
MPLE
& $[0.80, 1.23]$ & $[-1.27, -0.84]$
& $[0.33, 0.70]$ & $[-0.74, -0.38]$
\\
GS4Cox
& $[0.79, 1.22]$ & $[-1.25, -0.83]$
& $[0.35, 0.72]$ & $[-0.73, -0.39]$
\\
MH
& $[0.92, 1.09]$ & $[-1.13, -0.99]$
& $[0.46, 0.57]$ & $[-0.62, -0.50]$
\\
HMC
& $[0.94, 1.09]$ & $[-1.12, -0.99]$
& $[0.45, 0.58]$ & $[-0.61, -0.50]$
\\
NUTS
& $[0.96, 1.07]$ & $[-1.12, -0.99]$
& $[0.44, 0.57]$ & $[-0.62, -0.51]$
\\
MALA
& $-$ & $-$
& $-$ & $-$
\\
\hline
\hline
& $\beta_{5}$ & $\beta_{6}$ & $\beta_{7}$ & $\beta_{8}$
\\
\midrule
MPLE
& $[-0.06, 0.31]$ & $[-0.54, -0.18]$
& $[-0.08, 0.24]$ & $[-0.28, 0.08]$
\\
GS4Cox
& $[-0.06, 0.32]$ & $[-0.54, -0.18]$
& $[-0.09, 0.22]$ & $[-0.28, 0.07]$
\\
MH
& $[0.06, 0.18]$ & $[-0.42, -0.29]$
& $[0.01, 0.14]$ & $[-0.17, -0.03]$
\\
HMC
& $[0.07, 0.18]$ & $[-0.43, -0.30]$
& $[0.02, 0.13]$ & $[-0.16, -0.04]$
\\
NUTS
& $[0.07, 0.19]$ & $[-0.41, -0.32]$
& $[0.02, 0.13]$ & $[-0.14, -0.03]$
\\
MALA
& $-$ & $-$
& $-$ & $-$
\\
\bottomrule
\end{tabular}
\end{sc}
\end{small}
\end{center}
\vskip -0.1in
\end{table}

\subsection{Additional Actual Data Experiment Results}
\label{app-subsec:additional-actual-results}

First, Table~\ref{app-tab:actual-data-intervals} presents the 95\% confidence intervals and credible intervals for each method obtained from the experiment using the lung dataset from the survival R package with covariate standardization applied (Section~\ref{sec:actual-experiment}).
This shows that while the credible intervals of the other Bayesian methods deviated from confidence intervals corresponding to the MPLE, GS4Cox yielded intervals that were nearly identical to confidence intervals corresponding to the MPLE.

\begin{table}[htbp]
\caption{
95\% confidence intervals and credible intervals (Bayesian methods) for the lung dataset with covariate standardization applied.
The compared methods are MPLE, GS4Cox, MH, HMC, NUTS, MALA, and Cox-PG.
Coefficients are $\beta_{1}$ for age, $\beta_{2}$ for sex, $\beta_{3}$ for ph.ecog, $\beta_{4}$ for ph.karno, $\beta_{5}$ for pat.karno, $\beta_{6}$ for meal.cal, and $\beta_{7}$ for wt.loss.
}
\label{app-tab:actual-data-intervals}
\vskip 0.15in
\begin{center}
\begin{small}
\begin{sc}
\begin{tabular}{lrrrr}
\toprule
& $\beta_{1}$ & $\beta_{2}$ & $\beta_{3}$ & $\beta_{4}$
\\
\midrule
MPLE
& $[-0.11, 0.31]$ & $[-0.46, -0.08]$
& $[0.22, 0.86]$ & $[0.01, 0.57]$
\\
GS4Cox
& $[-0.10, 0.32]$ & $[-0.45, -0.09]$
& $[0.22, 0.87]$ & $[0.00, 0.56]$
\\
MH
& $[-0.13, 0.23]$ & $[-0.51, -0.12]$
& $[0.23, 0.83]$ & $[0.04, 0.60]$
\\
HMC
& $[-0.10, 0.32]$ & $[-0.49, -0.08]$
& $[0.29, 0.84]$ & $[0.05, 0.57]$
\\
NUTS
& $[-0.18, 0.22]$ & $[-0.44, -0.14]$
& $[0.40, 0.93]$ & $[0.07, 0.62]$
\\
MALA
& $[-0.08, 0.28]$ & $[-0.47, -0.08]$
& $[0.37, 0.76]$ & $[0.09, 0.54]$
\\
Cox-PG
& $[0.10, 0.22]$ & $[-0.27, -0.09]$
& $[0.36, 0.56]$ & $[0.19, 0.28]$
\\
\hline
\hline
& $\beta_{5}$ & $\beta_{6}$ & $\beta_{7}$ & 
\\
\midrule
MPLE
& $[-0.42, 0.06]$ & $[-0.20, 0.22]$
& $[-0.39, 0.01]$ &
\\
GS4Cox
& $[-0.41, 0.05]$ & $[-0.21, 0.22]$
& $[-0.39, 0.02]$ &
\\
MH
& $[-0.37, 0.03]$ & $[-0.22, 0.16]$
& $[-0.39, -0.07]$ &
\\
HMC
& $[-0.43, 0.05]$ & $[-0.17, 0.22]$
& $[-0.39, 0.00]$ &
\\
NUTS
& $[-0.33, -0.06]$ & $[-0.31, 0.23]$
& $[-0.43, -0.09]$ &
\\
MALA
& $[-0.39, 0.08]$ & $[-0.18, 0.20]$
& $[-0.40, 0.00]$ &
\\
Cox-PG
& $[-0.32, -0.02]$ & $[-0.09, 0.01]$
& $[-0.27, -0.16]$ &
\\
\bottomrule
\end{tabular}
\end{sc}
\end{small}
\end{center}
\vskip -0.1in
\end{table}

\begin{figure}[htbp]
\vskip 0.2in
\begin{center}
\centerline{\includegraphics[width=\textwidth]{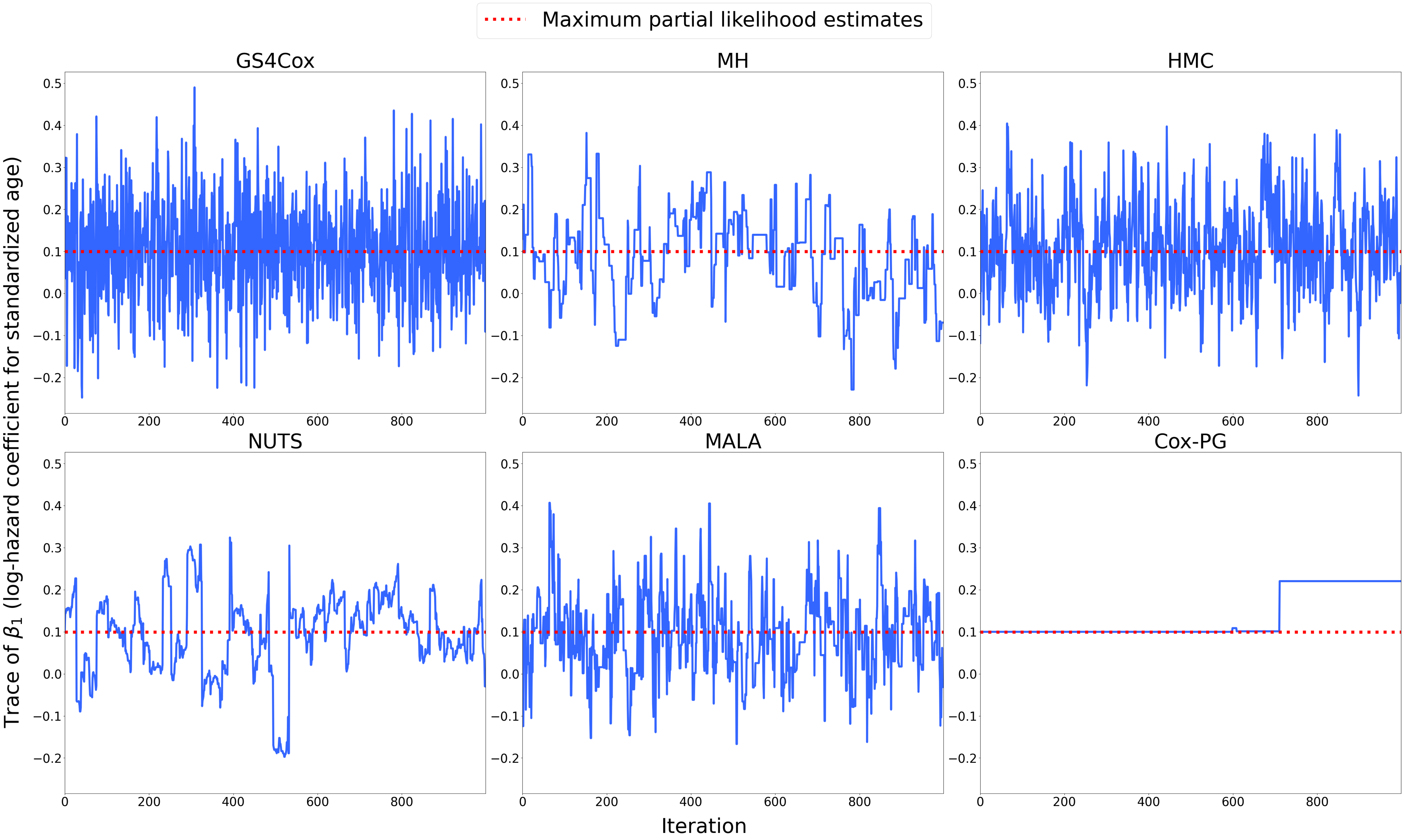}}
\caption{
Trace plots of the log-hazard coefficient for standardized age from the lung dataset experiments.
The dotted lines indicate the maximum partial likelihood estimates.
The compared methods are GS4Cox, MH, HMC, NUTS, MALA, and Cox-PG.
}
\label{app-fig:real-data-trace-plot}
\end{center}
\vskip -0.2in
\end{figure}

Second, Figure~\ref{app-fig:real-data-trace-plot} shows the trace plot for the regression coefficient for age in the experiment with the lung dataset from the survival R package, where covariate standardization was applied.
This suggests that GS4Cox, HMC, and MALA reached stationarity early, while MH, NUTS, and Cox-PG exhibited instability.

Finally, Table~\ref{app-tab:actual-data-intervals-raw} reports the 95\% confidence intervals and credible intervals for each method obtained from the experiment using the lung dataset from the survival R package, without covariate standardization.
The results suggest that GS4Cox, even without standardization, produced intervals that were very close to the confidence intervals corresponding to the MPLE, highlighting the effectiveness of our method.

\begin{table}[htbp]
\caption{
95\% confidence intervals and credible intervals (Bayesian methods) for the lung dataset.
The compared methods are MPLE, GS4Cox, MH.
Coefficients are $\beta_{1}$ for age, $\beta_{2}$ for sex, $\beta_{3}$ for ph.ecog, $\beta_{4}$ for ph.karno, $\beta_{5}$ for pat.karno, $\beta_{6}$ for meal.cal, and $\beta_{7}$ for wt.loss.
}
\label{app-tab:actual-data-intervals-raw}
\vskip 0.15in
\begin{center}
\begin{small}
\begin{sc}
\begin{tabular}{lrrrr}
\toprule
& $\beta_{1}$ & $\beta_{2}$ & $\beta_{3}$ & $\beta_{4}$
\\
\midrule
MPLE
& $[-0.01, 0.03]$ & $[-0.95, -0.16]$
& $[0.30, 1.18]$ & $[0.00, 0.04]$
\\
GS4Cox
& $[-0.01, 0.03]$ & $[-0.94, -0.17]$
& $[0.27, 1.20]$ & $[0.00, 0.05]$
\\
MH
& $[-0.01, 0.03]$ & $[-0.82, -0.17]$
& $[0.45, 1.06]$ & $[0.01, 0.04]$
\\
\hline
\hline
& $\beta_{5}$ & $\beta_{6}$ & $\beta_{7}$ & 
\\
\midrule
MPLE
& $[-0.03, 0.00]$ & $[-0.00, 0.00]$
& $[-0.03, 0.00]$ &
\\
GS4Cox
& $[-0.03, 0.00]$ & $[-0.00, 0.00]$
& $[-0.03, -0.00]$ &
\\
MH
& $[-0.03, 0.00]$ & $[-0.00, 0.00]$
& $[-0.02, -0.00]$ &
\\
\bottomrule
\end{tabular}
\end{sc}
\end{small}
\end{center}
\vskip -0.1in
\end{table}

\end{document}